\documentclass[12pt,a4paper,twoside]{book}
\usepackage[text={408pt,592pt},centering]{geometry}

\usepackage[T1]{fontenc}
\usepackage{lmodern} 
\usepackage[utf8]{inputenc}
\usepackage[english]{babel}
\usepackage{rsfso}

\usepackage{enumitem}
\usepackage{csquotes}
\usepackage[backend=biber,citestyle=numeric-comp,maxbibnames=5,sorting=none, sortcites=true,doi=false,url=true,abbreviate=true,giveninits=true]{biblatex}

\usepackage{layout}

\usepackage{pdfpages}

\usepackage{amsmath}
\usepackage{amsthm}
\usepackage{xparse}
\usepackage{graphicx}
\usepackage{xcolor}
\usepackage{amssymb}
\usepackage{physics}
\usepackage{cancel}
\usepackage{mathtools}
\usepackage[normalem]{ulem}

\usepackage{subcaption}

\usepackage{array}
\newcommand{\PreserveBackslash}[1]{\let\temp=\\#1\let\\=\temp}
\newcolumntype{C}[1]{>{\PreserveBackslash\centering}p{#1}}
\newcolumntype{R}[1]{>{\PreserveBackslash\raggedleft}p{#1}}
\newcolumntype{L}[1]{>{\PreserveBackslash\raggedright}p{#1}}

\definecolor{jyo}{RGB}{241,86,63}
\definecolor{jyb}{RGB}{0,41,87}
\definecolor{jyg}{RGB}{173,130,64} 

\usepackage[colorlinks=true]{hyperref}
\hypersetup{
	citecolor=[RGB]{241,86,63},
	linkcolor=[RGB]{0,41,87},
	urlcolor=[RGB]{173,130,64} 
}

\allowdisplaybreaks[1]  

\newcommand{\rt}{{\mathbf{r}}}
\newcommand{\kt}{{\mathbf{k}}}
\newcommand{\ktt}{\mathbf{k}_t}
\newcommand{\qt}{{\mathbf{q}}}
\newcommand{\pt}{{\mathbf{p}}}
\newcommand{\xt}{{\mathbf{x}}}
\newcommand{\cxt}{{\conj{\mathbf{x}}}}

\newcommand{\bt}{{\mathbf{b}}}
\newcommand{\pp}{\mathbf{P}}

\newcommand{\xij}[1]{\mathbf{x}_{#1}}

\newcommand{\vecp}[1]{\underline{p_{#1}}}
\newcommand{\Deltat}{{\boldsymbol{\Delta}}}

\newcommand{\ud}{\, \mathrm{d}}
\newcommand{\nc}{{N_\mathrm{c}}}
\newcommand{\nf}{{N_\mathrm{F}}}
\newcommand{\half}{\frac{1}{2}}
\newcommand{\cf}{C_\mathrm{F}}

\newcommand{\B}[1]{\mathbf{#1}}

\newcommand{\conj}[1]{\mathop{\overline{#1}}\nolimits}
\newcommand{\conz}{\conj{0}}
\newcommand{\cono}{\conj{1}}
\newcommand{\cont}{\conj{2}}
\newcommand{\contrip}{\conj{0} \conj{1} \conj{2}}

\renewcommand{\Im}{\operatorname{Im}}

\newcommand{\gev}{\ \textrm{GeV}}

\newcommand{\qs}{Q_\mathrm{s}}

\newcommand{\lqcd}{\Lambda_{\mathrm{QCD}}}
\newcommand{\as}{\alpha_{\mathrm{s}}}

\newcommand{\aem}{\alpha_{\mathrm{em}}}

\newcommand{\fig}{Fig.~}

\newcommand{\eq}{Eq.~}
\newcommand{\se}{Sec.~}
\newcommand{\eqs}{Eqs.~}

\newcommand{\xbj}{{x_\text{Bj}}}
\newcommand{\Yobk}{{Y_{0, \text{BK}}}}
\newcommand{\Yoif}{{Y_{0, \text{if}}}}

\newcommand{\Ytwoplus}{{Y_{2}^{+}}}

\newcommand{\xpom}{{x_\mathbb{P}}}
\newcommand{\sigmaltnlo}{\sigma_{L,T}^{\textrm{NLO}}}
\newcommand{\sigmaltdip}{\sigma_{L,T}^{\textrm{dip}}}
\newcommand{\sigmaltqg}{\sigma_{L,T}^{qg}}
\newcommand{\sigmaltqgs}{\sigma_{L,T}^{qg, \textrm{sub.}}}
\newcommand{\sigmaltqgu}{\sigma_{L,T}^{qg, \textrm{unsub.}}}
\newcommand{\sigmalt}[1]{\sigma_{L,T}^{#1}}
\newcommand{\qqbarg}{q \bar{q} g}
\newcommand{\zmin}{z_{2,\textrm{min}}}
\newcommand{\plusk}{k^{+}}
\newcommand{\plusq}{q^{+}}
\newcommand{\plusqpr}{{q'}^{+}}
\newcommand{\plum}[1]{#1^{+}}
\newcommand{\kzero}{k^{+}_0}
\newcommand{\kone}{k^{+}_1}
\newcommand{\ktwo}{k^{+}_2}
\newcommand{\qplus}{q^{+}}

\newcommand{\kplus}{k^{+}}
\newcommand{\kprimplus}{k'^{+}}
\newcommand{\pplus}{p^{+}}
\newcommand{\mx}{M_X}

\newcommand{\gamlam}{{\gamma^*_\lambda}}
\newcommand{\opse}{{\Hat{S}_E}}
\newcommand{\tree}{\textrm{tree}}

\newcommand{\ical}{\mathcal{I}}

\newcommand{\kcal}{\mathcal{K}}
\newcommand{\ocal}{\mathcal{O}}
\newcommand{\vcal}{\mathcal{V}}

\newcommand{\besj}{{\mathrm{J}}}
\newcommand{\besk}{{\mathrm{K}}}
\newcommand{\Li}[1]{\mathrm{Li}_2 \left({#1}\right)}

\newcommand{\der}{\mathrm{d}}

\makeatletter
\def\mathcolor#1#{\@mathcolor{#1}}
\def\@mathcolor#1#2#3{%
	\protect\leavevmode
	\begingroup
	\color#1{#2}#3%
	\endgroup
}
\makeatother

\renewbibmacro{in:}{}
\DeclareFieldFormat[article,inbook,incollection,inproceedings,thesis,unpublished]{title}{#1\isdot}
\DeclareFieldFormat{pages}{#1}
\DeclareNameAlias{sortname}{first-last}
\AtEveryBibitem{\ifentrytype{inproceedings}{\clearfield{year}}{}}
\AtEveryBibitem{\ifentrytype{unpublished}{\clearfield{year}}{}}
\AtEveryBibitem{\ifentrytype{online}{\clearfield{year}}{}}
\DefineBibliographyExtras{english}{%

}

\newbool{titlefirst}

\newbibmacro*{titlefirst}{%
	\ifboolexpr{
		test {\iffieldundef{title}}
		and
		test {\iffieldundef{subtitle}}
	}
	{}
	{\printtext[title]{%
			\bf%
			\printfield[titlecase]{title}%
			\setunit{\subtitlepunct}%
			\printfield[titlecase]{subtitle}}%
		\newunit}%
	\printfield{titleaddon}}

\AtEveryBibitem{
	\ifbool{titlefirst}{
		\renewbibmacro*{begentry}{%
			\usebibmacro{titlefirst}\newline\nopunct\newunit}
		\renewbibmacro*{title}{} 
	}
}

\DeclareFieldFormat{doilink}{\iffieldundef{doi}{#1}{\href{https://doi.org/\thefield{doi}}{#1}}}
\DeclareFieldFormat[article,periodical]{volume}{\mkbibbold{#1}}
\DeclareFieldFormat[article,periodical]{journaltitle}{#1}
\DeclareFieldFormat[article,periodical]{pages}{#1}

\usepackage{xpatch}
\xpatchbibdriver{article}
{\usebibmacro{journal+issuetitle}%
	\newunit
	\usebibmacro{byeditor+others}%
	\newunit
	\usebibmacro{note+pages}}
{\printtext[doilink]{%
		\usebibmacro{journal+issuetitle}%
		\newunit
		\usebibmacro{note+pages}}}
{}{}

\DeclareBibliographyCategory{paperit}
\addtocategory{paperit}{oma1,oma2,oma3}
\nocite{oma1,oma2,oma3}
\defbibnote{descr}{
	This thesis consists of an introductory part and of the following publications:
}
\defbibnote{authorcontr}{
	The author performed the numerical calculations, drew the figures, and participated in the writing and editing of the manuscript for the Article~\cite{oma1}.
	For the Article~\cite{oma2}, the author performed the numerical implementation and comparison of the cross sections calculated in~\cite{oma2} to the ones calculated in Refs.~\cite{Beuf:2016wdz, Beuf:2017bpd}.
	For the Article~\cite{oma3}, the author implemented and performed the fits, drew the figures, and wrote the first draft.
	The implementation used was based on a software component written by collaborator H. Mäntysaari.
	Chapter~\ref{ch:ddis} of this thesis is related to work done in collaboration with G.~Beuf, T.~Lappi, Y.~Mulian, and H.~Mäntysaari, which is in preparation for publication.
}

\begin{document}
	
	
	\frontmatter
	
	\includepdf[pages=-]{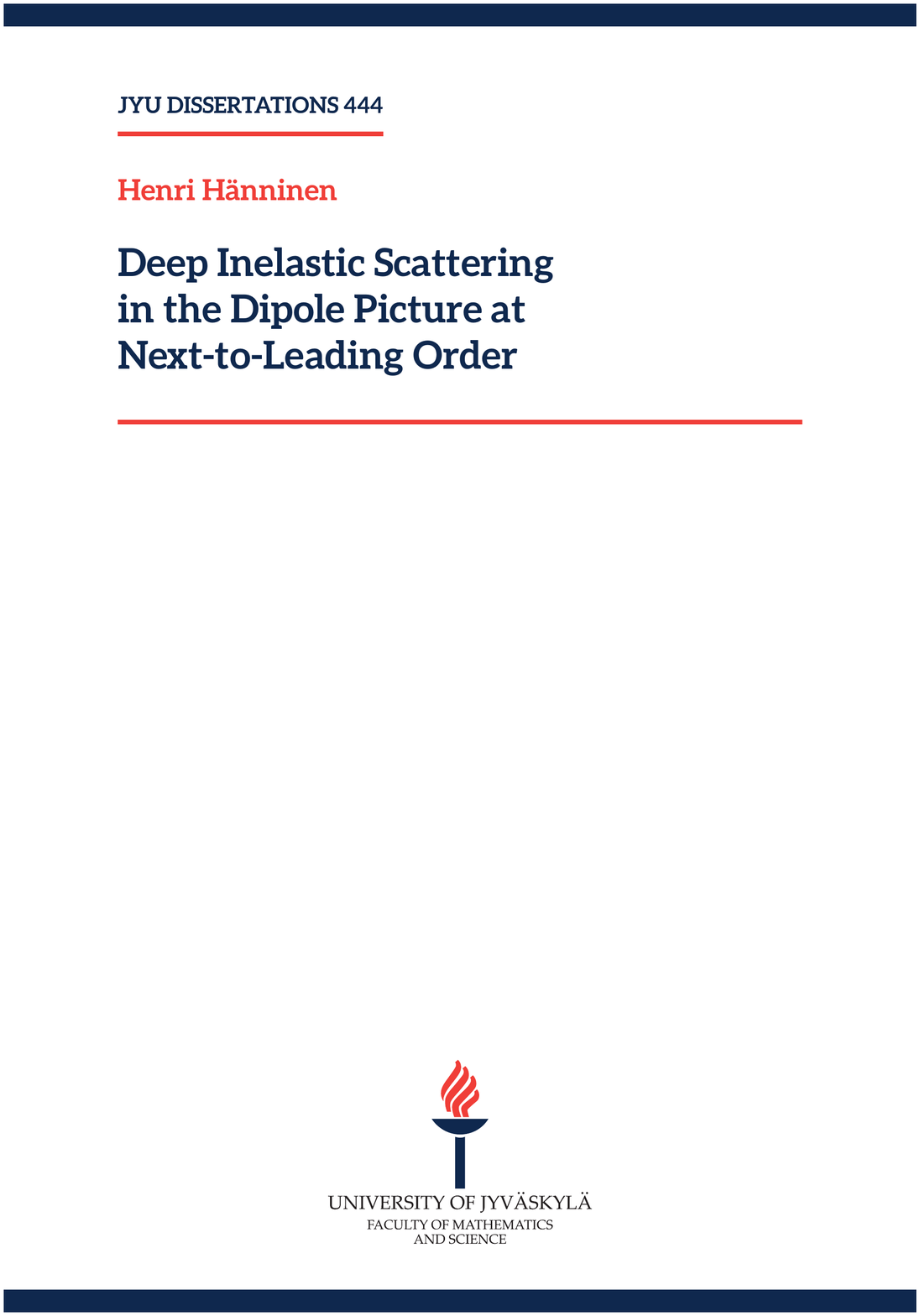}
	\setcounter{page}{1}
	
	\pagestyle{plain}
	
	\pdfbookmark{Abstract}{abstract}
	\phantom{.}
	\vfill
	\noindent\textbf{\LARGE Abstract}
	
	
	\bigskip\noindent 
	This thesis studies gluon saturation in hadronic matter at high energy by calculating next-to-leading order (NLO) corrections to inclusive and diffractive deep inelastic scattering cross sections in the Color Glass Condensate (CGC) effective field theory.
		
	We demonstrate that the large soft gluon logarithm is correctly factorized into the Balitsky--Kovchegov (BK) renormalization group equation by accurately connecting the NLO scattering kinematics to the rapidity scale of the dipole amplitude in the scattering. This brings the perturbative expansion under control and enables us to do precision comparisons between theory and data.
	We fit the initial condition of the BK evolution equation to HERA inclusive deep inelastic scattering data by	
	combining of the NLO accuracy inclusive cross sections with beyond leading order BK evolution prescriptions. This results in the state-of-the-art accuracy comparison between CGC theory and HERA data, and determination of the dipole amplitude initial shape which is a necessary input for all NLO CGC phenomenology. In the introductory part of this thesis, the effect of the NLO BK equation on the fits is assessed, and an alternative form for the NLO loop correction to the inclusive cross sections is derived which enables the consistent setting of the dipole amplitude rapidity scale in the NLO corrections.
	
	The underlying mechanism of diffraction in particle scattering is still unknown, with multiple competing pictures. Diffraction is studied in this thesis in the CGC formalism, and we calculate the tree-level $q \bar q g$ NLO contribution to the diffractive deep inelastic scattering structure functions where the $q \bar q g$ Fock state scatters off the target and becomes the diffractively produced system. This contribution has previously been known in the literature only in leading $\log(Q^2)$ accuracy valid at large $Q^2$, and only for the structure function $F_T^D$. The $q \bar q g$ contribution to both structure functions $F_T^D$ and $F_L^D$ are presented in full NLO accuracy.
	\vfill
	
	\cleardoublepage
	
	\pagestyle{plain}
	
	\pdfbookmark{Tiivistelmä}{tiiviste}
	\phantom{.}
	\vfill
	\noindent\textbf{\LARGE Tiivistelmä}
	
	\bigskip\noindent 
	Tässä väitöskirjassa tutkitaan hadronisen aineen gluonikyllästymistä korkealla energialla laskemalla toisen kertaluvun häiriöteorian korjauksia kaikenkattavaan ja diffraktiiviseen syvän epäelastisen sironnan vaikutusalaan efektiivisessä värila\-sikondensaatti-kenttäteoriassa.
	
	Näytämme, että suuri matalan pitkittäisliikemäärän gluoneista johtuva lo\-gartmi faktorisoituu oikein Balitsky--Kovchegov (BK) -renormalisaatio\-ryh\-mäyh\-tä\-löön, kun toisen kertaluvun sironnan kinematiikka yhdistetään tarkasti sironnan dipoliamplitudin rapiditeettiskaalaan. Tämä tuo häiriöteorian sarjakehitel\-män hallin\-taan ja mahdollistaa tarkkuusvertaamisen teorian ja kokeellisten tulosten välillä.
	Sovitamme BK-evoluutioyhtälön alkuehdon HERA-kokeen syvän epäelastisen sironnan kokonaisvaikutusalan mittaustuloksiin yhdistämällä toisen kertaluvun tarkkuuden vaikutusalalaskun yli johtavan kertaluvun tarkkuuden BK-evoluutio yhtälöiden kanssa.
	Tämä tuottaa huipputarkkuuden yhteensopi\-vuus\-testin väri\-lasikondensaatti-teorian ja mittaustulosten välillä, sekä määrityksen dipoliamplitudin evoluution alkumuo\-dolle, joka on välttämätön syöte hiukkastörmäysten kuvaamiseen värilasikondensaatti-teoriassa.
	Tässä väitöskirjan johdanto-osassa arvioidaan toisen kertaluvun tarkkuuden BK-yhtälön vaikutusta sovitustuloksiin, ja johdetaan vaih\-toehtoinen muoto toisen kertaluvun sil\-muk\-ka\-korjaukselle syvän epä\-e\-las\-ti\-sen si\-ronnan kokonaisvaikutusalaan, mikä mahdollistaa joh\-donmukaisen rapiditeetti\-skaa\-lan asettamisen toisen kertaluvun kor\-jauk\-sissa.
	
	Hiukkassironnassa tapahtuvan diffraktion perustavaa mekanismia ei vielä tunneta syvällisesti ja kilpailevia kuvia prosessille on useita. Diffraktiota tutki\-taan tässä väitöskirjassa värilasikondensaatti-teoriassa, jossa laskemme toisen kertaluvun puutason $q \bar q g$-korjauksen diffraktiivisen syvän epäelastisen sironnan rakennefunktioihin, missä $q \bar q g$ Fock-tila siroaa kohtiosta ja muodostaa diffraktiivisen systeemin. Kyseinen osuus vaikutusalasta on aiemmin tunnettu vain johtavan $\log(Q^2)$ tark\-kuu\-dessa ja vain $F_T^D$-rakennefunktiolle. Tämä $q \bar q g$-korjaus esitetään sekä $F_T^D$- että $F_L^D$-rakennefunktioille täydessä toisen kertaluvun tark\-kuu\-dessa.
	\vfill
	
	\cleardoublepage
	\phantom{.}
	\vfill
	\begin{description}[labelsep=0.25em,labelwidth=2.5cm,leftmargin=4.0cm,itemindent=0.0cm]
		\item[Author]
		Henri Hänninen\\
		Department of Physics\\
		University of Jyväskylä\\
		Finland
		\item[Supervisors]
		Prof.~Tuomas~Lappi\\
		Department of Physics\\
		University of Jyväskylä\\
		Finland
		
		\vspace{\itemsep}
		Dr.~Heikki~Mäntysaari\\
		Department of Physics\\
		University of Jyväskylä\\
		Finland
		\item[Reviewers]
		Prof.~Anna~Sta\'sto\\
		Department of Physics\\
		Penn State University\\
		USA
		
		\vspace{\itemsep}
		Prof.~Bo-Wen~Xiao\\
		School of Science and Engineering\\
		The Chinese University of Hong Kong\\
		China
		\item[Opponent]
		Prof.~Krzysztof~Golec-Biernat\\
		Division of Theoretical Physics\\
		Institute of Nuclear Physics PAN\\
		Poland
	\end{description}
	\vfill
	
	\cleardoublepage
	
	\pdfbookmark{Preface}{preface}
	\chapter*{Preface}
	The research reported in this thesis has been carried out at the University of Jyväskylä from April 2017 to September 2021.
	This work was supported by the European Union’s Horizon 2020 research and innovation programme by the European Research Council (ERC, grant agreement No. ERC-2015-CoG-681707), and in the very final stages by the Academy of Finland (project 321840).
	The computing resources provided by CSC – IT Center for Science in Espoo, Finland, and the Finnish Grid and Cloud Infrastructure (persistent identifier urn:nbn:fi:research-infras-2016072533) were crucial for the completion of this work.
	
	I have been delighted and fortunate to have had the opportunity to work in the excellent supervision of Prof. Tuomas Lappi and Dr. Heikki Mäntysaari, both of whom I thank for their expert guidance and support.
	Secondly, I thank Dr. Guillaume Beuf for all the enlightening discussions and correspondence that helped making many aspects of this thesis possible.
	I also wish to thank Dr.~Bertrand Ducloué, Dr.~Yair Mulian, Dr.~Risto Paatelainen, and Dr.~Yan Zhu for fruitful collaboration. 
	I am also thankful to Prof Kari J. Eskola, Prof. Tero Heikkilä, and Prof. Kimmo Kainulainen for their lessons in theoretical physics and for their guidance.
	I thank Prof. Anna Sta\'sto and Prof. Bo-Wen~Xiao for reviewing this manuscript, and Prof. Krzysztof~Golec-Biernat who has agreed to act as my opponent.
	
	Friends and colleagues have had an important, if less direct, part in the realization of this thesis. Special thanks go to Kalle Kansanen and Mikko Kuha for sharing this journey of becoming a young scientist and for their friendship. I had the privilege to begin this journey, in the beforetimes, in the grad student office YFL 353 known as Holvi. I thank the Holvi collaborators Lotta Jokiniemi, Mikko Kivekäs, Mikko Kuha, Miha Luntinen, Topi Löytäinen, Petja Paakkinen, Jani Penttala, Pekka Pirinen, and Oskari Saarimäki
	for the engaging and inspiring \sout{working environment} group chat.
	I also thank Toni Ikonen, Joonas Niinikoski, and Timo Schultz for the adventures into advanced mathematics and the camaraderie during the tumultuous first years under the accelerated physics study programme.
	
	I thank my mother and father for their support in all my endeavors. My father's book recommendations of Gamow's and Feynman's works are among the earliest memories of interest in, and excitement about, theoretical physics. I also thank my grandfather Osmo for always nurturing my interest in the natural sciences. Finally, with the deepest appreciation, I thank Laura, Petrus, and Aurora for their love and support.

	\vspace{0.5cm}
	\rightline{Henri Hänninen, September 2021, Jyväskylä}
	
	\vfill
	\clearpage
	\booltrue{titlefirst}
	{
	\printbibliography[category=paperit,title={	\pdfbookmark{List of Publications}{lop} List of Publications},prenote=descr,postnote=authorcontr]
	}
	\boolfalse{titlefirst}
	
	\cleardoublepage
	
	\setcounter{tocdepth}{3}
	{
	\hypersetup{linkcolor={black}}
	\pdfbookmark{\contentsname}{Contents}
	\tableofcontents
	}

	\mainmatter
	
	\chapter{Introduction}

There are some things we know about the proton~\cite{ParticleDataGroup:2020ssz}. Since its discovery just over 100 years ago by Rutherford~\cite{Rutherford:doi:10.1080/14786440608635919,Romer:proton}, we have learned that it is not an elementary particle which means it has an internal structure. Thus, the known properties of the proton must arise from this internal structure. For example the charge, mass, size, and spin of the proton somehow emerge from this structure and the interactions of the constituents of the proton.

We have learned that the proton is composed of three elementary particles called quarks: two up quarks and one down quark. These quarks interact with each other via the strong nuclear force, which is mediated by a massless electric-chargeless gauge boson called the gluon. Quarks, gluons, and their interactions are described by the quantum field theory Quantum Chromodynamics (QCD).
QCD predicts that the gluons exchanged between the quarks can temporarily split into new quark-antiquark pairs, which can produce further gluon emissions and splittings. Thus the proton is expected to have a background presence of quarks and gluons on top of the three valence quarks. The densities of these sea quarks and gluons depend on the energy scale the proton is studied at, and these densities have been determined experimentally as the parton distribution functions. Since the gluons and sea quark-antiquark pairs do not have net electric charge, the charge $+1e$ of the proton is indeed the sum of the charges of the up and down quarks.

How about the rest mass of the proton, does it compose straightforwardly from the rest masses of the up and down quarks? Well, yes, to the extent that around \textit{one percent} of the proton mass is from the rest masses of the three quarks. The remaining $99\%$ of the mass arises from the dynamics --- movement and confinement --- of the quarks and gluons~\cite{Walker-Loud:2018gvh}. A qualitative theory understanding of the decomposition of the missing mass has been known~\cite{Ji:1994av, Walker-Loud:2018gvh}, but only recently quantitative theory calculations for these components of the mass have been calculated with lattice QCD~\cite{Yang:2018nqn}. The proton mass arises from four contributions: the quark condensate $(\sim 9\%)$, the quark energy $(\sim 32\%)$, the gluonic field strength $(\sim 37\%)$, and the anomalous gluonic contribution $(\sim 23\%)$. The first and smallest of the contributions arises from the masses of the valence and sea quarks, and only this contribution would vanish if the quarks were massless. The quark energy and gluonic field strength contributions arise from the kinetic energies of the quarks and gluons, and the final anomalous component is a quantum effect~\cite{Walker-Loud:2018gvh}.

Well maybe the size of the proton as a charged composite particle is uncomplicated to measure and understand? Yes, in the sense that in the last decade there has been more than a $7\sigma$ disagreement
on the charge radius of the proton~\cite{Bernauer:2020ont}. This so-called proton radius puzzle originates from a $4\%$ difference between a newer experiment using muonic hydrogen spectroscopy and older measurements done with regular hydrogen spectroscopy and low-energy electron-proton scattering, both of which were in agreement previously. New experiments are being planned to understand the origin of the discrepancy and to settle the correct charge radius~\cite{Bernauer:2020ont}. 

Surely at least the spin of the proton is comprised as the sum of the spins of the three valence quarks? Alas, no. With global QCD analyses of spin-dependent data --- such as longitudinally polarized proton-proton collision data~\cite{Aschenauer:2013woa, Aschenauer:2015eha} --- it has been determined that the quark spin contribution is roughly $30\%$ of the proton spin~\cite{Aschenauer:2013woa, Aschenauer:2015eha}. The remaining spin arises from gluons and the orbital angular momentum of the quarks and gluons, the determination of which has been more uncertain. Current estimates for the quark and gluon contributions are $\sim 30\% - 40\%$, and $\sim 26\% - 52\%$, respectively, which fall short to produce the proton spin of $\half$~\cite{Aschenauer:2013woa, Aschenauer:2015eha, Kovchegov:2021lvz}. Both new experimental data of polarized beams, and improved theory understanding of the high-energy behavior of these spin contributions~\cite{Leader:2013jra, Kovchegov:2018znm, Kovchegov:2021lvz} will be needed to find the remaining spin.

The work done for this thesis is concerned with the theoretical understanding of the internal structure of the proton at high energy. In this energetic regime QCD predicts that more and more gluons are emitted by the quarks and gluons in the proton. However, as the gluon density rises dramatically with growing energy, at some point the reabsorption of gluons becomes preferred over new emissions, which halts the increase of the gluon density. This phenomenon is known as \textit{saturation}, and in this work it is studied in the Color Glass Condensate (CGC) effective field theory, which describes QCD at very high energies.
The CGC formalism has been used to describe strong interactions in a large variety of scattering processes in e+p, e+A, p+A, and A+A collisions~\cite{Iancu:2003xm, Weigert:2005us, Gelis:2010nm, Albacete:2014fwa, Blaizot:2016qgz, Morreale:2021pnn}.
Further, it has the power to \textit{ab initio} describe thermalization in heavy ion collisions, and the initial conditions of thermalized quark gluon plasma.
The collective high-density behavior of gluons described in the CGC has been discovered to have a surprising connection to gravity, which has recently been proposed as the CGC--Black Hole correspondence~\cite{Dvali:2021ooc}.
More specifically regarding this thesis, state-of-the-art next-to-leading order (NLO) accuracy theory results for proton structure functions are evaluated numerically and used for data comparison. Furthermore, a new analytical calculation of diffractive proton structure functions in NLO accuracy is presented.
These accuracy improvements of the CGC framework theory calculations are necessary to rise to the challenge brought on by the upcoming Electron-Ion Collider (EIC), which will provide state-of-the-art precision measurements of the structure of the proton and nuclei~\cite{Accardi:2012qut, Aschenauer:2017jsk, AbdulKhalek:2021gbh}.

The internal structure of this thesis is as follows. Chapter~\ref{ch:cgc} reviews the basics of Color Glass Condensate effective field theory needed to describe particle scattering processes.
The following chapters discuss the extension of CGC formalism calculations both analytically and numerically to next-to-leading order in perturbative QCD.
In Chapter~\ref{ch:dis} are discussed the recent NLO accuracy CGC framework results for the inclusive deep inelastic scattering (DIS) structure functions of the proton.
Chapter~\ref{ch:fits} considers the application of these proton structure functions to make comparisons between theory and experimental data.
The calculation of diffractive DIS structure functions at NLO in the CGC formalism is explored in Chapter~\ref{ch:ddis}.
Finally, in Chapter~\ref{ch:conclusions} we conclude on the work done for this thesis, including previously unpublished results. The Articles~\cite{oma1},~\cite{oma2}, and~\cite{oma3} are joined as appendices to the thesis.
	\chapter{The Color Glass Condensate effective field theory}
	\label{ch:cgc}

\section{Proton structure at high energy}	
	
	The structure of the proton as seen by a probe particle in a scattering --- and more generally the physical picture of any scattering process --- can vary dramatically in different reference frames and gauges~\cite{Mueller:2001fv}. In the introduction we considered the proton in the parton model where it consisted of quarks and gluons. The parton model picture is valid in the infinite momentum frame. In a perturbation theory calculation of a probe scattering off the parton model proton, the perturbative expansion takes place on the target side of the scattering which amounts to seeing more partonic detail\footnote{For example in deep inelastic scattering, quarks are seen at leading order whereas gluons start to be seen at next-to-leading order.} at higher orders in perturbation theory. In this picture saturation arose as a limit on the occupation number of quarks and gluons in the proton. In the picture of CGC on the other hand, it manifests as a unitarity limit on the scattering from the target. The CGC effective field theory has been established as a well-suited theory tool to study saturation phenomena in QCD~\cite{Gelis:2010nm, Albacete:2014fwa}.
	
	The CGC formalism considers scattering processes in a different picture built on light-front quantization of the probe particle, and the target is seen as a collective force field of its constituents~\cite{Bjorken:1970ah, Brodsky:1997de}. In this case the incoming probe particle becomes perturbatively calculable which gives it a picture of internal structure~\cite{Bjorken:1970ah}, for example a virtual photon consists of color-neutral states of partons and leptons. Simultaneously, the target proton or nucleus is seen to be composed of a strong color-field, which contains the non-perturbative QCD physics of the scattering. 
	These color-fields emerge from the large density of gluons present in the target at high energy. The gluon occupation number is much larger than the commutators between the gluon creation and annihilation operators, which permits the semiclassical description of the color-fields~\cite{Albacete:2014fwa}. Thus the probe particle sees the target as an incredibly strong color-field, which is Lorentz contracted into a thin pancake.
	At high energy where saturation manifests, the target system enters a non-linear weakly-coupled regime of QCD and the energy evolution of the color-field becomes calculable in perturbation theory.
	
	This chapter first discusses in Sec.~\ref{sec:lcpt} the description of the projectile particle in light-front perturbation theory. Scattering off the target color-field and the energy dependence of this scattering process are described in Secs.~\ref{sec:dipole} and~\ref{sec:bk-evol}.

\section{Light-front perturbation theory}
	\label{sec:lcpt}

	To begin delving into the mathematics of the Color Glass Condensate, we must first consider the coordinate system that is conventionally used. Instead of the familiar frame of Minkowskian flat spacetime with the metric $g \! = \! \mathrm{diag}(1,\mkern-4mu -1,\mkern-4mu -1,\mkern-4mu -1)$, light-front perturbation theory (LFPT) calculations are done in coordinates where the $t$- and $z$-axes are on the light-front; this is expressed by
	the light-front metric~\cite{Brodsky:1997de}:
	\begin{equation}
		\tilde{g}  = 
		\begin{pmatrix}
			0 & 0 & 0 & 1 \\
			0 & -1 & 0 & 0 \\
			0 & 0 & -1 & 0 \\
			1 & 0 & 0 & 0			
		\end{pmatrix}.
	\end{equation}
	Together with the light-front metric, the components of a four-vector $x^\mu$ are
	\begin{align*}
		x &\coloneqq (x^+ , \xt, x^-),
		\\
		x^\pm & \coloneqq \frac{1}{\sqrt{2}} (x^0 \pm x^3),
		\\
		\xt & \coloneqq (x^1, x^2),
	\end{align*}
	i.e. the time and $x^3$-components of $x^\mu$ are mixed, and the transverse components are unchanged.
	With the above, the inner product under the light-front metric is: $x \cdot y = x_\mu y_\nu \tilde{g}^{\mu \nu} = x^+ y^- + x^- y^+ - \B{x} \cdot \B{y}$. In these light-front coordinates, the $x^+$ component is called light-front time, the $x^-$ component is light-front position and in momentum space the $k^-$ component is light-front energy~\cite{Brodsky:1997de}.

	Next, we will collect the essential definitions and conventions of the LFPT calculations done in Refs.~\cite{Beuf:2016wdz,Beuf:2017bpd}, which will be used throughout this thesis and especially in the calculations performed in Ch.~\ref{ch:ddis}. The required definitions are used in LFPT calculations that are performed in the mixed space of transverse positions $\xt$, as seen above, and of longitudinal momenta $\kplus$; this phase space comes up in the Fourier transformation of transverse momenta: $(\kt, \kplus) \to (\xt, \kplus)$. For a total and rigorous definition of the full light-front quantization and light-front perturbation theory calculation rules and conventions, we refer the reader to the Refs.~\cite{Beuf:2016wdz,Beuf:2017bpd, oma2}.
	
	We begin by considering the description of the physical state of a photon on the light-front, which is the probe particle in the scattering processes considered in this thesis. Specifically, in Chs.~\ref{ch:dis} and~\ref{ch:ddis} we discuss calculations of virtual photon-proton scattering cross sections using the formalism reviewed in this chapter. In light-front perturbation theory the photon state is written as a Fock state expansion using light-front wavefunctions (LFWF)\footnote{For conciseness, this is shortened to wavefunction many times in this thesis.} $\tilde{\Psi}_{\gamma^{*}_\lambda \rightarrow X}$ as follows:
	\begin{multline}
		\label{eq:fock-gamma}
		\ket{\gamma^*_\lambda(\qplus, \qt; Q^2)_H \vphantom{\se}}
		=
		\sqrt{Z_{\gamlam}}
		\Bigg\lbrace
		\textrm{Non-QCD Fock states}
		\\
		+
		\widetilde{\sum_{q_0 \bar{q}_1 ~ \textrm{F. states}}}
		\tilde{\Psi}_{\gamma^{*}_\lambda \rightarrow q_0 \Bar{q}_1}
		\tilde{b}_0^\dagger \tilde{d}_1^\dagger \ket{0}
		\\
		+
		\widetilde{\sum_{q_0 \bar{q}_1 g_2 ~ \textrm{F. states}}}
		\tilde{\Psi}_{\gamma^{*}_\lambda \rightarrow q_0 \Bar{q}_1 g_2}
		\tilde{b}_0^\dagger \tilde{d}_1^\dagger \tilde{a}_2^\dagger \ket{0}
		+ \cdots
		\Bigg\rbrace,
	\end{multline}
	for a photon with four-momentum $q$ and virtuality $Q^2 = -q^2$. The subscript $H$ denotes that the dressed state is in the Heisenberg picture~\cite{Beuf:2016wdz}. The non-QCD basis states are composed of colorless particles such as photons and leptons and so can be neglected, since they will not scatter off the color-field of the target --- this will be discussed in the next section. The remaining two expansion contributions that are shown start at different orders in perturbation theory: the quark-antiquark contributions denoted by $q\bar q$ start at the order of the electromagnetic coupling $e$, and the quark-antiquark-gluon ($q \bar q g$) contributions at order $e g$, where $g$ is the coupling of the strong interaction. Thus only the former contributes at leading order (LO) in the perturbative expansion, and the latter starts at next-to-leading order (NLO). The remaining terms in the Fock basis expansion that are shortened to dots ($\cdots$) start only at order $e g^2$, i.e. next-to-next-to-leading order (NNLO), and do not contribute to the calculations discussed in this thesis. The photon LFWF normalization is of the order $Z_{\gamma^*} = 1 + \ocal(e^2)$, and so can be dropped~\cite{Beuf:2017bpd}.

	The LFWFs are calculated using conventional quantum mechanical perturbation theory, the light-front rules for which can be found in Refs.~\cite{oma2, Beuf:2016wdz, Beuf:2017bpd}, and in a more pedagogical detail in Refs.~\cite{Brodsky:1997de, Kovchegov:2012mbw}. For example, the LFWF for the virtual photon splitting into a quark-antiquark dipole is calculated as:
	\begin{equation}
		\tilde{\Psi}_{\gamma^{*}_\lambda \rightarrow q_0 \Bar{q}_1}
		=
		\frac{\bra{q_0 \bar q_1} {\hat V}_I(0) \ket{\gamma^*} }{k_{\gamma^*}^- - k^-_{q_0 \bar q_1} + i \varepsilon}
		+ \cdots
		,
	\end{equation} 
	where ${\hat V}_I(0)$ is the interaction operator at the moment of the scattering $x^+ = 0$, $k_{\gamma^*}^-$ and $k^-_{q_0 \bar q_1}$ are the light-front energies of the incoming $\ket{\gamma^*}$ and outgoing $\ket{q \bar q}$ states, and beyond leading order contributions are represented by the dots.

	The $\ket{q \bar q}$ and $\ket{q \bar q g}$ contribution expressions of Eq.~\eqref{eq:fock-gamma} are further composed as follows. The creation operators of the quark, antiquark and gluon are $\tilde{b}^\dagger, \tilde{d}^\dagger,$ and $\tilde{a}^\dagger$, respectively, using the shorthand $\tilde{b}_0^\dagger \coloneqq \tilde{b}^\dagger(\kplus_0, \xt_0, h_0, \alpha_0)$.
	The notation $\widetilde{\sum}$ denotes the sum over the quantum numbers of each parton in the Fock state and a phase-space integration~\cite{Beuf:2016wdz}:
	\begin{align}
		\widetilde{\sum_{q_0 \bar{q}_1 ~ \textrm{F. state}}}
		\coloneqq &
		\sum_{h_0,\alpha_0,f_0} \sum_{h_1,\alpha_1,f_1}
		\prod_{i=0}^{1} \left[
		\int_{-\infty}^{\infty} \frac{\ud \kplus_i}{2 \pi} \frac{\theta(\kplus_i)}{2 \kplus_i}
		\int \ud^2 \xt_i
		\right],
		\\
		\widetilde{\sum_{q_0 \bar{q}_1 g_2 ~ \textrm{F. state}}}
		\coloneqq &
		\sum_{h_0,\alpha_0,f_0} \sum_{h_1,\alpha_1,f_1} \sum_{\lambda_2, a_2}
		\prod_{i=0}^{2} \left[
			\int_{-\infty}^{\infty} \frac{\ud \kplus_i}{2 \pi} \frac{\theta(\kplus_i)}{2 \kplus_i}
			\int \ud^2 \xt_i
		\right],
	\end{align}
	where $h_i$ are helicities of the quarks and antiquarks, $\alpha_i$ their colors, and $f_i$ their flavors. For the gluon we have its polarization $\lambda_2$ and color $a_2$. Any additional internal partons present in the $\ket{q \bar q}$ and $\ket{q \bar q g}$ Fock states, such as a gluon loop and corresponding internal quark propagators, will need sums and integrals of their own. 

	Finally, the two light-front wavefunctions shown in Eq.~\eqref{eq:fock-gamma} are defined as
	\begin{align}
		\widetilde{\Psi}_{\gamma^{*}_\lambda \rightarrow q_0 \Bar{q}_1}
		& =
		(2 \qplus) 2\pi \delta(\kplus_0 + \kplus_1 - \qplus)
		e^{i \frac{\qt}{\qplus} \cdot (\kplus_0 \xt_0 + \kplus_1 \xt_1)} {\bf 1}_{\alpha_0 \alpha_1} 	\widetilde{\psi}_{\gamma^{*}_\lambda \rightarrow q_0 \Bar{q}_1}
		\label{eq:gamma-lfwf-qqbar}
		\\
		\widetilde{\Psi}_{\gamma^{*}_\lambda \rightarrow q_0 \Bar{q}_1 g_2}
		& =
		(2 \qplus) 2\pi \delta(\kplus_0 + \kplus_1 + \kplus_2 - \qplus)
		e^{i \frac{\qt}{\qplus} \cdot (\kplus_0 \xt_0 + \kplus_1 \xt_1 + \kplus_2 \xt_2)}
		\nonumber
		\\
		&\quad
		\hspace{6.5cm}
		\times
		{t}^{a_2}_{\alpha_0 \alpha_1} \widetilde{\psi}_{\gamma^{*}_\lambda \rightarrow q_0 \Bar{q}_1 g_2} ,
		\label{eq:gamma-lfwf-qqbarg}
	\end{align}
	where one defines the \emph{reduced wavefunctions} $\widetilde{\psi}_{\gamma^{*}_\lambda \rightarrow q_0 \Bar{q}_1}$ and $\widetilde{\psi}_{\gamma^{*}_\lambda \rightarrow q_0 \Bar{q}_1 g_2}$ by factorizing out the color factors and the photon transverse momentum $\qt$ dependence. Here we note the only difference in convention to Refs.~\cite{Beuf:2016wdz,Beuf:2017bpd}: the factorization of $(2 \qplus)$ is new, and will allow for neater calculations in Ch.~\ref{ch:ddis}. The reduced wavefunctions contain the perturbative physics of the virtual photon fluctuating into a given Fock state, such as the $q \bar q$ or $q \bar q g$ states.

	The (anti-)commutation relations for the quark, antiquark and gluon creation and annihilation operators are in the mixed-space~\cite{Beuf:2016wdz}:
	\begin{align}
		\left\lbrace
		b(\kplus_0, \xt_0, h_0, \alpha_0), b^\dagger(\kplus_1, \xt_1, h_1, \alpha_1)
		\right\rbrace
		= \, &
		(2 \kplus_0) (2\pi) \delta(\kplus_0 - \kplus_1)
		\nonumber
		\\
		& \times
		\delta^{(2)}(\xt_0 - \xt_1) \delta_{h_0, h_1} \delta_{\alpha_0, \alpha_1},
		\label{eq:op-com-anticom-b}
		\\
		\left\lbrace
		d(\kplus_0, \xt_0, h_0, \alpha_0), d^\dagger(\kplus_1, \xt_1, h_1, \alpha_1)
		\right\rbrace
		= \, &
		(2 \kplus_0) (2\pi) \delta(\kplus_0 - \kplus_1)
		\nonumber
		\\
		& \times
		\delta^{(2)}(\xt_0 - \xt_1) \delta_{h_0, h_1} \delta_{\alpha_0, \alpha_1},
		\label{eq:op-com-anticom-d}
		\\
		\left[
		a(\kplus_0, \xt_0, \lambda_0, a_0), a^\dagger(\kplus_1, \xt_1, \lambda_1, a_1)
		\right]
		= \, &
		(2 \kplus_0) (2\pi) \delta(\kplus_0 - \kplus_1)
		\nonumber
		\\
		& \times
		\delta^{(2)}(\xt_0 - \xt_1) \delta_{\lambda_0, \lambda_1} \delta_{a_0, a_1}.
		\label{eq:op-com-a}
	\end{align}
	These will be needed in Ch.~\ref{ch:ddis} to calculate an overlap of Fock states.

	To apply the above Fock state decomposition of the virtual photon to calculations of scattering processes in the CGC formalism, we need to understand how the \emph{bare} quark, antiquark, and gluon scatter off the color-field of the target. The fact that only bare particles take part in the scattering off the color-field is a fundamental feature of the CGC formalism, which was originally developed in Ref.~\cite{Bjorken:1970ah} in the context of an electromagnetic field. This is discussed in the next section.

\section{Dipole amplitude}
	\label{sec:dipole}
	
	In the high-energy limit where gluon densities in hadronic matter grow to be enormous, QCD is described by the Color Glass Condensate effective field theory (EFT). At this limit the density of the gluons is so substantial that other structures of the proton or nucleus are overshadowed by their presence; so much so that the strong interactions of the target are described in the CGC formalism by strong semiclassical color-fields instead of individual gluons.
	For reviews of the CGC EFT, see for example Refs.~\cite{Iancu:2003xm, Weigert:2005us, Gelis:2010nm, Albacete:2014fwa, Blaizot:2016qgz}.
	To compute the cross section of a particle scattering off this strong color-field of the target, one must have an understanding of how the individual Fock basis states interact with the color-field.	
	
	Early calculations using light-front perturbation theory did not have precise theory tools to describe the scattering of the $q \bar q$ and $q \bar q g$ Fock states off the target, and so phenomenologically motivated models were used. One of the most well-known models is the GBW model of the dipole amplitude by Golec-Biernat and Wusthoff, which describes the scattering of a quark-antiquark dipole off the target color-field. It is~\cite{GolecBiernat:1998js,GolecBiernat:1999qd}
	\begin{equation}
		\sigma_{\textrm{GBW}}(\xbj, \rt) = \sigma_0 \left[1-e^{-\frac{1}{4} \qs^2(\xbj) \rt^2}\right],
		\quad
		\qs^2(\xbj) \coloneqq Q_0^2 \left(\frac{x_0}{\xbj}\right)^\lambda,
	\end{equation}
	where $\sigma_0$ is related to the target size, $\rt$ is the transverse size of the quark-antiquark dipole, $\xbj$ is the Bjorken-$x$, and $Q_s$ is the saturation scale. These quantities and the success of the GBW model are discussed in more detail in Ch.~\ref{ch:dis}.
	
	The progress towards precision theory calculations of the dipole amplitude needed two key advancements.
	One is the semiclassical approximation by He\-beck\-er et al.~\cite{Buchmuller:1995mr, Buchmuller:1996xw, Hebecker:1997gp, Buchmuller:1998jv, Hebecker:1999ej} in which the dipole amplitude is described as the correlator of color-fields.
	The other realization, that evolved through the works of McLerran et al.~\cite{McLerran:1993ni, McLerran:1993ka, McLerran:1994vd}, Jalilian-Marian et al.~\cite{JalilianMarian:1996xn, JalilianMarian:1997jx, JalilianMarian:1997gr, Jalilian-Marian:1997ubg}, Balitsky~\cite{Balitsky:1995ub}, Kovchegov~\cite{Kovchegov:1999yj, Kovchegov:1999ua}, and Iancu et al.~\cite{Iancu:2000hn, Iancu:2001md, Iancu:2001ad, Ferreiro:2001qy}, is that the saturation of low-transverse-momentum gluon density can be given a microscopic picture in terms of gluon fields. This gluon saturation prevents the unrealistic growth of the gluon density at large energy.
	These ideas would lead to the framework which became called the Color Glass Condensate.
	
	Now, the scattering of a quark or an antiquark off the color-field proceeds as follows. As the quark propagates into the $x^+$-direction through the color-field it can interact with the field multiple times. In each of these interactions with the field, the transverse displacement of the quark is suppressed at high energy: the displacement is of the order $\Delta x_\bot \sim L k_\bot/E$, where $L$ is the size of the target along the quark path, $k_\bot$ the change of transverse momentum in the interaction, and $E$ the energy of the quark in the target rest frame. The approximation that the displacement in the transverse position of the quark can neglected, is known as the eikonal approximation. The propagator that takes the quark through the color-field including arbitrarily many interactions is known as the Wilson line~\cite{Balitsky:1995ub}, which is in the eikonal approximation\footnote{The Wilson line has been derived without the eikonal approximation as well~\cite{Balitsky:1995ub, Weigert:2005us}, which amounts to permitting non-trivial quark paths in the resulting path integral.}~\cite{Beuf:2017bpd, Kovchegov:2012mbw}:
	\begin{equation}
		\label{eq:wilson-line}
		U_R(\xt) = \mathcal{P} e^{\mbox{\normalsize $ - i g \int_{-\infty}^\infty \ud x^+ T^a_R \mathcal{A}_a^-(x^+,\xt) $ }}.
	\end{equation}
	Here the path-ordering operator $\mathcal{P}$ enforces the path ordering in the integral over the particle path, and $T^a_R$ are the color generators. The identifier $R$ is either $F$ or $A$ for the fundamental or adjoint representations of $SU(\nc)$ which depends on the particle: $F$ for the quarks and $A$ for the gluon. Lastly, $\mathcal{A}_a^-(x^+,\xt)$ is the semiclassical color-field of the target. Thus, in the eikonal approximation, the quark only picks up a rotation in color phase space in the scattering. The propagation of a gluon through the target proceeds analogously, and only affects the representation of the Wilson line~\eqref{eq:wilson-line} produced, as mentioned above.
 	
 	With the Wilson line we can quantify the effect of the scattering on the bare quark, antiquark or gluon state. This is formulated using the eikonal scattering operator $\opse$: it acts on the creation operators of the quark, antiquark, and gluon as:
	\begin{align}
		\hat{S}_E\; \tilde{b}^{\dagger}(k^+,\xt,h,\alpha)
		&=  U_{F}(\xt)_{\beta \alpha}\;
		\tilde{b}^{\dagger}(k^+,\xt,h,\beta)\; \hat{S}_E ,
		\label{com_b_dag_SE}
		\\
		\hat{S}_E\; \tilde{d}^{\dagger}(k^+,\xt,h,\alpha)
		&= \left[U_{F}^{\dag}(\xt)\right]_{\alpha \beta}\;
		\tilde{d}^{\dagger}(k^+,\xt,h,\beta)\; \hat{S}_E , 
		\label{com_d_dag_SE}
		\\
		\hat{S}_E\; \tilde{a}^{\dagger}(k^+,\xt,\lambda,a)
		&= U_{A}(\xt)_{b a}
		\; \tilde{a}^{\dagger}(k^+,\xt,\lambda,b)\; \hat{S}_E.
		\label{com_a_dag_SE}
	\end{align}
	The Fock vacuum is invariant under the action of the eikonal scattering operator: $\opse \ket{0} = \ket{0}$.
	
	The above can be applied to describe the scattering of the Fock basis states discussed in the previous section. This allows us to formulate the scattering amplitude of the quark-antiquark dipole off the color-field in terms of Wilson lines:
	\begin{equation}
		S_{01} \coloneqq \frac{1}{\nc} \Tr \left( U_F(\xt_0) U_F^\dagger(\xt_1) \right),
	\end{equation}
	where $\xt_0$ and $\xt_1$ are the transverse positions of the quark and antiquark, and $\nc$ is the number of colors.
	To be able to compute observables depending on $S_{01}$, one must average over the color charge density configurations of the target~\cite{Venugopalan:1999wu, Beuf:2017bpd}. In terms of scattering observables, this amounts to the replacement $S_{01} \to \langle S_{01} \rangle$, where the angle brackets denote the average over the classical gluon field configurations.
	Often $S_{01}$, which is technically a scattering matrix that includes the case that nothing happens in the scattering, is substituted with an alternative definition that subtracts the identity operator corresponding to no scattering:
	\begin{equation}
		N_{01} \coloneqq 1 - S_{01}.
	\end{equation}
	This is called the \emph{dipole amplitude}, and it is the forward elastic scattering amplitude of the dipole scattering off the color-field. It is a non-perturbative quantity and as such cannot be calculated from the first principles using QCD perturbation theory. However, it has an implicit dependence on the scattering energy which is calculable perturbatively; this is discussed in more detail in the next section. The scattering amplitude of the quark-antiquark dipole will be used in the following chapters of this thesis to compute cross sections of electron-proton scattering processes.
	
	We have thus far discussed only the scattering of quarks, gluons, and Fock basis states from the target color-field, which were described by the Wilson lines and the dipole amplitude. However, this means that in the Color Glass Condensate formalism the dipole amplitude is a universal component of scattering processes. Any scattering process where the projectile can be decomposed in the QCD Fock state basis of quarks and gluons is then described by the dipole amplitude and analogous higher order correlators related to larger Fock states. The process dependent questions then include but are not limited to the calculation of the coefficient functions in the Fock state decomposition, such as we saw for the photon~\eqref{eq:fock-gamma}. Currently, the determination of the dipole amplitude requires a comparison to data --- which is the topic of Article~\cite{oma3} and is discussed in Ch.~\ref{ch:fits} --- but once it is known it can be used to make predictions for other scattering processes. For example, determining the dipole amplitude with a fit to electron-proton total cross section data allows one to make theory predictions of a number other observables, such as the longitudinal $F_L$, diffractive $F_2^D$, charm $F_2^c$, and bottom $F_2^b$ structure functions of the proton, exclusive production of vector mesons, and deeply virtual Compton scattering~\cite{Gelis:2010nm}. This of course requires theory calculations of the scattering processes in the CGC formalism, which are plentiful --- see Refs.~\cite{Iancu:2003xm, Weigert:2005us, Gelis:2010nm, Albacete:2014fwa, Blaizot:2016qgz, Morreale:2021pnn} for reviews of phenomenological studies of different observables.
	

\section{Perturbative energy evolution of the dipole amplitude}
	\label{sec:bk-evol}
	
	The derivation of the energy dependence of the dipole amplitude begins by considering an emission of a gluon from the quark-antiquark Fock state before the scattering.
	At higher energy the viable phase space for a gluon emission from the $q \bar q$-dipole grows larger, i.e. the emission becomes more likely.
	The emitted gluon can either be considered to be a part of the scattering $q \bar q g$ Fock state, or it can be taken to be a part of the dense gluon population of the target.
	The requirement that these two pictures of the same scattering lead to a single description of the scattering process eventually yields	
	the leading order Balitsky--Kovchegov (BK) equation~\cite{Balitsky:1995ub,Kovchegov:1999yj}, which in the large-$\nc$ approximation is:
	\begin{equation}
		\label{eq:bk-evolution}
		\frac{\partial \langle S_{01} \rangle_{Y}}{\partial Y} =  \int \der^2 \xij{2} {\cal K}_\text{BK}
		(\xij{0}, \xij{1}, \xij{2})
		[\langle S_{02} \rangle_{Y} \langle S_{21} \rangle_{Y} - \langle S_{01} \rangle_{Y}].
	\end{equation}
	The leading order kernel
	\begin{equation}
		{\cal K}_\text{BK}
		=
		\frac{\nc \as}{2\pi^2}
		\frac{\xij{01}^2}{\xij{12}^2 \xij{02}^2}
	\end{equation}
	is proportional to the probability to emit a gluon at $\xt_2$ from the quark-antiquark dipole of size $\xt_{01}$, where the notation used is $\xt_{ij} \coloneqq \xt_i - \xt_j$. The evolution of the dipole scattering amplitude is parametrized in the evolution variable $Y$, which will be discussed in a moment. The coupling $\as$ is fixed in the LO BK equation.

	Implementing running coupling into the BK equation corresponds to a partial inclusion of NLO effects, which are taken as a part of the running coupling prescription. With the addition of the running coupling corrections according to the widely used Balitsky prescription~\cite{Balitsky:2006wa}, the kernel becomes
	\begin{multline}
		\label{eq:bk-rc-balitsky}
		K_\text{BK}(\xij{0},\xij{1}, \xij{2}) = \frac{\nc \as(\xij{01}^2)}{2\pi^2} \left[
		\frac{\xij{01}^2}{\xij{12}^2 \xij{02}^2} \right.
		+ \frac{1}{\xij{02}^2} \left( \frac{\as(\xij{02}^2)}{\as(\xij{12}^2)} -1 \right) 
		\\
		+ \left. \frac{1}{\xij{12}^2} \left( \frac{\as(\xij{12}^2)}{\as(\xij{02}^2)} -1 \right)
		\right].
	\end{multline}
	In comparison to the LO BK equation with fixed coupling, with the Balitsky running coupling the BK equation leads to a slower $Y$-evolution of the dipole amplitude, which is more consistent with experimental data. The effect of the Balitsky prescription in phenomenological applications is further discussed in Sec.~\ref{sec:lo-fits}, and in Ch.~\ref{ch:fits} in regards to the work done in Article~\cite{oma3}.
	
	Once the BK equation is applied to phenomenology, i.e. used to drive the energy evolution of the dipole amplitude which then is used to calculate observables, the relation of the evolution variable to the kinematics of the dipole becomes important. Generally, the evolution variable must be proportional to the logarithm of the squared center-of-mass energy, however at NLO the precise definition of the variable begins to matter.
	At leading order the evolution variable $Y$ is a rapidity-like quantity conventionally defined as
	\begin{equation}
		\label{eq:Y}
		Y \coloneqq \ln \left( \frac{\qplus}{P^+}\right) = \ln \frac{W^2}{Q_0^2},
	\end{equation}
	where $\qplus$ is the plus-momentum of the incoming $q \bar q$-dipole, $P^+$ is a plus-momen\-tum scale related with the target, $Q_0^2$ is a non-perturba\-ti\-ve momentum-scale characteristic to the target, and $W^2 = 2 \qplus P_0^-$ is the center-of-mass energy of the dipole-target system. Evolution in $Y$ is known as the projectile momentum fraction or rapidity picture, or alternatively the plus-momentum ordering picture --- the latter refers to the fact that in $Y$-evolution the successive gluon emissions are strongly ordered by their plus-momenta.
	Many leading order phenomenological studies have used also the definition $Y = \ln \frac{1}{\xbj}$, which becomes problematic beyond leading order since the evolution is in this case parameterized in target momentum fraction --- this is discussed further in a moment.
	At NLO with the additional gluon in the Fock state, the above definition \eqref{eq:Y} of $Y$ becomes inaccurate. Taking into account the plus-momentum of the gluon becomes crucial, and the evolution rapidity is defined as
	\begin{equation}
		Y \coloneqq \ln \left( \frac{\kplus_2}{P^+}\right) = \ln z_2 + \ln \frac{\qplus}{P^+},
	\end{equation}
	where $\kplus_2$ is the longitudinal momentum of the emitted gluon. The momentum fraction of the gluon is defined as $z_2 \coloneqq \frac{\kplus_2}{\qplus}$. This $z_2$-dependence of $Y$ has a key role in NLO phenomenology, which is the main point in Article~\cite{oma1} and will be discussed in Sec.~\ref{sec:dis-nlo}.
	
	Even though this projectile momentum fraction picture is a rather natural way to parametrize the BK evolution --- in the sense that the longitudinal momenta of the quarks and gluons in the Fock state stay constant in the scattering off the color-field --- it is not quite problem free. In the derivation of the BK equation successive gluon emissions from the projectile dipole are considered, and their lifetimes should be strongly ordered: each daughter gluon should have a shorter lifetime than its parent. However, without further work the $Y$ evolution includes emissions which violate this lifetime hierarchy, which leads to an instability of the evolution.
	For a more in-depth discussion of the challenges with $Y$ evolution, we refer the reader to Ref.~\cite{Ducloue:2019ezk}.
	
	To cure the time-ordering problem in $Y$-evolution, different techniques have been used to include related higher-order corrections to the BK equation --- this has been done by resumming radiative corrections which are enhanced by double transverse logarithms to all orders. In~\cite{Beuf:2014uia} this resummation produces a BK equation non-local in $Y$ with a kinematical constraint that enforces the ordering; in later parts of this thesis this will be called the KCBK equation, following the convention of~\cite{oma3}. Another approach~\cite{Iancu:2015vea} resums these same corrections in a way that produces a local BK equation. This formulation has further been improved~\cite{Iancu:2015joa}, by including a resummation of terms enhanced by single transverse logarithms, which arise from DGLAP physics and one-loop running coupling corrections; this formulation will be called the ResumBK equation, as in~\cite{oma3}. These resummations are able to capture a substantial subset of the NLO contributions to the BK equation~\cite{Lappi:2016fmu}, and in Ref.~\cite{Albacete:2015xza} it was shown that the two approaches resumming the double-log contributions produce very comparable evolutions of the dipole amplitude --- the difference between the resummation prescriptions should be of the order of $\ocal(\as^2)$.
		
	In Ref.~\cite{Ducloue:2019ezk} an alternative approach is proposed to resolve some of the issues with the $Y$-formulation of the BK equation. Their approach recasts the BK equation as a function of the target momentum fraction. This naturally enforces proper time-ordering in the evolution, since the ordering of the emissions in the minus-momentum is equivalent with their ordering in lifetime. The new formulation of the BK evolution in the target rapidity $\eta$, which is defined by~\cite{Ducloue:2019ezk}
	\begin{equation}
		\eta \coloneqq Y - \rho = Y - \ln \frac{Q^2}{Q_0^2} = \ln \frac{W^2}{Q^2} = \ln \frac{1}{\xbj},
	\end{equation}
	is derived by performing this change of variables from $Y$ to $\eta$, which is a non-perturbative operation that mixes terms of all orders in the perturbative expansion. The target rapidity formulation of the BK equation was found to have a milder instability caused by large double logarithms of a different kind, which arise in the collinear emission limit of the gluon. An all-order resummation of these contributions was performed, and this resummed formulation will be discussed later in this thesis under the name TBK as in Article~\cite{oma3}.
	
	The derivation of the NLO corrections to the BK equation considers two gluon emissions from the incoming dipole, which scatter off the target via multiple gluon exchange. The NLO BK equation has been derived both in $Y$~\cite{Balitsky:2008zza, Lappi:2020srm,} and $\eta$~\cite{Ducloue:2019ezk} evolution pictures, and the former has been solved numerically including resummations as discussed above~\cite{Lappi:2015fma, Lappi:2016fmu}. Calculations of scattering processes in the CGC formalism at full NLO accuracy require the usage of an NLO accuracy evolution equation --- the usage of the NLO BK in such calculations is discussed in Ch.~\ref{ch:fits}.

	While the BK equation was derived first, it was later discovered that it corresponds to a large-$\nc$ mean-field approximation of a more general evolution equation: the JIMWLK (Jalilian-Marian, Iancu, McLerran,
	Weigert, Leonidov and Kovner) equation~\cite{JalilianMarian:1996xn, JalilianMarian:1997jx, JalilianMarian:1997gr, Jalilian-Marian:1997ubg, Iancu:2001md, Iancu:2000hn, Ferreiro:2001qy, Iancu:2001ad}. The JIMWLK equation describes the energy dependence of the probability distribution of the Wilson lines.
	Next-to-leading order corrections have been derived to the JIMWLK equation~\cite{Balitsky:2013fea, Kovner:2013ona}, and analogous resummations of higher order corrections to those of the BK equation discussed above have been studied as well~\cite{Hatta:2016ujq}.
	A recent analysis performs detailed comparisons of numerical solutions of the JIMWLK equation, studying the effects of the numerical implementation and running coupling prescriptions~\cite{Cali:2021tsh}.
	The JIMWLK equation has been used in phenomenology to describe scattering processes, some studies are discussed in Sec.~\ref{sec:lo-fits}.

	\chapter{Deep Inelastic Scattering}
	\label{ch:dis}

\section{Probing the internal structure of the proton}
	\label{sec:dis-general}

	We begin with an overview of deep inelastic scattering (DIS), which probes the hadronic structure of a nucleon or an atomic nucleus with a high-energy lepton. DIS had an important role in the early development of QCD, the first experiments taking place at the Stanford Linear Accelerator Center (SLAC)~\cite{Bloom:1969kc,Breidenbach:1969kd} contemporaneously with the development of the parton model in the late 1960s. More recently the high-energy precision measurements of electron-proton DIS at DESY-HERA~\cite{Adloff:1997mf,Aid:1996au,Breitweg:1997hz,Derrick:1995ef,Breitweg:2000yn,Chekanov:2001qu,Adloff:2000qk,Aaron:2009aa,Abramowicz:2015mha,H1:2018flt,Abramowicz:1900rp,Andreev:2013vha,Abramowicz:2014jak} have sparked a keen interest in the low-$\xbj$ physics of proton structure.
	
\begin{figure}
		\centering
		\includegraphics[width=0.5\textwidth]{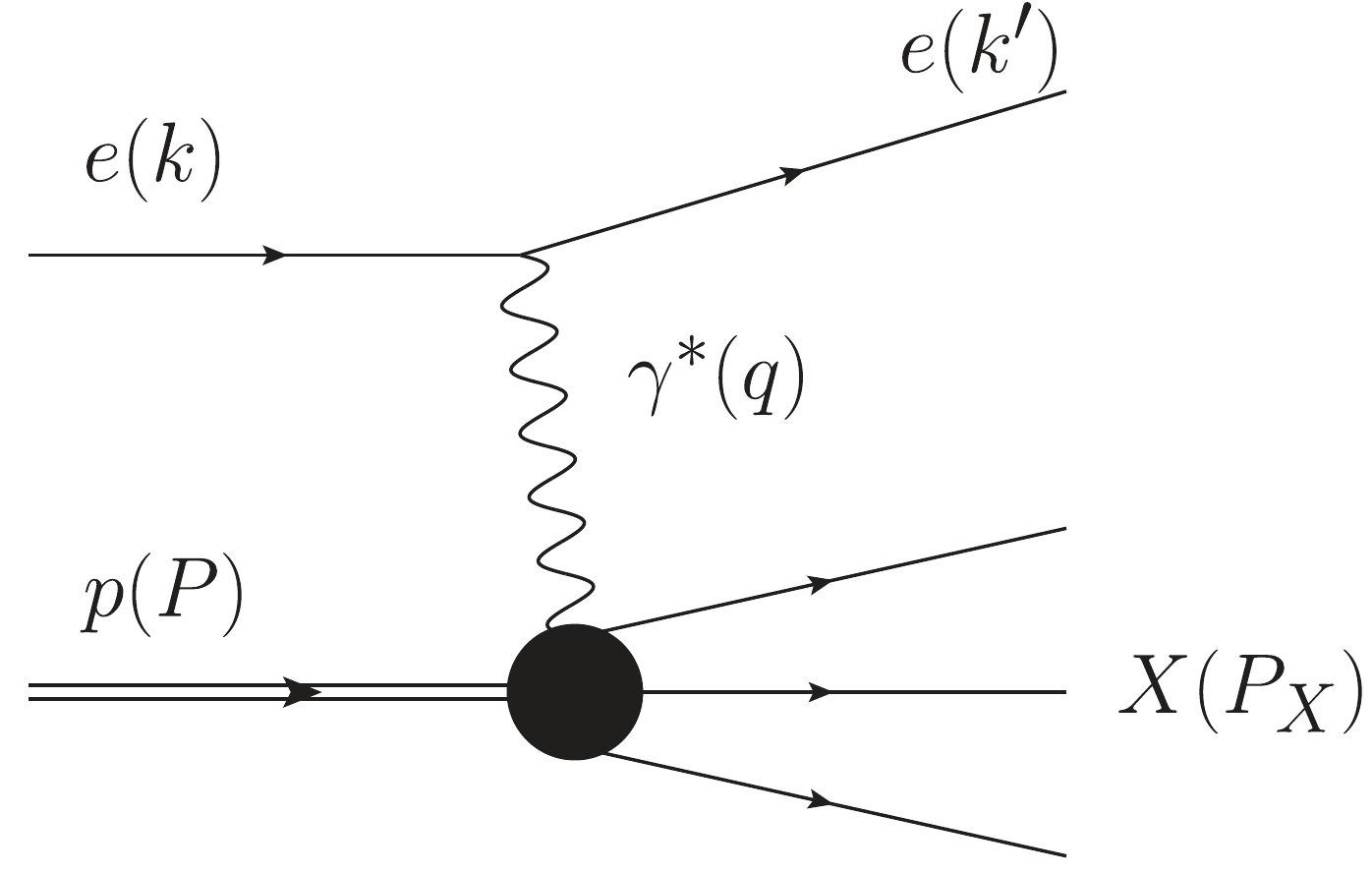}
		\caption{Deep inelastic scattering of an electron off a proton.}
		\label{fig:dis}
	\end{figure}

	As a specific example, Fig.~\ref{fig:dis} shows the deep inelastic scattering of an electron off a proton, where the electron-proton interaction takes place as the exchange of a highly virtual photon $\gamma^*$. In the parton model, the virtual photon kicks out a parton from the proton as it scatters breaking up the target and therefore probing its structure. The lepton is a particularly suitable probe particle since it does not have internal structure: the lepton current can be separated from the hadronic part, providing a clear window into the hadronic structure. From the theory point of view we are left to describe the virtual photon-proton scattering. See Ref.~\cite{Kovchegov:2012mbw} for an in-depth discussion of DIS and a derivation of the parton model using light-front perturbation theory. 

	The kinematics of the virtual photon-proton deep inelastic scattering is completely described by two quantities\footnote{Electron-proton DIS has a third parameter --- the inelasticity $y$ --- which will be discussed in Ch.~\ref{ch:fits}.} --- the virtuality of the photon $Q^2$, and the Bjorken-$x$ --- which are defined as:
	\begin{align}
		Q^2 & \coloneqq -q^2 = - (k - k')^2, \\
		\xbj & \coloneqq \frac{Q^2}{2 P \cdot q} = \frac{Q^2}{W^2+Q^2-M^2} \simeq \frac{Q^2}{W^2+Q^2} \simeq \frac{Q^2}{W^2},
		\label{eq:xbj}
	\end{align}
	where $W^2 \coloneqq (P+q)^2$ is the center-of-mass energy\footnote{Not to be confused with the electron-proton c.o.m. energy $s = (P+k)^2$.}, and the momenta are illustrated in Fig.~\ref{fig:dis}. The first of the approximative equalities is valid at large enough energies where the proton mass $M^2$ can be neglected, and the second in the proper high-energy regime, $W^2 \gg Q^2$, also known as the Regge limit. From Eq.~\eqref{eq:xbj} we see why the low-$\xbj$ regime corresponds to the high-energy limit. In the laboratory frame one has $Q^2 = 2 E E' (1-\cos \theta)$, where $E, E'$ are the initial and final energy of the electron, and $\theta$ the scattering angle. This tells us that $Q^2$ is a measurable quantity, and therefore $\xbj$ is as well.
	
	These quantities can receive further interpretation. Since the virtual photon can interact with partons inside the proton with transverse momentum at most of the order $\kt^2 \sim Q^2$, this means through the Heisenberg uncertainty principle that the partons are localized on a scale $\rt^2 \sim 1/Q^2$~\cite{Taels:2017shj}. Thus $Q^2$ gives the resolution at which the target structure is probed. On the other hand in the infinite momentum frame, compared to the large virtuality $Q^2$, the virtuality of the parton inside the target can be taken to be negligible and so the parton is on-shell. If we then assume the parton carries a fraction of the target momentum $k^\mu = x P^\mu$, we have for the real outgoing parton after the interaction:
	\begin{equation}
		(x P^\mu + q^\mu)^2  = k'^2 = 0,
	\end{equation}
	which yields,
	\begin{equation}
		x = \frac{Q^2}{2 P \cdot q} \equiv \xbj,
	\end{equation}
	i.e. the kinematically defined Bjorken-$x$ can be interpreted as the momentum fraction of the hit parton in the frame where the proton longitudinal momentum is very large.
	
	The total virtual photon-proton cross section can be written in the high-energy limit as \cite{Kovchegov:2012mbw}:
	\begin{equation}
		\sigma^{\gamma^*p}_{\textrm{tot}} = \frac{(2\pi)^2 \aem}{Q^2} F_{2}(x, Q^2),
	\end{equation}
	which is also called the (fully) inclusive cross section, as in inclusive of all processes that produce any final state.\footnote{This is in contrast with the exclusive cross sections that \textit{exclusively} consider a single process producing a specific final state, such as the production of a given particle, like a vector meson.}
	The proton structure function $F_2$ encodes the unknown information about the internal structure of the proton. It is related to the structure functions $F_T$ and $F_L$:
	\begin{equation}
		F_{2}(x, Q^2) = F_{T}(x, Q^2) + F_{L}(x, Q^2),
	\end{equation}
	where the photon polarization specific structure functions are related to the corresponding total $\gamma^*_{L,T} p$ cross sections as:
	\begin{equation}
		F_{T,L}(x, Q^2) = \frac{Q^2}{(2\pi)^2 \aem} \sigma_{L,T}^{\gamma^* p}(x, Q^2).
	\end{equation}
	Here $L$ and $T$ are referring to the longitudinal and transverse polarizations of the virtual photon. The longitudinal structure function $F_L$ is sensitive to the gluonic structure of the proton, since in the case that the proton only contained spin-1/2 particles, the Callan-Gross relation would state that $F_L \equiv 0$~\cite{Kovchegov:2012mbw}.

\section{DIS in dipole picture at leading order}
	\label{sec:dis-lo}

	\begin{figure}
		\centering
		\includegraphics[width=0.6\textwidth]{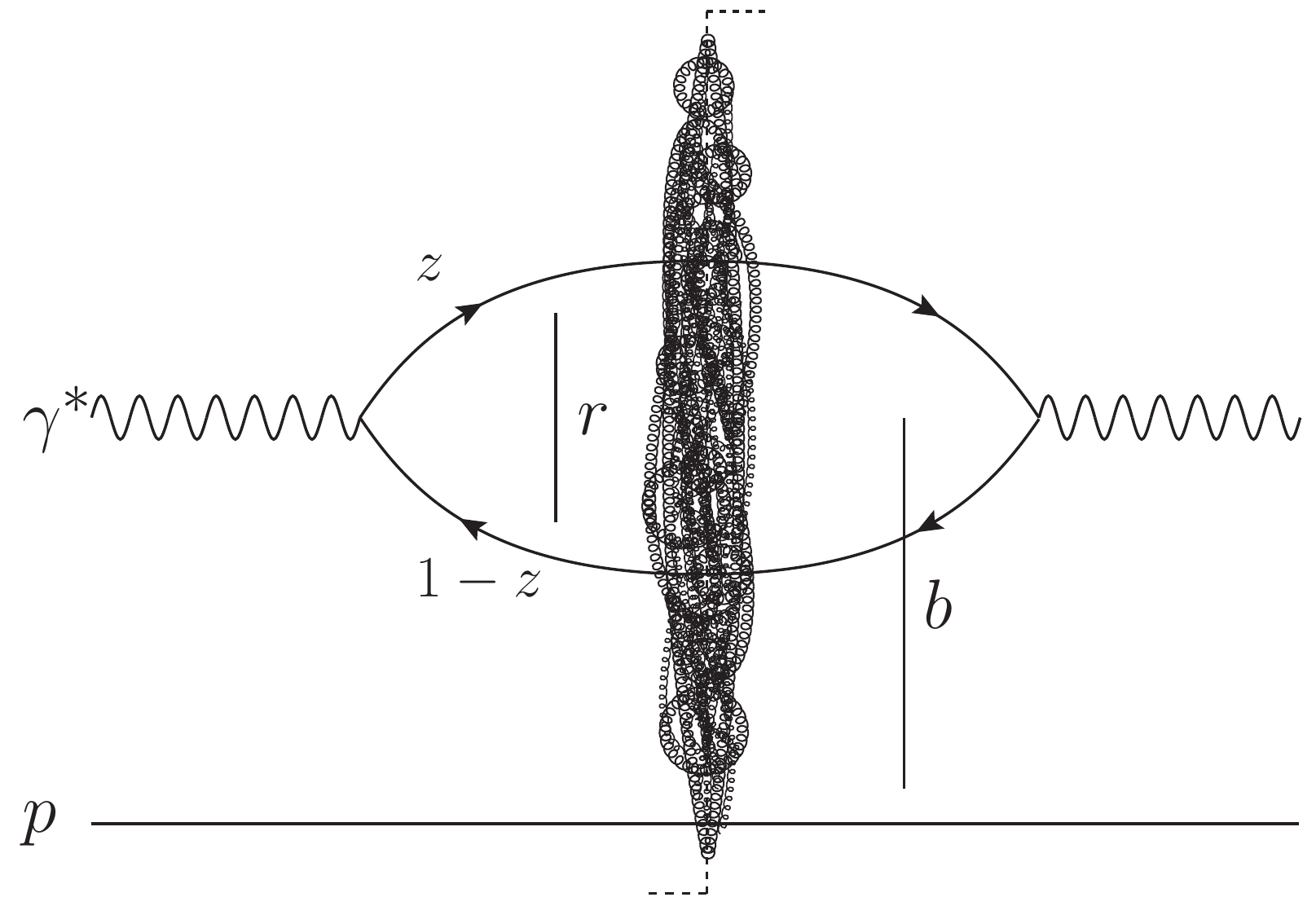}
		\caption{Deep inelastic scattering at leading order in the dipole picture. The diagram depicts the calculation of the elastic scattering amplitude ${\cal M}^{\mathrm{fwd}}_{\gamma_\lambda^{*} \rightarrow \gamma_\lambda^{*}}$, to be used with the optical theorem \eqref{eq:optical-theorem} to get the total cross section.}
		\label{fig:lo-dipole}
	\end{figure}

	In contrast to the description of deep inelastic scattering in the previous section, in the high-energy regime a different picture of the scattering can be constructed. At low-$\xbj$ the target is full of low momentum fraction gluons that form semi-classical gluon fields, which flatten into a shockwave due to Lorentz contraction. In this high-energy regime, DIS is described as the scattering of the virtual photon from the color-field of the target. In the target rest frame this proceeds by the incoming --- color-chargeless --- virtual photon fluctuating into a Fock state which has color-charged constituents that then can scatter off the color-field of the target. This is the dipole picture of DIS, which builds on the ideas of Bjorken, Kogut, and Soper~\cite{Bjorken:1970ah} who conceptualized scattering at high-energy as the scattering from a force-field, originally the electromagnetic field.
	
	The leading contributing state in the dipole picture of DIS is the quark-antiquark pair that the photon can form in a QED pair production, depicted in Fig.~\ref{fig:lo-dipole}. Then, in the spirit of a leading order calculation in the high-energy limit, the transverse positions of the quark and antiquark can be taken to be fixed in the scattering off the shockwave. This is justified since the relative transverse momentum the quarks pick up in the scattering is suppressed in powers of the scattering energy \cite{Kovchegov:2012mbw}. This assumption of transverse immobility is the so-called eikonal approximation. Thus, in the eikonal approximation, the quark-dipole only picks up a color rotation in the scattering off the color-field, as described in Sec.~\ref{sec:dipole}. The optical theorem~\cite{Beuf:2017bpd} then connects this elastic scattering amplitude to the inclusive cross section for the virtual photon-proton scattering.

	One of the first dipole model depictions of DIS was derived by Nikolaev and Zakharov \cite{Nikolaev:1990ja}. In their wake, the process has been calculated in light-front perturbation theory \cite{Dosch:1996ss}. Many of the Color Glass Condensate framework based theory descriptions of saturation phenomenology in HERA DIS data are based on the dipole picture of DIS, for reviews see Refs.~\cite{Iancu:2003xm,Weigert:2005us,Gelis:2010nm,Albacete:2014fwa}.

	\begin{figure}
		\centering
		\quad \quad\quad
		\includegraphics[width=0.555\textwidth]{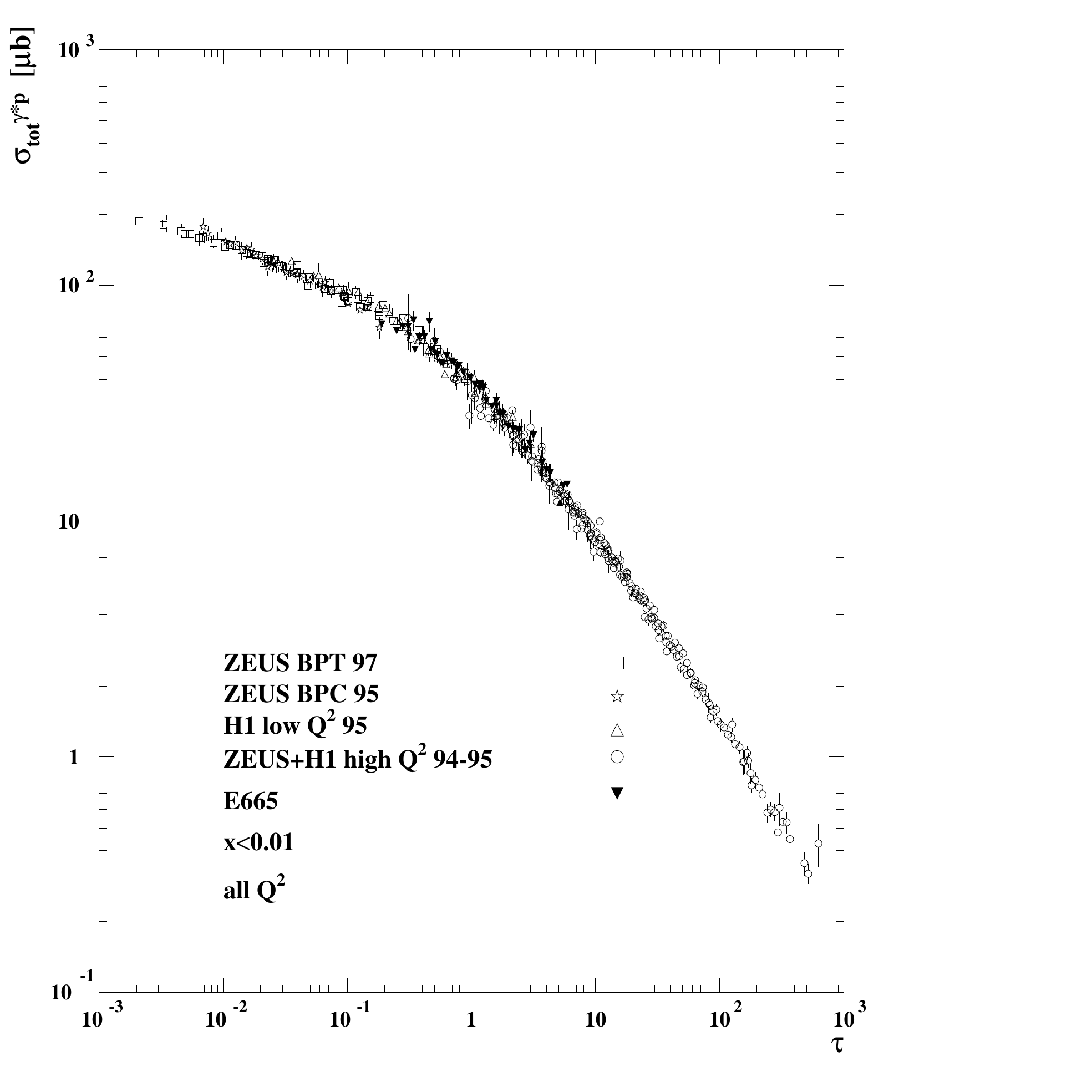}
		\caption{Geometric scaling of the photon-proton inclusive HERA data as the function of $\tau \coloneqq Q^2/Q_s^2(x)$~\cite{Stasto:2000er}.
		Reprinted figure with permission from
		A. M. Stasto, K. Golec-Biernat, and J. Kwiecinski,
		Phys. Rev. Lett., 86, 596, 2001.
		Copyright (2001) by the American Physical Society.
		}
		\label{plot:geometric-scaling}
	\end{figure}

	One final thing to consider is when the dipole picture of DIS is valid. Specifically, when is it a good depiction of photon-proton DIS --- how small does $\xbj$ need to be ---, and is the dipole picture related to gluon saturation in the proton.
	In the target rest frame, the formation time of the quark-antiquark dipole is $\tau_{q\bar{q}} \sim \frac{1}{m_p \xbj}$ which much longer than the typical interaction time $\tau_{\textrm{int.}} \sim R_p$ when $\xbj$ is small~\cite{Barone:2002cv,Kovchegov:2012mbw}; The quantities $m_p$ and $R_p$ are the mass and radius of the proton. Thus the dipole picture only requires the limit $\xbj \ll 1$.
	This ties into the manifestation of saturation physics which takes place at high-energy, which is equivalent with the very small $\xbj$ required by the dipole picture. Consequently, the dipole picture is well suited to be combined with a theory description of saturation, as is done in the Color Glass Condensate effective field theory.
	To get more insight into the correct regime in $\xbj$, we look at the HERA $\gamma^*p$ data shown in Fig.~\ref{plot:geometric-scaling}. The plot shows that the inclusive photon-proton cross section scales as the function of $\tau \coloneqq Q^2/Q_s^2(x)$, where $Q_s^2(x)$ is an emergent semi-hard scale, the saturation scale, in the scattering. This scaling phenomenon is called geometric scaling, which can be indicative of non-linear saturation physics \cite{Iancu:2002aq}, though there are other possible mechanisms \cite{Gelis:2010nm}.
	Based on empirical observations such as the geometric scaling, $\xbj \lesssim 0.01$ is conventionally taken to be the valid regime of the dipole picture.
	A further point supporting the applicability of the dipole picture to $\gamma^*p$ DIS is that it works phenomenally well: HERA data is described very well at low $\xbj$ --- important analyses will be discussed in Sec.~\ref{sec:lo-fits}.

	\subsection{Inclusive deep inelastic scattering cross section at leading order}

		Let us discuss the high-level steps to be taken in the calculation of the leading order inclusive virtual photon-proton DIS cross sections. First we need to pick out the relevant contributions to the dressed virtual photon Fock state --- discussed in Sec.~\ref{sec:lcpt} --- that contribute at leading order. Only the quark-antiquark state contributes and so at leading order:
		\begin{multline}
			\ket{\gamma^*_\lambda(\qplus, \qt; Q^2)_H \vphantom{\opse}}
			=
			\sqrt{Z_{\gamlam}}
			\left\lbrace
			\widetilde{\sum_{\substack{q_0 \bar{q_1} \\ \textrm{F. states}}}}
			\tilde{\Psi}_{\gamma^{*}_\lambda \rightarrow q_0 \bar{q}_1}
			\tilde{b}_0^\dagger \tilde{d}_1^\dagger \ket{0}
			+
			\textrm{negl.\,Fock\,states}
			\vphantom{\widetilde{\sum_{q_0 \bar{q_1} ~ \textrm{F. states}}}}
			\right\rbrace .
		\end{multline}
		One then calculates the forward elastic scattering amplitude for the virtual photon-shockwave scattering, which is defined in light-front quantization as~\cite{Beuf:2017bpd}:
		\begin{equation}
			\label{eq:lo-s-amplitude}
			\bra{\gamma_\lambda^{*} (q')_H \vphantom{\hat S_E} } \left(\hat S_E - 1 \right) \ket{\gamma_\lambda^{*} (q)_H \vphantom{\hat S_E}} = (2 \plusq) 2 \pi \delta (\plusqpr - \plusq) i {\cal M}^{\mathrm{fwd}}_{\gamma_\lambda^{*} \rightarrow \gamma_\lambda^{*}},
		\end{equation}
		where $\opse$ is the scattering operator that acts on the creation operators of the quark and antiquark. With the forward elastic scattering amplitude, one can then use the optical theorem~\cite{Beuf:2017bpd} to relate the amplitude to the total inclusive cross section of the virtual photon-proton scattering:
		\begin{align}
			\label{eq:optical-theorem}
			\sigma^{\gamlam p \to X}
			= 2 \Im {\cal M}^{\mathrm{fwd}}_{\gamma_\lambda^{*} \rightarrow \gamma_\lambda^{*}}
			= 2 \Re \left(-i {\cal M}^{\mathrm{fwd}}_{\gamma_\lambda^{*} \rightarrow \gamma_\lambda^{*}} \right).
		\end{align}
		The last missing piece are the wavefunctions $\tilde{\Psi}_{\gamma^{*}_\lambda \rightarrow q_0 \bar{q}_1}$ for the virtual photon splitting into a quark-antiquark dipole, which have been computed in light-front perturbation theory by many authors, see for example Refs.~\cite{Dosch:1996ss, Beuf:2011xd}\footnote{For a $D$-dimensional derivation, see Refs.~\cite{Beuf:2016wdz,oma2}. Dimensional regularization of the LO result is necessary in the derivation of the complete NLO wavefunctions.}. The squares of these wavefunctions can be interpreted as the probability for the virtual photon to fluctuate into the quark-antiquark dipole. Summed over helicities and photon transverse polarizations, they are~\cite{Kowalski:2006hc}
		\begin{align}
			\label{eq:lfwf-sq-lo-dis-t}
			\left| \Psi_{\gamma_{T}^{*} \to q \bar{q}} \right|^2 & = \frac{2 \nc}{\pi} \aem e_f^2 \left\lbrace \left[ z^2 + (1-z)^2 \right] \varepsilon^2 \textrm{K}_1^2(\varepsilon r) + m_f^2 \textrm{K}_0^2(\varepsilon r) \right\rbrace \, ,
			\\
			\label{eq:lfwf-sq-lo-dis-l}
			\left| \Psi_{\gamma_{L}^{*} \to q \bar{q}} \right|^2 & = \frac{8 \nc}{\pi} \aem e_f^2 Q^2 z^2 (1-z)^2 \textrm{K}_0^2(\varepsilon r) \, ,
		\end{align}
		where $\varepsilon^2 = z(1-z) Q^2 + m_f^2$, $z \coloneqq \frac{\plusk}{\plusq}$ is the longitudinal momentum fraction of the quark, $r$ is the size of the $q \bar{q}$ dipole as shown in \fig\ref{fig:lo-dipole}, $f$ and $m_f$ are the flavor and mass of the quark, and $\textrm{K}_0, ~ \textrm{K}_1$ are modified Bessel functions of the second kind.
		
		Following the outlined calculation one derives the total virtual photon-proton deep inelastic scattering cross sections, which are~\cite{Kowalski:2006hc, Kovchegov:2012mbw}:
		\begin{equation}
			\label{eq:lo-dis-cs}
			\sigma_{L,T}^{\gamma* p}(x, Q^2)
			=
			\sum_f \int \der^2 \rt \int_{0}^{1} \frac{\der z}{4 \pi} \left| \Psi_{\gamma_{L,T}^{*} \to q \bar{q}} \right|^2 \sigma_{q \bar{q}}(x,r),
		\end{equation}
		where $r \coloneqq \lVert \rt \rVert$, and the quark-antiquark dipole scattering amplitude is defined as
		\begin{equation}
			\sigma_{q \bar{q}}(x,r) = \int \der^2 \bt \, 2 \left[1 - \Re S(x, r, \bt)\right] ,
		\end{equation}
		and $S(x, r, \bt)$ is the scattering matrix for the dipole-gluon shockwave scattering, where $\bt$ is the transverse separation of the $q \bar q$ dipole from the target, as shown in \fig\ref{fig:lo-dipole}. Sec.~\ref{sec:dipole} discusses how the scattering matrix arises in the calculation of the scattering amplitude~\eqref{eq:lo-s-amplitude}. In Eq.~\eqref{eq:lo-dis-cs} we see the explicit factorization between the virtual photon wavefunction and the dipole amplitude --- the cross section is composed of two independent pieces, a piece with the perturbative QED physics, and a piece with the non-perturbative QCD physics. This feature of the LO dipole picture cross sections is called dipole factorization.

	\subsection{Comparisons of leading order DIS cross sections and measurements}
		\label{sec:lo-fits}
		
		Some of the most exhaustive searches for saturation effects have been done by studying deep inelastic scattering. These analyses rely on the dipole picture of DIS \cite{Nikolaev:1990ja, Dosch:1996ss} and introduce theory description of saturation effects through the dipole amplitude, which the DIS cross sections depend on. The description of these saturation effects can be roughly divided in two eras: pre-BK and BK era. In this section we discuss some of the key saturation physics analyses and data comparisons done using LO dipole picture DIS structure functions.
	
		The first analysis of DIS data incorporating saturation physics was done using the GBW model by Golec-Biernat and Wusthoff~\cite{GolecBiernat:1998js} --- an analytic parametrization of the dipole amplitude, discussed in Sec.~\ref{sec:dipole} --- and it achieved a reasonably good description of the old HERA data \cite{Aid:1996au, Adloff:1997mf, Derrick:1995ef, Breitweg:1997hz}. This model was modified to include DGLAP evolution~\cite{Bartels:2002cj}, which improved the description of the total DIS cross sections, especially at large $Q^2$. Another analytic model was constructed by Iancu, Itakura and Munier (IIM) \cite{Iancu:2003ge} to improve upon on the success of GBW by including features of BK evolution \cite{Balitsky:1995ub, Kovchegov:1999yj}. The IIM model has been updated to include impact parameter dependence, leading to the bCGC model \cite{Kowalski:2006hc,Watt:2007nr}. These updated models provide improved agreement with the then new and much more accurate HERA data \cite{Breitweg:2000yn, Chekanov:2001qu,Adloff:2000qk}, in comparison to the simple GBW model.

		One more important analytic parametrization to capture saturation effects is the Impact Parameter Saturation (IP-Sat) model proposed by Kowalski and Teaney \cite{Kowalski:2003hm}. In the IP-Sat model the dipole amplitude evolution is induced by the LO DGLAP $Q^2$-evolution of the gluon distribution, and the $\xbj$-dependence is parametrized. The original IP-sat implementation provided a good description of HERA data in a much wider $Q^2$-range than the GBW model was capable of. Since then, the model has been updated \cite{Rezaeian:2012ji} with more precise data \cite{Aaron:2009aa}, and another work \cite{Mantysaari:2018nng} updates the model with the most recent HERA data \cite{Aaron:2009aa,Abramowicz:1900rp,Abramowicz:2015mha,H1:2018flt}, and performs a comparison between Impact Parameter models with and without saturation, finding that they are comparably capable of describing the available data.

		The BK era was brought about by the AAMS and AAMQS global fits \cite{Albacete:2009fh,Albacete:2010sy}, which were the first fits to use the BK equation to drive the small-$x$ dependence of the dipole amplitude. This was an important upgrade in terms of theory precision, since now the only non-perturbative input would be the dipole amplitude at the initial scale of the BK evolution. Using the translationally invariant running coupling BK \cite{Balitsky:2006wa,Kovchegov:2006vj}, the AAMQS fits found remarkably good agreement with the inclusive structure functions. A similar analysis using the running coupling BK was done with a slightly different parametrization \cite{Lappi:2013zma}, and in \cite{Kuokkanen:2011je} energy conservation corrections are implemented to the running coupling BK, yielding excellent agreement with inclusive data.
		
		Steps have been taken to improve the theoretical precision of the BK approach by including important beyond leading order contributions to the BK equation. One approach~\cite{Kwiecinski:1997ee} enhances the momentum space BK equation by including contributions from DGLAP evolution, which together with a consistency constraint on real gluon emissions produce a large part of higher order corrections. A good description of HERA data was found by an analysis~\cite{Kutak:2012rf} using this evolution equation incorporating aspects of both BK and DGLAP evolutions.
		
		Another approach to enhance the BK equation is the resummation of beyond leading order contributions to the evolution that are enhanced by large logarithms of $1/\xbj$  --- see the discussion of different approaches in Sec.~\ref{sec:bk-evol}. The first of such improved BK equations to be used in data analysis is the ResumBK\footnote{\label{footnote:BK-acronyms}ResumBK, KCBK and TBK are acronyms coined in~\cite{oma3} for the formulations of BK evolution derived in Refs.~\cite{Iancu:2015joa},~\cite{Beuf:2014uia}, and~\cite{Ducloue:2019ezk}, respectively.}~\cite{Iancu:2015joa}, where they find a good description of HERA data even at fairly high-$Q^2$. The resummation of the single transverse log -enhanced contributions has a minor effect on the fit quality, yielding slightly higher values of $\chi^2$, but nevertheless too small a change to be considered in the accuracy of the analysis.
		A study~\cite{Albacete:2015xza} comparing the ResumBK and KCBK~\cite{Beuf:2014uia} formulations of the double log resummed BK evolutions found both approaches equally capable of describing HERA DIS data well, and that more theoretical work is needed to distinguish a preferred prescription of resummation and running coupling in the BK equation. 
		A more recent data comparison~\cite{Ducloue:2019jmy} fits inclusive HERA data using BK evolution which is formulated as a function of the target rapidity~\cite{Ducloue:2019ezk} (TBK in~\cite{oma3}), as opposed to the projectile rapidity picture used in the above prescriptions.
		The effects of the resummations of the single and double transverse log enhanced contributions are considered comprehensively, and they find that the inclusion of the single log resummation has a substantial effect on the fit quality in this scheme --- an excellent description of the HERA data is found when both resummations are included, even up to reasonably high $Q^2$.
		
		Another branch of BK evolution improvements implements some form of impact parameter dependence. In~\cite{Berger:2011ew} a comparison to HERA data is done with a fairly good agreement, however a full fit was not performed due to the computational cost involved with the impact parameter dependent BK. Similar work has been done using the leading order JIMWLK evolution \cite{Mantysaari:2018zdd}, where a quite good description of the inclusive HERA data was found at small and moderate $Q^2$. Recent work incorporating the collinearly improved BK kernel into the impact parameter dependent evolution \cite{Bendova:2019psy} finds a good description of HERA data.
		
		In conclusion, saturation effects have been captured by varied approaches which are able to describe inclusive deep inelastic scattering data at small-$x$. Recent developments have explored two important areas: running coupling BK evolution and its resummation corrections are central in precision theory calculations, whereas the impact parameter dependence in other approaches is obligatory for the description of exclusive diffractive processes. Impact parameter dependence is needed for exclusive processes since the total momentum transfer with the target is measured, and it is the Fourier conjugate of the impact parameter. This makes the process sensitive to the transverse structure of the target. The recent developments on the impact parameter dependent BK--JIMWLK evolution have promise to bring precision calculations to exclusive processes, and on the other hand target transverse profile sensitivity to inclusive precision calculations.

\section{DIS in the dipole picture at next-to-leading order} \label{sec:dis-nlo}

	\begin{figure}
		\centering
		\begin{subfigure}[t]{0.465\textwidth}
			\includegraphics[width=0.95\textwidth]{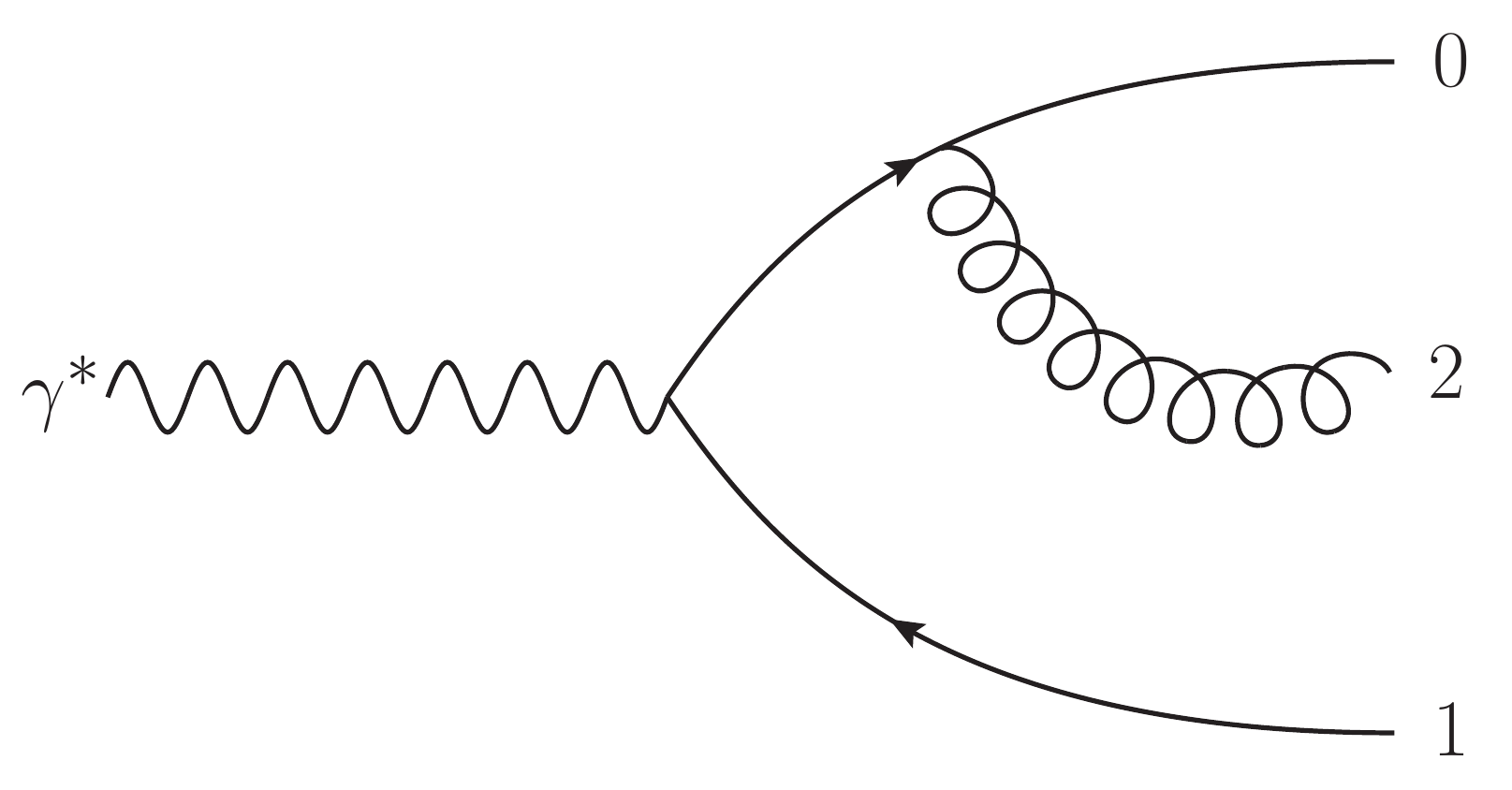}
			\caption{Example of a tree-level diagram: a quark emits a gluon.}
			\label{fig:nlo-dis-diag-tree}
		\end{subfigure} \hfill
		\begin{subfigure}[t]{0.465\textwidth}
			\includegraphics[width=0.95\textwidth]{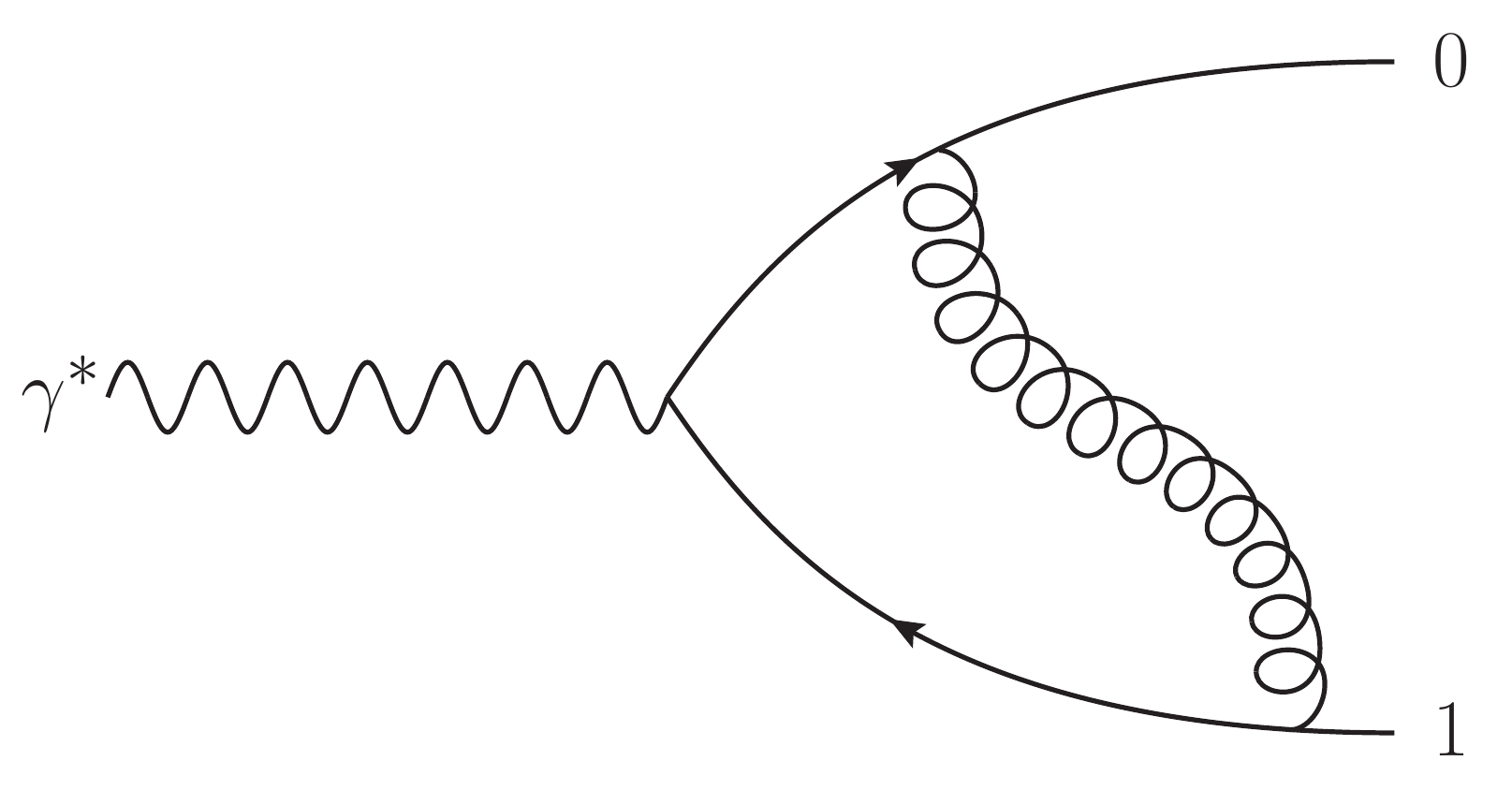}
			\caption{Example of a loop diagram: a gluon is emitted and reabsorbed.}
			\label{fig:nlo-dis-diag-loop}
		\end{subfigure}
		\caption{Two general types of diagrams contribute to the photon wavefunction at next-to-leading order: tree-level and loop diagrams.}
		\label{fig:nlo-dis-diagrams}
	\end{figure}

	At next-to-leading order (NLO) in the dipole picture, a gluon contributes to the virtual photon wavefunction in addition to the quark-antiquark dipole. The gluon is emitted by either the quark or the antiquark, and then can participate in the scattering from the target shockwave with the dipole, or it can be reabsorbed by one of the quarks. Contributions of the former type are tree-level at NLO, and the latter are loop contributions --- \fig\ref{fig:nlo-dis-diagrams} shows an instance for either type of contribution. In addition to this normal emission of the gluon, a transversely polarized photon can instantaneously split into a quark-antiquark-gluon tripole, which introduces additional tree-level and loop diagrams at NLO~\cite{Beuf:2011xd,Beuf:2016wdz,Beuf:2017bpd}.
	
	The first calculations of the photon impact factors, i.e. the photon splitting wavefunctions squared and summed over relevant particle helicities, were performed at NLO in momentum space and in a BFKL evolution context \cite{Bartels:2000gt, Bartels:2001mv, Bartels:2002uz, Bartels:2004bi}. Another computation of the impact factors that includes saturation effects has been done in full coordinate space \cite{Balitsky:2010ze}, however the results were not presented in the dipole factorized form which complicates their use in phenomenology. They were rewritten in NLO photon impact factor form after a linearization \cite{Balitsky:2012bs}, making them compatible with BFKL evolution, but not the gluon saturation regime.
	
	In this section we will discuss the recent derivation of the NLO photon impact factors done in mixed phase space \cite{Beuf:2011xd,Beuf:2016wdz,Beuf:2017bpd}, which incorporates gluon saturation effects, and where the results are written in a more suitable form for phenomenological studies. These impact factors were independently derived and verified in \cite{oma2}.

	\subsection{Inclusive deep inelastic scattering cross section at next-to-leading order}
	\label{sec:dis-nlo-crosssections}
	
		The calculation of the next-to-leading order DIS cross sections proceeds analogously to the LO derivation discussed in the previous section. One first derives from light-front perturbation theory the NLO corrections to the virtual photon Fock state --- normal and instantaneous gluon emissions and loop contributions to the quark-antiquark state. Together these give the virtual photon splitting wavefunctions for the transverse and longitudinal polarizations to NLO accuracy. With these, the optical theorem is used to get the total inclusive DIS cross section from the forward elastic scattering amplitude, like at leading order. Both the tree-level and loop contributions are separately UV-divergent at NLO and must be regulated: conventional dimensional regularization was used to derive the loop corrections to the photon Fock state in \cite{Beuf:2016wdz}, which were combined with the tree-level contributions in \cite{Beuf:2017bpd} for the final result, and the UV divergences canceled in the end. The total NLO cross sections were found to be
		\begin{equation}
			\sigmaltnlo = \sigmalt{\textrm{IC}} + \sigmaltdip + \sigmaltqg .
		\end{equation}
		The first term is a lowest order contribution --- in a strict perturbative expansion sense --- to the cross section, where the dipole amplitude is evaluated at the initial scale of the evolution:
		\begin{align}
			\label{eq:sigma_ic}
			\sigmalt{\textrm{IC}} (Y_0, Q^2)
			= 4 \nc \aem \frac{\as \cf}{\pi} \sum_f e_f^2 \int_0^1 \ud z_1
			\int\displaylimits_{\xt_0, \xt_1}
			&
			\kcal_{L,T}^\text{LO}
			\!
			\left(z_1,\xt_0,\xt_1\right)
			\nonumber
			\\
			& \times
			\left( 1 - \langle S(\xij{01}) \rangle_{Y_0} \right),
		\end{align}
		where the polarization specific kernels are
		\begin{align}
			\kcal_{L}^\text{LO}(z_1, \xt_0, \xt_1) 
			& = 4 Q^2 z_1^2 (1-z_1)^2
			\mathrm{K}_0^2(Q X_2),
			\label{eq:lo-impactfactor-L}
			\\[2ex]
			\kcal_{T}^\text{LO}(z_1, \xt_0, \xt_1)
			& = Q^2 z_1 ( 1 \! - \! z_1 ) \! \left( z_1^2 \! + \! (1 \! - \! z_1)^2\right)
			\mathrm{K}_1^2(Q X_2),
			\label{eq:lo-impactfactor-T}
		\end{align}
		with $X_2 \coloneqq z_1(1-z_1) \xt_{01}^2$, $\xij{ij} \coloneqq \xt_i - \xt_j$, and $S(\xij{ij}) = S(\xij{ij}, \bt)$, and the shorthand $\int_{\xt_i} \coloneqq \int \frac{\der^2 \xt_i}{2 \pi}$ was defined.
		The variables $\xt_0, \xt_1$ are the transverse coordinates of the quark and antiquark, and $z_1$ the momentum fraction of the antiquark. Even though \eqref{eq:lo-impactfactor-L} and \eqref{eq:lo-impactfactor-T} are the squared LO wavefunctions, i.e. the same as the Eqs.~\eqref{eq:lfwf-sq-lo-dis-t} and \eqref{eq:lfwf-sq-lo-dis-l} but with massless quarks, this is not the full leading order cross section \eqref{eq:lo-dis-cs} since the dipole amplitudes are not evolved\footnote{In a strict perturbative expansion sense, $\sigma^{\textrm{IC}}$ is the leading contribution in $\as$. However, the BK evolution produces a contribution of the order $\as \ln \frac{1}{\xbj} \sim 1$, which is conventionally included in the leading order cross section, upgrading it technically to leading logarithm precision.}. This ties into a discussion about soft gluons that we will look at in a moment. As for the NLO contributions, the 'dipole' term is:		
		\begin{align}
			\sigmaltdip
			&= 4 \nc \aem \frac{\as \cf}{\pi} \sum_f e_f^2 \int_0^1 \ud z_1
			\int\displaylimits_{\xt_0, \xt_1} 
			\kcal_{L,T}^\text{LO}
			\!
			\left(z_1, \xt_0 ,\xt_1 \right) 
			\!
			\nonumber
			\\ \label{eq:NLO_dip}
			&
			\hspace{2.6cm}
			\times
			\left( 1 - \langle S(\xij{01}) \rangle_{Y} \right)
			\left[\frac{1}{2}\ln^2\!\left(\!\frac{z_1}{1\!-\!z_1}\!\right)\!-\!\frac{\pi^2}{6}\!+\!\frac{5}{2}\right],
		\end{align}
		and the '$qg$' term is:
		\begin{align}
			\sigmaltqg & = 
			\notag
			8 \nc \alpha_\text{em} \frac{\as \cf}{\pi} \sum_f e_f^2
			\int_0^1 \der z_1 \int_{\zmin}^{1-z_1} \frac{\der z_2}{z_2}
			\\
			&
			\hspace{3.5cm}
			\times
			\int\displaylimits_{\xt_0, \xt_1, \xt_2}
			\mathcal{K}_{L,T}^{\textrm{NLO}}(z_1, z_2, \xt_0, \xt_1, \xt_2, Y).
			\label{eq:NLO_qg_nonsubtr}
		\end{align}
		The expressions for $\sigmaltdip$ or $\sigmaltqg$ above are not unique, but their sum is. This is due to the UV divergences that the respective contributions contain originally, and which are canceled between the two contributions~\cite{Beuf:2016wdz,Beuf:2017bpd}.
		The NLO kernels present in the $qg$ contribution are:	
	    \begin{align}
	    	\nonumber
			&\kcal_L^\text{NLO}(z_1, z_2, \xt_0, \xt_1, \xt_2, Y)
			=
			4 Q^2 z_1^2 (1-z_1)^2
				\\ \nonumber
				&
				\hspace{0.2cm}
				\times \bigg\{
				P \! \left(\frac{z_2}{1-z_1}\right) \! \frac{\xt_{20}}{\xt_{20}^2} \!\cdot\! 		\left(\frac{\xt_{20}}{\xt_{20}^2}-\frac{\xt_{21}}{\xt_{21}^2}\right) \!
			\left[ \mathrm{K}_0^2(Q X_3) \left( 1 - \langle S_{012} \rangle_{Y} \right)-(\xt_2 \to \xt_0) \right]
			\\
			& \hspace{1cm}
			+
				\left(\frac{z_2}{1-z_1}\right)^2 \frac{\xt_{20} \cdot \xt_{21}}{\xt_{20}^2 \xt_{21}^2} \mathrm{K}_0^2(Q X_3) 	\left( 1 - \langle S_{012} \rangle_{Y} \right)
			\bigg\} ,
		\end{align}
		\begin{align}
			&\kcal_T^\text{NLO}(z_1, z_2, \xt_0, \xt_1, \xt_2, Y) =
			Q^2 z_1(1-z_1) 
				\nonumber
				\\
				&\hspace{0.2cm}
				\times \bigg\{ 
				P\left(\frac{z_2}{1-z_1}\right) \left(z_1^2+(1-z_1)^2\right) \frac{\xt_{20}}{\xt_{20}^2} \cdot 			\left(\frac{\xt_{20}}{\xt_{20}^2}-\frac{\xt_{21}}{\xt_{21}^2}\right)
				\nonumber
				\\
				&\hspace{1.3cm}
				\times
				\Big[\mathrm{K}_1^2(Q X_3) \left( 1 - \langle S_{012} \rangle_{Y} \right)-(\xt_2 \to \xt_0) \Big]
				\nonumber
				\\
				&\hspace{1cm}
				+ \left(\frac{z_2}{1-z_1}\right)^2 \left[ \left(z_1^2+(1-z_1)^2\right) \frac{\xt_{20} \cdot
				\xt_{21}}{\xt_{20}^2 \xt_{21}^2} + 2 z_0 z_1 \frac{\xt_{20} \cdot \xt_{21}}{\xt_{20}^2 X_3^2} - 			\frac{z_0(z_1+z_2)}{X_3^2} \right]
				\nonumber
				\\
				&\hspace{1.3cm}
				\times
				\mathrm{K}_1^2(Q X_3) \left( 1 - \langle S_{012} \rangle_{Y} \right)
			\bigg\} .
		\end{align}
		Here $\xt_0, \xt_1, \xt_2$, and $z_0, z_1, z_2$ are the transverse positions and longitudinal momentum fractions of the quark, antiquark and gluon; the indexing is illustrated in Fig~\ref{fig:nlo-dis-diag-tree}. The fractions satisfy ${\sum_i z_i = 1}$. The definitions $X_3^2 \coloneqq z_0 z_1 \xt_{01}^2 + z_0 z_2 \xt_{02}^2 + z_2 z_1 \xt_{21}^2$ and $ P(z) \coloneqq 1 + (1-z)^2$ are made. The arrow notation denotes that the corresponding limit is to be taken of the preceding term: $f(a,b,c \to a) = f(a,b,a)$. The $\qqbarg$ state-target scattering Wilson line operator is \cite{Beuf:2017bpd, oma2}
		\begin{align}
			S_{012} 
			& \coloneqq
				\frac{1}{\nc \cf} \tr
					\left(t^b U_F(\xt_{0}) t^a U_F^\dagger(\xt_{1})\right) U_A(\xt_{2})_{ba}
			\nonumber
			\\
			& =
				\frac{1}{2 \nc \cf}
					\left[
						\tr \left( U_F(\xt_{0}) U_F^\dagger(\xt_{2}) \right) \tr \left( U_F(\xt_{2}) U_F^\dagger(\xt_{1}) \right)
						\right.
						\nonumber
						\\
						& \hspace{2.5cm} \left.
						- \frac{1}{\nc} \tr \left( U_F(\xt_{0}) U_F^\dagger(\xt_{1}) \right)
					\right]
			\nonumber
			\\
			& \equiv \frac{\nc}{2\cf}
				\left(
				S(\xij{02})S(\xij{21}) - \frac{1}{\nc^2}S(\xij{01})
				\right),
		\end{align}
		where in the first step the identity~\cite{Field:1989uq} used is:
		\begin{equation}
			t^a_{i j} t^a_{kl}
			=
			\half \left(
				\delta_{il} \delta_{jk} - \frac{1}{\nc} \delta_{ij} \delta_{kl}
			\right).
		\end{equation}

		Two details are yet to be determined: the rapidity scales $Y$ related to the dipole and $qg$ contributions --- since they are not provided by the perturbation theory calculation --- and the lower limit $\zmin$ of the gluon momentum fraction integration that regulates the logarithmically divergent integral. These turn out to be connected, and are the topic of the work done for Article~\cite{oma1}, which will be discussed in the next section.

	\subsection{Summary of Article \texorpdfstring{\cite{oma1}}{[I]}: factorization of the soft gluon large logarithm}
	\label{sec:summary-oma1}	
	
		The crux to be resolved by Article~\cite{oma1} was the negativity problem of NLO cross sections calculated in the Color Glass Condensate formalism, which was expected to afflict the NLO DIS cross sections as well. For context, in the case of single inclusive forward hadron production in $p+A$ it had been found that the NLO cross sections would become negative when the produced hadron would have a transverse momentum of the order of a few GeV \cite{Stasto:2013cha}. A resolution to this issue for single inclusive hadron production was proposed \cite{Iancu:2016vyg}, which was demonstrated to be effective \cite{Ducloue:2017mpb}. The topic of Article \cite{oma1} was the first numerical computation of the NLO DIS cross sections, whereby we demonstrated the presence of the negativity problem, and the application of the aforementioned solution to correct it. Resolving the negativity problem for the NLO DIS cross sections would bring the perturbative expansion under control, and make precision comparisons between theory and data possible in the future.
		
		The root cause of this problem is the handling of the large logarithm induced by the $z_2$-integration in the $qg$ contribution \eqref{eq:NLO_qg_nonsubtr} discussed in the previous section. Qualitatively, it needs to be factorized and absorbed into the BK renormalization group evolution. In practice, this means that the large logarithm that arises from the $qg$ contribution provides the rapidity evolution for the unevolved lowest order cross section \eqref{eq:sigma_ic}, upgrading it from LO to the full LL (leading log) cross section. If this factorization is done inaccurately, the negative $qg$ contribution has a leftover large logarithm, that can make the NLO cross sections negative. The development of an accurate subtraction scheme for NLO DIS cross sections was the principal goal of Article~\cite{oma1}. Once this was under control, we were able to quantify the importance of the NLO corrections.
		
		This issue would be corrected by accurately connecting the lower limit $\zmin$, and the rapidity scale of the Wilson-line operators in the $qg$ term, which were discussed in the previous section. It was understood that at NLO the $qg$ contribution drives the exact amount of evolution through $\zmin$. In connection to this, the $qg$ term dipole amplitudes must be evaluated at a rapidity scale that depends on the longitudinal momentum fraction of the gluon $z_2$: the evolution rapidity in this context is defined as $Y \coloneqq \ln z_2$. We derive in~\cite{oma1} that the $qg$ term rapidity scale should be
		\begin{equation}
			\label{eq:oma1:Yqg}
			Y_{qg} = \ln \frac{z_2}{\zmin} = \ln \frac{x_0}{X(z_2)} = \ln \frac{z_2 x_0 Q^2}{\xbj Q_0^2} \approx \ln \frac{z_2 x_0}{\xbj},
		\end{equation}
		where $x_0$ is the initial scale of the evolution, and $Q_0^2$ is some hadronic low transverse momentum scale. The last approximation was done since the $Q^2$-dependence introduces some complexity, since it affects the subtraction of the soft gluon large logarithm. The quantity $X(z_2)$ is a target momentum fraction scale similarly to $x_0$ and $\xbj$. We showed that by using the relation \eqref{eq:oma1:Yqg} instead of the simple $Y=\ln x_0/\xbj$ the negativity problem is resolved.
		
		The above procedure also naturally regulates the soft gluon large logarithm correctly, leading to the subtraction scheme dubbed 'unsub' scheme:
		\begin{equation}
			\sigmaltnlo = \sigmalt{\textrm{IC}} + \sigmaltdip + \sigmaltqgu ,
		\end{equation}
		where the first term is \eqref{eq:sigma_ic}, the NLO dipole contribution $\sigmaltdip$ is \eqref{eq:NLO_dip} for which the rapidity scale is chosen to be the same as in the LL cross sections $Y=\ln x_0/\xbj$, and the $qg$ term is
		\begin{align}
			\sigma_{L,T}^{qg,\textrm{unsub,\,approx.}} & = 
			\notag
			8 \nc \alpha_\text{em} \frac{\as \cf}{\pi} \sum_f e_f^2
			\int_0^1 \der z_1 \int_{\frac{\xbj}{x_0}}^{1-z_1} \frac{\der z_2}{z_2}
			\\
			&
			\hspace{2.5cm}
			\times
			\int\displaylimits_{\xt_0, \xt_1, \xt_2}
			\mathcal{K}_{L,T}^{\textrm{NLO}}(z_1, z_2, \xt_0, \xt_1, \xt_2, \ln \frac{z_2 x_0}{\xbj}).
			\label{eq:NLO_qg_unsub_oma1}
		\end{align}
		This formulation was found to yield physical cross sections. Without the approximation the $qg$ contribution writes
		\begin{align}
			\sigmaltqgu & = 
			\notag
			8 \nc \alpha_\text{em} \frac{\as \cf}{\pi} \sum_f e_f^2
			\int_0^1 \der z_1 \int_{\zmin}^{1-z_1} \frac{\der z_2}{z_2}
			\\
			&
			\hspace{3.5cm}
			\times
			\int\displaylimits_{\xt_0, \xt_1, \xt_2}
			\mathcal{K}_{L,T}^{\textrm{NLO}}(z_1, z_2, \xt_0, \xt_1, \xt_2, Y_{qg}),
			\label{eq:NLO_qg_unsub_oma3}
		\end{align}
		which is the form that was used in Article~\cite{oma3}, with the reparametrization $x_0 \equiv e^{-\Yoif}$.
		More details, and a discussion of how an explicit subtraction scheme for the soft gluon large logarithm is derived are found in Article \cite{oma1}. The factorization of the large logarithm is done with an extra term that subtracts the large log from the $qg$-term, and resums it into the BK evolution of the $\sigma^{\text{IC}}$ term, upgrading it to the LL accuracy cross section $\sigma^{\text{LO}}$. This subtraction was called the 'sub' scheme, and it writes
		\begin{equation}
			\sigmaltnlo = \sigmalt{\textrm{LO}} + \sigmaltdip + \sigmaltqgs ,
		\end{equation}
		where the LL rapidity scale is $Y=\ln x_0/\xbj$, and
		\begin{align}
			\sigma_{L,T}^{qg,\textrm{sub,\,approx.}} & = 
			\notag
			8 \nc \alpha_\text{em} \frac{\as \cf}{\pi} \sum_f e_f^2
			\int_0^1 \der z_1 \int_{\frac{\xbj}{x_0}}^{1} \frac{\der z_2}{z_2}
			\\
			&
			\hspace{0.4cm}
			\times
			\int\displaylimits_{\xt_0, \xt_1, \xt_2}
			\Bigg[
			\theta(1-z_1-z_2)
			\mathcal{K}_{L,T}^{\textrm{NLO}}\left(z_1, z_2, \xt_0, \xt_1, \xt_2, \ln \frac{z_2 x_0}{\xbj}\right)
			\notag
			\\
			&
			\hspace{2.5cm}
			-
			\mathcal{K}_{L,T}^{\textrm{NLO}}\left(z_1, 0, \xt_0, \xt_1, \xt_2, \ln \frac{z_2 x_0}{\xbj}\right)
			\Bigg].
			\label{eq:NLO_qg_sub_oma1}
		\end{align}
		
		One relevant detail about the 'unsub' and 'sub' schemes, that is not discussed in~\cite{oma1}, is that the equivalence of the two subtraction schemes is broken once a beyond-LO BK evolution is introduced. This creates a finite difference between the schemes of the order of NNLO. Relatedly, selecting between two different beyond-LO BK equations with the same subtraction scheme also introduces a finite difference of the order of NNLO. Sec.~\ref{sec:fits-uncert} discusses the estimation of the significance of these effects, and a similar effect concerning the running coupling is discussed in Sec.~\ref{sec:improvements}.
	
		Analogous subtraction schemes for DIS are put forth in Ref.~\cite{Beuf:2017bpd}, with some minor differences. In the unsubtracted scheme, the dipole amplitude in the NLO dipole term $\sigmaltdip$ is left unevolved like the lowest order contribution. This is valid at this precision of the calculation, since input for the correct scale should arise only at NNLO precision of the cross sections. In Article~\cite{oma1} our preference would have been to evaluate the NLO dipole term $\sigmaltdip$ at the same rapidity as the $qg$-term, i.e. at $Y_{qg}$ in Eq.~\eqref{eq:oma1:Yqg}, but this was not possible since the integration over the loop momentum fraction $z_2$ has been performed analytically in Eq.~\eqref{eq:NLO_dip}. The next section discusses the reversion of this integration.
	
	\subsection{Undoing the loop integration of \texorpdfstring{$\sigmaltdip$}{the NLO dipole contribution}} \label{sec:dip-z2}
	
	As is discussed in \cite{oma1}, the most natural thing to do is to evaluate both $\sigmaltdip$ and $\sigmaltqg$ at the same rapidity scale. However, this was not possible in \cite{oma1} since the loop momentum integration was performed analytically in \cite{Beuf:2016wdz}, and the available results were independent of the gluon longitudinal momentum fraction $z_2$. In this section we propose a form for $\sigmaltdip$, where this integration has been undone, and discuss its features. The derivation here uses the conventions and intermediate results of Ref.~\cite{Beuf:2016wdz}.
	
	To begin the undoing of the $z_2$-integrations done in the derivation of \eqref{eq:NLO_dip}, let us first revert some manipulations:
	\begin{equation}
		\half \log^2 \left( \frac{z_1}{1-z_1}\right) - \frac{\pi^2}{6} + \frac{5}{2}
		=
		-\Li{-\frac{z_1}{1-z_1}} -\Li{-\frac{1-z_1}{z_1}} - \frac{\pi^2}{3} + \frac{5}{2},
	\end{equation}
	where the relation used is
	\begin{equation}
		\Li{z} +\Li{\frac{1}{z}} = - \half \log^2(z) - \frac{\pi^2}{6}.
	\end{equation}
	The result is the sum of contributions symmetric in the exchange of the quark and antiquark, so in fact we are looking for one of these halves, which is
	\begin{equation} \label{eq:li2-pi2-const}
		-\Li{-\frac{1-z_1}{z_1}} - \frac{\pi^2}{6} + \frac{5}{4} + \half \half,
	\end{equation}
	where the last $\frac{1}{4}$ is a half of the UV scheme dependent finite leftover, that will cancel in the full NLO cross section. This contribution arises in a different way for T and L polarizations.
	
	In the case of the longitudinal photon, using the notations and definitions of Ref.~\cite{Beuf:2016wdz}, these terms are found in the following sum of contributions~\cite{Beuf:2016wdz}:
	\begin{equation}
		\vcal^L_A +\vcal^L_1 + \vcal^L_{3a} = \cdots -\Li{-\frac{\kzero}{\kone}} - \frac{\pi^2}{6} + \frac{3}{2} + \cdots,
	\end{equation}
	where the unrelated terms have been omitted. To discover the unintegrated form, we need to look closely at the integral forms of these contributions. To this end, it is useful to be aware of the following integral relations \cite{Beuf:2016wdz}:
	\begin{align}
		\int_0^1 \frac{\der \xi}{\xi} \log(1+R\xi)
		&= -\Li{-R},
		\label{eq:int-li}
		\\
		\int_0^1 \frac{\der \xi}{\xi} \log(1-\xi)
		&= -\frac{\pi^2}{6}.
		\label{eq:int-pi}
	\end{align}
	The integral forms of $\vcal^L_A$ and $\vcal^L_1 + \vcal^L_{3a}$ are:
	\begin{eqnarray}
		{\cal V}_{A}^L = {\cal V}_{A}^T &=& -\int_{0}^{k_0^+}\!\!\frac{\ud {k_2^+}}{k_{0}^+}
		\left\{ \mathcolor{jyg}{2} \left(\frac{\mathcolor{jyg}{k_0^+}\!-\!k_2^+}{\mathcolor{jyg}{k_2^+}}\right) +\frac{k_2^+}{k_0^+} \right\} \!
		\Bigg\{
		\Gamma \!  \left(2\!-\! \frac{D}{2}\right) \!
		\left[\frac{-2\, k_0^+ (ED_{\mathrm{LO}})}{4\pi\, \mu^2}\right]^{\frac{D}{2}-2}
		\nonumber\\
		&& \hspace{1cm}
		- \log\left(\frac{k_2^+}{k_0^+}\right)
		\mathcolor{jyg}{- \log\left(\frac{k_0^+\!-\!k_2^+}{k_0^+}\right)}\Bigg\}
		\nonumber\\
		&& \hspace{2cm}
		+\int_{0}^{k_0^+}\!\!\frac{\ud {k_2^+}}{k_{0}^+}\; \left(\frac{k_2^+}{k_0^+}\right)\;\;
		+ O\left(D\!-\!4\right) 
		\nonumber\\
		&=& 2 \left[\log\left(\frac{k^+_{\min}}{k_0^+}\right) +\frac{3}{4} \right]\;
		\Gamma\!\left(2\!-\! \frac{D}{2}\right)\;
		\left[\frac{-2\, k_0^+ (ED_{\mathrm{LO}})}{4\pi\, \mu^2}\right]^{\frac{D}{2}-2}
		\nonumber\\
		&& -\left[\log\left(\frac{k^+_{\min}}{k_0^+}\right)\right]^2
		\mathcolor{jyg}{-\frac{\pi^2}{3}} + 3+\frac{1}{2}
		+ O\left(D\!-\!4\right)\, .
		\label{eq:VA_3}
	\end{eqnarray}
	and
	\begin{eqnarray}
		{\cal V}_{1}^{L}+ {\cal V}_{3a}^{L}&=& \int_{0}^{k_0^+}\!\! \frac{\ud{k_2^+}}{k_0^+}\;
		\Bigg\{ \mathcolor{jyo}{\Bigg(} \mathcolor{jyg}{\frac{k_0^+}{k_2^+}} \mathcolor{jyo}{\Bigg)}
		-\frac{k_1^+}{q^+}
		-\frac{k_2^+}{q^+}
		\Bigg\}\;
		\nonumber\\
		&&
		\quad
		\times
		\Bigg\{
		\Gamma\!\left(2\!-\! \frac{D}{2}\right) \!
		\left(\frac{\overline{Q}^2}{4\pi\, \mu^2}\right)^{\frac{D}{2}-2}
		\! \! \! \!
		\mathcolor{jyo}{-\log\left(\frac{k_1^+\!+\!k_2^+}{k_1^+}\right)}
		\!
		\mathcolor{jyg}{-\log\left(\frac{k_0^+\!-\!k_2^+}{k_0^+}\right)}
		\! \!
		\Bigg\}
		\nonumber\\
		&& 
		+\int_{0}^{k_0^+}\!\! \frac{\ud{k_2^+}}{k_0^+}\;
		\Bigg\{ \mathcolor{jyo}{\frac{2k_0^+}{k_2^+}} +\left(\frac{k_0^+\!-\!k_1^+}{k_1^+}\right)
		-\frac{k_2^+}{k_1^+}
		\Bigg\}\;
		\nonumber\\
		&&
		\quad
		\times
		\Bigg\{
		- \log\left(\frac{k_2^+}{k_0^+}\right)
		\mathcolor{jyo}{ + \log\left(\frac{k_1^+\!+\!k_2^+}{k_1^+}\right)}
		- \log\left(\frac{q^+}{k_1^+}\right)
		\Bigg\}
		\nonumber\\
		&& 
		-\int_{0}^{k_0^+}\!\! \frac{\ud{k_2^+}}{k_0^+}\;
		\left(\frac{k_0^+\!-\!k_2^+}{q^+}\right)
		+ O\left(D\!-\!4\right)
		\nonumber\\
		&=&
		- \left[\log\left(\frac{k^+_{\min}}{k_0^+}\right) +\frac{1}{2}+\frac{k_1^+}{2q^+} \right]\,
		\Gamma\!\left(2\!-\! \frac{D}{2}\right)\,
		\left(\frac{\overline{Q}^2}{4\pi\, \mu^2}\right)^{\frac{D}{2}-2}
		\nonumber\\
		&&
		+\left[2\log\left(\frac{k^+_{\min}}{k_0^+}\right)+\frac{3}{2}\right]\,
		\log\left(\frac{q^+}{k_1^+}\right)
		\nonumber\\
		&&
		+\left[\log\left(\frac{k^+_{\min}}{k_0^+}\right)\right]^2
		\mathcolor{jyg}{+\frac{\pi^2}{6}} \mathcolor{jyo}{-\textrm{Li}_2\left(-\frac{k_0^+}{k_1^+}\right)}
		-2+\frac{k_0^+}{2q^+}
		\nonumber\\
		&&
		-\frac{k_0^+}{2q^+}
		+ O\left(D\!-\!4\right)
		\label{eq:V1L_plus_V3aL_2}
		\, ,
	\end{eqnarray}
	where the terms contributing to $\frac{\pi^2}{6}$ and $\mathrm{Li_2}$ have been highlighted in gold and orange, respectively, and the two-colored factor contributes to both.
	Now we may collect from \eqref{eq:VA_3} and \eqref{eq:V1L_plus_V3aL_2} the terms contributing to the polylogarithm and $\frac{\pi^2}{6}$ terms in \eqref{eq:li2-pi2-const}:
	\begin{align}
		-\Li{-\frac{1-z_1}{z_1}} & = -\textrm{Li}_2\left(-\frac{k_0^+}{k_1^+}\right)
		=
		\int_{0}^{k_0^+}\!\! \frac{\ud{k_2^+}}{k_2^+}\; \log \left( \frac{\kone + \ktwo}{\kone} \right) \, ,
		\\
		-\frac{\pi^2}{6} & = \int_{0}^{k_0^+}\!\! \frac{\ud{k_2^+}}{k_2^+}\; \log \left( \frac{\kzero - \ktwo}{\kzero} \right) \, ,
	\end{align}
	where the temporary variable $\xi \coloneqq \frac{\ktwo}{\kzero}$ is used in \eqref{eq:int-li}, \eqref{eq:int-pi}. Then, adding this together with the quark-antiquark exchange terms, we can write:
	\begin{align}
		\half & \log^2 \left( \frac{z_1}{1-z_1}\right) - \frac{\pi^2}{6}
		=
		\left( \vcal^L_A +\vcal^L_1 + \vcal^L_{3a} \right) \bigg\rvert_{-\mathrm{Li}_2 - \frac{\pi^2}{6}} + \left(\kzero \leftrightarrow \kone \right)
		\nonumber
		\\
		&= \int_{0}^{k_0^+}\!\! \frac{\ud{k_2^+}}{k_2^+}\; \left\lbrace \log \left( \frac{\kone + \ktwo}{\kone} \right) + \log \left( \frac{\kzero - \ktwo}{\kzero} \right) \right\rbrace + \left(\kzero \leftrightarrow \kone \right)
		\nonumber
		\\
		&=
		\int_{0}^{1-z_1} \ud z_2 
		\left[
			\frac{1}{z_2}
			\left(
				\log \left( 1 + \frac{z_2}{z_1} \right) + \log \left( 1 - \frac{z_2}{1-z_1} \right)
			\right)
		\right]
		\nonumber
		\\
		& \quad +
		\int_{0}^{z_1} \ud z_2 
		\left[
			\frac{1}{z_2}
			\left(
				\log \left( 1 + \frac{z_2}{1-z_1} \right) + \log \left( 1 - \frac{z_2}{z_1} \right)
			\right)
		\right] \, ,
		\label{eq:int-log2-pi2}
	\end{align}
	where the vertical bar notation refers to picking only the relevant terms, and in the last equality the change of variables $z_i \coloneqq \frac{\plum{k}_i}{\plusq}$ was used together with the identity\footnote{The loop momentum fraction $z_2$ is not constrained by the plus-momentum conservation, unlike in the case of the tree-level $q \bar q g$ contributions discussed in Sec.~\ref{sec:dis-nlo-crosssections}.} $z_0 + z_1 = 1$.
	
	The derivation of the respective result for a transverse photon proceeds in an analogous way, however now the contributions come from the diagrams $\vcal^T_A + \vcal^T_1 + \left(\kzero \leftrightarrow \kone \right)$. The integral form \eqref{eq:int-log2-pi2} found for the longitudinal photon is found for the transverse polarization as well.
	
	Now, we are left with the constant $\frac{5}{2}$ that has been excluded in the previous discussion, since its derivation is different in a few key ways. First, even though the constant is the same for both transverse and longitudinal photons, it arises from different contributions. While the polylogarithm and $\pi^2$ terms were completely UV regularization scheme independent, the $3 = \frac{5}{2} + \half$ constant seen in the result of the loop contribution receives UV scheme dependent contributions from the quark and antiquark self-energy and total vertex corrections. These will cancel with the UV scheme dependent terms from the $q \bar q g$-contribution~\cite{Beuf:2017bpd}. In addition to these scheme dependent contributions, the constant receives contributions from a large number of integrals, some of which are IR divergent, and many of which have substantial cancellations at the integrated level. The IR divergences are regulated with an IR cut-off, and the cut-off dependent terms cancel out completely~\cite{Beuf:2016wdz}. This means that some of the integrals that contribute to the $\frac{5}{2}$ we are interested in would need to be IR regulated, which could make them precarious for numerical evaluation and the cancellation of the divergences. Lastly, while a half of the constant arises from both of the quark-antiquark symmetric partitions independently for the longitudinal photon, this is not the case for the transverse photon, and it would be necessary to have all the integrals from $\vcal^T_A + \vcal^T_1 + \left(\kzero \leftrightarrow \kone \right)$ separately in order to get the correct cancellations. Due to these complications we chose to keep the final UV scheme independent constant at the integral level as well, as an averaged effect for these contributions.
	
	Now with \eqref{eq:int-log2-pi2} and including the $\frac{5}{4}$ as discussed above, we may rewrite \eqref{eq:NLO_dip} back into an integral form:
	\begin{align} \label{eq:NLO_dip_z2}
		\notag
		\sigma_{L,T}^{\text{dip},z_2}
		=
		& 4 \nc \aem \frac{\as \cf}{\pi} \sum_f e_f^2 
		\int_{\zmin}^{1-\zmin} \ud z_1 
		\int_{\xt_0, \xt_1} \!\! {\kcal}_{L,T}^\text{LO}(z_1,\xt_0,\xt_1)
		\\ \notag
		&
		\times
		\left\lbrace
		\int_{\zmin}^{z_1} \ud z_2 
		\left[
		\frac{1}{z_2}
		\left(
		\log \left( 1 + \frac{z_2}{1-z_1} \right) + \log \left( 1 - \frac{z_2}{z_1} \right)
		\right)
		+
		\frac{1}{z_1} \frac{5}{4}
		\right]
		\right.
		\\ \notag
		&
		\quad \quad
		\times
		\left[
		1 - \langle S_{01} \rangle_{Y_2^{+}}
		\right]
		\\ \notag
		&
		+
		\int_{\zmin}^{1-z_1} \ud z_2 
		\left[
		\frac{1}{z_2}
		\left(
		\log \left( 1 + \frac{z_2}{z_1} \right) + \log \left( 1 - \frac{z_2}{1-z_1} \right)
		\right)
		+
		\frac{1}{1-z_1} \frac{5}{4}
		\right]
		\\ \notag
		&
		\quad \quad
		\times
		\left.
		\left[
		1 - \langle S_{01} \rangle_{Y_2^{+}}
		\right]
		\vphantom{\int_{\zmin}^{z_1}} \right\rbrace
		\\ \notag
		=
		& 8 \nc \aem \frac{\as \cf}{\pi} \sum_f e_f^2 
		\int_{\zmin}^{1-\zmin} \ud z_1 
		\int_{\xt_0, \xt_1} \!\! {\kcal}_{L,T}^\text{LO}(z_1,\xt_0,\xt_1)
		\\ \notag
		&
		\times
		\int_{\zmin}^{z_1} \ud z_2 
		\left[
		\frac{1}{z_2}
		\left(
		\log \left( 1 + \frac{z_2}{1-z_1} \right) + \log \left( 1 - \frac{z_2}{z_1} \right)
		\right)
		+
		\frac{1}{z_1} \frac{5}{4}
		\right]
		\\
		&
		\quad \quad
		\times
		\left[
			1 - \langle S_{01} \rangle_{Y_2^{+}}
		\right]
		,
	\end{align}
	where in the second equality the two terms symmetric in the quark-antiquark exchange are combined. When the lower limit $\zmin$ is introduced for $z_2$ in a subtraction scheme, the phase space of the $z_1$-integration must be consistently limited as well. The upper limits $z_1$ or $1-z_1$ cannot become smaller than $\zmin$. As we wish to give the dipole amplitude a $z_2$ dependent rapidity scale, we move the dipole amplitude inside the integral make the replacement of the evaluation scale
	\begin{equation*}
		\left[
		1 - \langle S_{01} \rangle_{Y^{+}_f}
		\right]
		\rightarrow
		\left[
		1 - \langle S_{01} \rangle_{Y_2^{+}}
		\right] \, ,
	\end{equation*}
	where $Y^{+}_f$ is the rapidity scale of the LO+LL DIS cross sections~\cite{Beuf:2017bpd}, and the new rapidity scale is
	\begin{equation}
		\label{eq:Y2plus}
		\Ytwoplus \coloneqq \log \left( \frac{z_2}{\zmin} \right) = \log \left( z_2 \frac{x_0 Q^2}{\xbj Q_0^2}\right),
	\end{equation}
	which is the same as was used for the $qg$ contribution in~\cite{oma3}, and in the approximation $Q^2/Q_0^2 \sim 1$ the same as was used in~\cite{oma1}.

	In conclusion, we have derived an alternative, and previously unpublished, form~\eqref{eq:NLO_dip_z2} for the NLO dipole contribution Eq.~\eqref{eq:NLO_dip}. The new form has been rewritten back into a loop-integral form, which permits the consistent usage of the same $z_2$-dependent rapidity scale \eqref{eq:Y2plus} in both the $q \bar q g$ term~\eqref{eq:NLO_qg_nonsubtr} and the new NLO dipole term~\eqref{eq:NLO_dip_z2}. This would have been the preferred rapidity scale for the NLO dipole term in the works~\cite{oma1,oma3}.

	\subsection{Summary of Article \texorpdfstring{\cite{oma2}}{[II]}: NLO DIS cross sections in the four-dimensional helicity scheme}
				
		The principal purpose of the Article~\cite{oma2} is to develop new methodology for light-front perturbation theory loop calculations, where the key feature is that the elementary vertices are written in a helicity basis. Specifically, in this article the calculation rules originally introduced in~\cite{Lappi:2016oup} are reformulated in a better way, and a correction to the scheme is made. A key point in the new scheme is the proper handling of the Kronecker-deltas of different dimensionalities. As a demonstration, this calculation scheme is then used to derive the virtual photon splitting wavefunctions at next-to-leading order in the dipole picture. These photon wavefunctions are then used to write the NLO DIS cross sections, which are then compared with --- and found equivalent to --- the ones derived in Refs.~\cite{Beuf:2016wdz,Beuf:2017bpd}.
	
		The usage of the explicit helicity basis for the quarks and gluons is naturally combined with the regularization of the UV divergences in the four-dimensional helicity (FDH) scheme. As opposed to the conventional dimensional regularization (CDR) used in Refs.~\cite{Beuf:2016wdz,Beuf:2017bpd} --- where all vectors and momenta are continued to $d$ dimensions --- the FDH scheme keeps all observed particles in four dimensions and continues unobserved particle momenta to $d>4$ dimensions, and spins/polarizations of unobserved internal states are $d_s > d$ dimensional.
		
		This new calculation of the NLO DIS cross sections provides a verification of the UV regularization scheme independence of the results in Ref.~\cite{Beuf:2017bpd}, since while the intermediate steps differ in FDH from CDR, the final results were shown to agree both analytically and numerically. Specifically, the numerical test verifies that the different functional form used for the UV subtraction produces equivalent results with the one used in Ref.~\cite{Beuf:2017bpd}.

	\chapter{Next-to-Leading Order DIS fits to HERA data}
\label{ch:fits}

\section{Fundamentals}

	This section briefly describes the steps needed to go from the NLO cross sections \eqs\eqref{eq:NLO_dip}, \eqref{eq:NLO_qg_nonsubtr} to their numerical evaluation and comparison to data. Section \ref{sec:nlodis-numerics} discusses the solution of the BK equation to get the dipole amplitude functional form over a range of rapidity, and the numerical evaluation of the NLO DIS cross sections, including their dimensional reduction for numerical evaluation, and \se\ref{sec:fitting} covers the basics of the fitting methodology, along with few remaining pieces of the computation needed to be able to make comparisons to data.

	\subsection{Numerical evaluation of the NLO DIS cross sections}
		\label{sec:nlodis-numerics}
		Here we discuss the steps that are taken in order to evaluate the NLO DIS cross sections \eqref{eq:NLO_dip}, \eqref{eq:NLO_qg_unsub_oma3} numerically, and efficiently, leading to the results first presented in Article~\cite{oma1}. The discussion uses the notations and conventions from Sec.~\ref{sec:dis-nlo}.
		
		The first object we need to understand in our implementation is the scattering amplitude of the dipole, $\langle S_{01} \rangle$. After the BK evolution described in \se\ref{sec:bk-evol} that started from a given initial functional form, and assuming that it is not dependent on the impact parameter or the orientation of the $q\bar{q}$ dipole, it is a scalar function of two parameters: $\langle S_{01} \rangle_Y \equiv S(\left|\xt_{01}\right|, Y)$. Once we have a numerical solution for $S_{01}$ from the BK equation that is interpolatable both in $\left|\xt_{01}\right|$ and $Y$, over both of which we will need to integrate, we may begin calculating the cross sections.
		
		Looking at the expressions \eqref{eq:NLO_dip}, \eqref{eq:NLO_qg_unsub_oma3}, we see that the former has similar structure to the LO cross section \eqref{eq:sigma_ic}, and the same dimensionality of integration phase space. The latter, on the other hand, has a larger phase space to integrate over. In order to improve the efficiency of the numerical implementation, we wish to reduce the dimension of the integration phase space as far as possible, which uses the assumed symmetry properties of $S_{01}$ discussed above. Secondly it would be ideal to write the vector dot products in terms of scalar quantities in order to have a scalar numerical implementation.
		
		Let us first go through the dimensional reduction that takes place for the leading order contribution, and therefore for the loop contribution as well. Originally the transverse structure of the impact factors comprises of the transverse positions of the quark and antiquark, and after a change of variables, the dipole size $\rt \coloneqq \xt_{0} - \xt_{1}$ and the impact parameter $\bt \coloneqq (\xt_0+\xt_1)/2$, which is a four-dimensional phase space. Now, since the photon splitting wavefunctions \eqref{eq:sigma_ic}, and $S_{01}$ are independent of the impact parameter it can be integrated over the target shape profile yielding the target size, which we take to be a constant factor $\sigma_0/2$. Furthermore, since the dipole amplitude was assumed to be agnostic of the quark-antiquark dipole orientation, the cross sections are independent of the angle of $\rt$, which can be also integrated analytically. This means that of the transverse integrations, only the integral over $\left|\rt\right|$ remains --- the transverse phase space has been reduced from four to one dimension.
		
		Now we want to apply similar simplifications to the $qg$ contribution \eqref{eq:NLO_qg_unsub_oma3}, to apply the same symmetry assumptions, and to achieve a similar reduction in the dimensionality, since this term will absolutely need it in terms of the integration efficiency. Out of the box the $qg$-term transverse phase space is six dimensional: integrations take place over the quark-antiquark-gluon transverse positions $\xt_0, \xt_1, \xt_2$.
		A similar reparametrization to impact factor variables, with the addition $\B{z} = \xt_2 - \bt$, is one way to extract an independent transverse integral that can be recognized as the target size.
		The scalar parametrization used in this work uses the lengths $x_{01},~ x_{02}$, and the angle $\angle(\xt_{01},\xt_{02}) \eqqcolon \phi$ --- with these the third dipole size is $x_{21}^2 = x_{01}^2 + x_{02}^2 - 2 x_{01} x_{02} \cos \phi$.
		With these scalar quantities the $qg$ impact factors are sufficiently parametrized. This leaves one free angular and one transverse planar integral that can be done analytically, given the above assumptions about the dipole amplitude. The latter yields again the target size $\sigma_0/2$.
		This leaves us with a scalar form of the $qg$ contribution with a three-dimensional transverse phase space.
		
		With the above considerations we are at the stage to implement the remaining integrations numerically. In the work for \cite{oma1, oma2, oma3}, this was done in C++ using the Cuba library for multidimensional numerical integration \cite{Hahn:2004fe,Hahn:2014fua}. Specifically Cuba provides powerful methods for Monte Carlo integration, which becomes more efficient in higher phase space dimensions than deterministic integration, as is the case with the $qg$ contribution. The Monte Carlo methods implemented by Cuba use importance sampling and globally adaptive subdivision, which substantially improve the efficiency of the evaluation of complex high-dimensional integrals. Even with all the possible simplifications done the $qg$ contribution is the hardest by far to compute, especially so for the transversely polarized photon, which is possibly caused by the behavior of the integrands in the aligned jet limit $z_0 \to 0, 1$.

	\subsection{Extraction of the BK evolution initial amplitude shape from data}
		\label{sec:fitting}

		Based on the previous section we know how to numerically compute the NLO DIS cross sections and structure functions. This can be used to extract the initial condition for the BK evolution given an ansatz functional form for the amplitude. This is done by fitting the initial condition for the BK evolution through the comparison of the calculated cross sections to experimental data. DIS total cross section measurements are typically reported either for the structure functions, or for the reduced cross section, which is defined as
		\begin{equation}
			\sigma_r(\xbj, Q^2, y) = F_2(\xbj, Q^2) - \frac{y^2}{1+(1-y)^2} F_L(\xbj, Q^2).
		\end{equation}
		Here $y$ is the so-called inelasticity of the scattering, which is fraction of energy in target rest frame the photon gains from the electron.
		
		Now to fit the initial condition of the BK evolution to data, an ansatz for the functional form is needed, and some initial guess for its parameters. A widely used form is the McLerran-Venugopalan model for the dipole amplitude \cite{McLerran:1993ni}:
		\begin{equation}
			\label{eq:mv-model}
			S(\xij{ij}, Y = \Yobk) =
				\exp 
				\left[
					- \frac{\xij{ij}^2 Q_{s,0}^2}{4} \right.
					\left. \ln \left( \frac{1}{|\xij{ij}| \Lambda_\text{QCD}}
					+ e \right)
				\right],
		\end{equation}
		where there is only one free shape parameter, $Q_{s,0}$, which is the saturation scale at the earliest rapidity to be considered in the analysis. A modified version of the McLerran-Venugopalan model is the MV-$\gamma$ model, which writes		
		\begin{equation}
			\label{eq:mv-gamma-model}
			S(\xij{ij}, Y = \Yobk) =
			\exp 
			\left[
				- \frac{\left(\xij{ij}^2Q_{s,0}^2\right)^\gamma}{4} \right.
				\left. \ln \left( \frac{1}{|\xij{ij}| \Lambda_\text{QCD}}
				+ e \right)
			\right] \, ,
		\end{equation}
		where the new parameter $\gamma$ controls the steepness of the amplitude tail at small dipole sizes. The MV-$\gamma$ model has been used in leading order DIS fits \cite{Albacete:2009fh, Albacete:2010sy, Lappi:2013zma}, and so was chosen as the initial shape for the fits done in \cite{oma3}.
		
		On top of the two free parameters discussed above, the running coupling scheme used in \cite{oma3} introduces a third parameter --- the used strong coupling constant in coordinate space is
		\begin{equation}
			\label{eq:alphas}
			\as(\xij{ij}^2) = \frac{4 \pi}{ \beta_0 \ln \left[ \left( \frac{\mu_0^2}{\lqcd^2}\right)^{1/c} + \left(\frac{4 C^2}{\xij{ij}^2 \lqcd^2}\right)^{1/c} \right]^c },
		\end{equation}
		with $\beta = (11 \nc - 2 \nf)/3$ and $\nf=3$, $\lqcd=0.241\gev$. The parameter $C^2$ sets the scaling of the running coupling in coordinate space, i.e. it is the scale in the connection $\as(k^2) \sim \as(C^2/r^2)$. Theory calculations give it the value $C^2 = e^{-2 \gamma_E}$ \cite{Kovchegov:2006vj, Lappi:2012vw}, however it can be used as free parameter to absorb theoretical uncertainty related to non-perturbative or higher-order contributions. In \cite{oma3} we use it as a fit parameter in this functionality, as has been done in previous LO DIS analyses \cite{Albacete:2010sy, Lappi:2013zma}. The remaining fixed parameters $c$ and $\mu_0$ control the infrared freezing of the coupling, see~\cite{oma3} for more details.
		
		Two schemes of selecting the dipole size going into the running coupling \eqref{eq:alphas} were used in the analysis. First, the simple one, is the parent dipole running coupling where the coupling strength is always set by the quark-antiquark dipole size $\rt= \xt_{01} = \xt_{0}-\xt_{1}$, i.e. $\as = \as(\rt^2)$. The second coupling uses the Balitsky prescription~\cite{Balitsky:2006wa} in the LO BK kernel, and the smallest dipole prescription in the resummation kernels and DIS impact factors. In the smallest dipole scheme the smallest daughter dipole sets the coupling strength
		\begin{equation}
			\alpha_{\textrm{s}, \mathrm{sd}} \left( \xij{01}^2, \xij{02}^2, \xij{21}^2 \right)
			= 
			\as \left( \min \left\lbrace \xij{01}^2, \xij{02}^2, \xij{21}^2 \right\rbrace \right).
		\end{equation}
		This is motivated by the observation that the typical scale that sets the coupling strength is the largest momentum scale, which in position space corresponds to the smallest length scale.
		One important feature of the second scheme is that the Balitsky coupling reduces to the smallest dipole coupling in the limit that one of the daughter dipoles is much smaller than the others.
		
		With the initial condition and running coupling schemes determined, we may proceed to make comparisons between theory calculations and data. In the process of fitting a quality function for the agreement of the theory calculation and data is needed, for which then a global minimum or maximum is searched, which ever is the extremum corresponding to the best fit. One such quality function is the $\chi^2$, which is
		\begin{equation}
			\chi^2 = \sum_{\text{data points} ~ i} \left( \frac{\sigma_{\text{th}}(i) -\sigma_{\text{exp}}(i) }{\epsilon_{\text{exp}}(i)} \right)^2 \, ,
		\end{equation}
		where the measured data set is indexed with $i$, $\sigma_{\text{exp}}(i)$ is the measurement for the datapoint, $\sigma_{\text{th}}(i)$ the theoretical calculation at the datapoint, and $\epsilon_{\text{exp}}(i)$ the total error of the measurement for datapoint $i$. Perfect agreement between the theory and data would give $\chi^2 = 0$, and the larger $\chi^2$ is the worse the fit quality is. Due to the random nature of experimental errors, theory cannot account for them point by point, and so $\chi^2=0$ is not actually a good fit. Typically in a well predicted region the difference between the theory calculation and the measurement should be of the order of the experimental error, which gives $\chi^2 \sim N$, or as is typically reported $\chi^2/N \sim 1$, where $N$ is the number of datapoints considered in the analysis. Then in the fitting process inputs are parametrizations of the initial condition and outputs are values of $\chi^2/N$, for which a global minimum is found. This gives the initial condition preferred by the data. One final note is that a more sophisticated analysis could take into account the correlations of experimental errors. These are available~\cite{Aaron:2009aa} for the combined HERA data, but they were not included in the analysis done in~\cite{oma3}.

		Now that we have gone through the procedure of computing cross sections from the theory and comparing them to data, we can move on to summarize the work and results of Article \cite{oma3}.

\section{Summary of Article \texorpdfstring{\cite{oma3}}{[III]}: Fitting NLO DIS cross sections to HERA data}
	\label{sec:summary-oma3}

	The work done in Article~\cite{oma3} brings together the state-of-the-art dipole picture calculation of the next-to-leading order DIS cross sections \cite{Beuf:2016wdz, Beuf:2017bpd}, and the soft gluon large logarithm resummation work done for NLO DIS in \cite{oma1}, covered in the Sections \ref{sec:dis-nlo-crosssections} and \ref{sec:summary-oma1}, respectively. Together these works provide a scheme for a stable perturbative expansion of the DIS cross section in the dipole picture up to next-to-leading order, which makes NLO precision comparisons to data possible for the first time. The work done in \cite{oma3} uses these theory results to determine the initial condition to the BK evolution using NLO accuracy fits to the combined HERA data~\cite{Aaron:2009aa}. The theoretical uncertainty of the calculation is gauged by running fits using alternative prescriptions for the initial condition, BK evolution, and running coupling.
	
	\begin{figure}[t]
		\centering
		\includegraphics[width=0.75\textwidth]{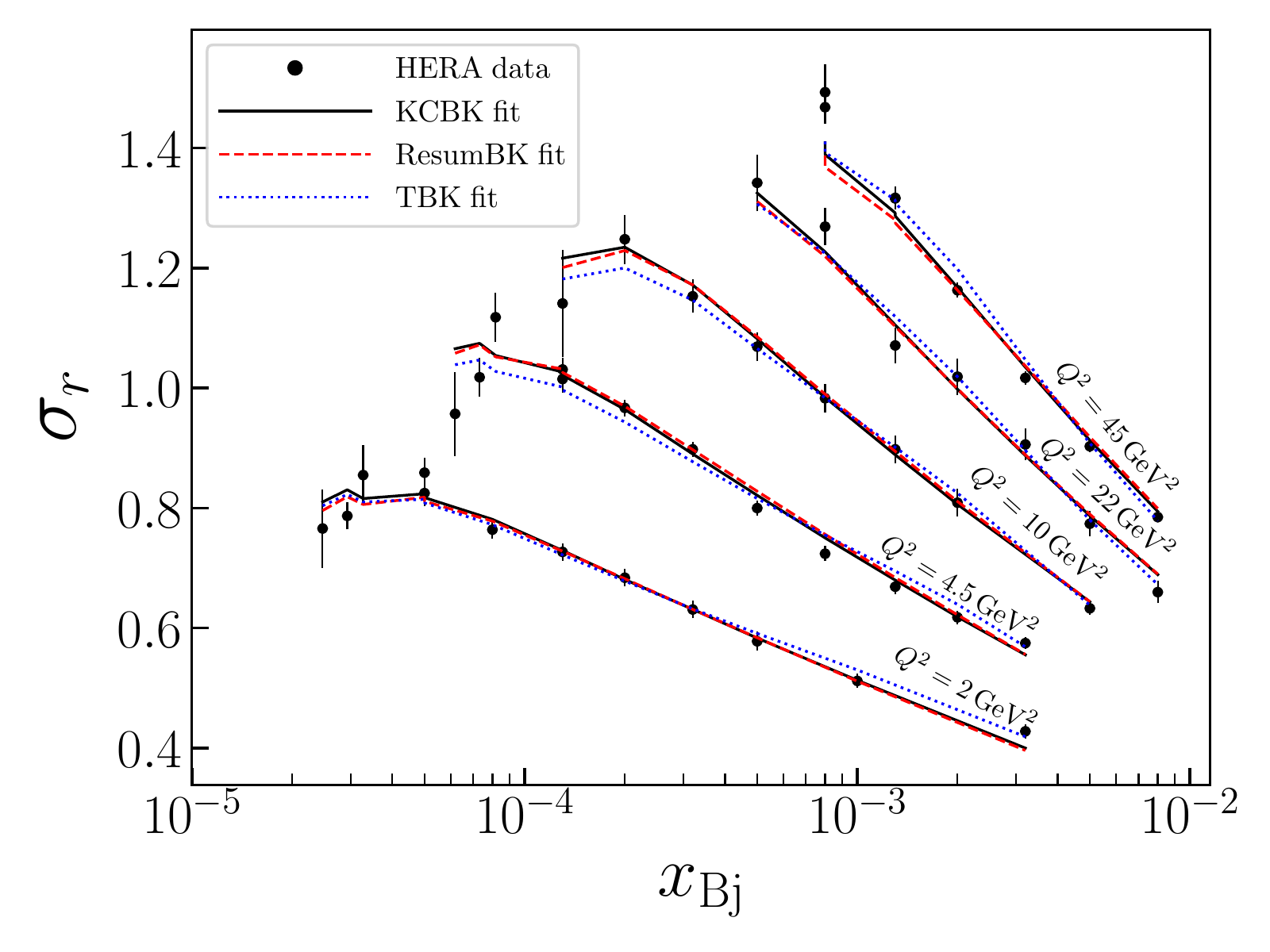}
		\caption{HERA data compared to reduced cross section calculated based on three fits, one with each of the enhanced BK evolutions. Balitsky + smallest dipole running coupling and $\Yobk = \ln \frac{1}{0.01}$ initial scale were used. From \cite{oma3}.}
		\label{fig:fits-hera}
	\end{figure}

	The fit setups are as follows. For the BK evolution we compare the enhanced BK equations: ResumBK, KCBK, and TBK, which are described in Sec. \ref{sec:bk-evol}. These were used instead of the full NLO BK since it is numerically very demanding, though solvable in principle~\cite{Lappi:2015fma,Lappi:2016fmu}. For the running coupling we use a combination of Balitsky and smallest dipole prescription as the 'realistic' coupling, and compare this to the simple parent dipole coupling. We found that upgrading the simplistic lower limit $\zmin = \xbj/x_0$ used in \cite{oma1}, to $\zmin = (\xbj Q_0^2)/(x_0 Q^2)$ was necessary in order to get a good agreement with the data. This more accurate lower limit is discussed in \cite{Beuf:2014uia, Beuf:2017bpd, oma1}. Lastly, two options for the shape of the BK evolution initial condition were considered. The key difference between the two initial condition schemes is the starting rapidity scale of the BK evolution. In the first setup we take as the initial scale a rapidity corresponding to a reasonably small $\xbj$ to begin the evolution, and in the second we take the rapidity scales of the BK evolution and the impact factors to be the same. Both schemes have their merits, details are found in \cite{oma3}.

	The datasets used in this analysis are the combined HERA data of the H1 and ZEUS experiments for the reduced cross section \cite{Aaron:2009aa}, and the charm and bottom quark contributions to the inclusive cross sections \cite{Abramowicz:1900rp, H1:2018flt}.
	While a newer combined dataset including the data from the HERA-II run would have been available~\cite{Abramowicz:2015mha}, the two datasets result in very comparable fits~\cite{Mantysaari:2018nng} at low $x$ and moderate $Q^2$. We verified this with our fit setup as well, where with the fit parametrization using the KCBK equation, Bal+SD coupling, and $\Yobk=\ln 1/0.01$ we found that the fit quality receives a minor change from $\chi^2/N = 1.89$ to $\chi^2/N = 1.60$.
	
	The heavy quark dataset is used to generate a light-quark-only dataset by subtracting the charm and bottom contributions from the inclusive reduced cross section, see Sec.~\ref{sec:fits-uncert} for details. The motivation for this is the fact that the theory calculation was done for massless quarks, and the heavy quark contributions in the inclusive data are non-negligible. The generated light-quark-only reduced cross section data is then used in fits in order to gain insight into the compatibility of the massless quark cross sections and the available data.
	
\begin{figure}
		\centering
		\includegraphics[width=0.75\textwidth]{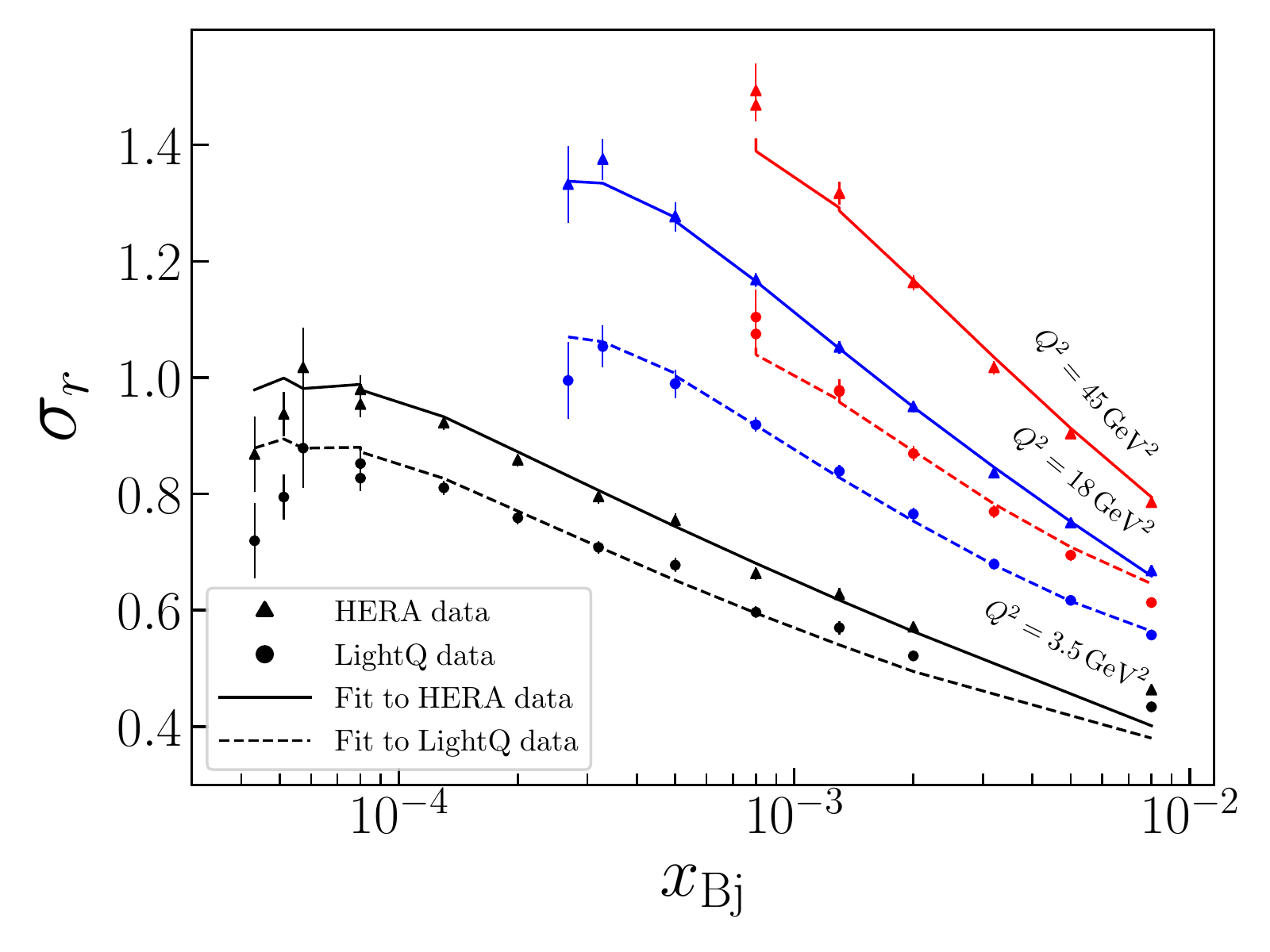}
		\caption{Inclusive and light-quark-only reduced cross sections calculated based on KCBK fits to the corresponding dataset compared to their respective datasets. Balitsky + smallest dipole running coupling and $\Yobk = \ln \frac{1}{0.01}$ initial scale were used. From \cite{oma3}.}
		\label{fig:fits-lightq}
	\end{figure}

	As for the results, we find that the NLO cross sections are able to describe the HERA data very well, one comparison is shown in \fig\ref{fig:fits-hera}. Even the combined HERA data cannot properly differentiate between the enhanced BK equations or running coupling schemes: all setups describe the data comparably. Though we did see that the setups using Bal+SD running coupling on average performed slightly worse in terms of $\chi^2/N$ than those using the parent dipole coupling. Description of the light-quark-only data is found to be good as well. The Figure \ref{fig:fits-lightq} shows a fit to the light-quark-only data, and a fit with the same BK equation and $\as$ prescription to the HERA data. We see that the same setup can fit both of the datasets separately, and that the fit parametrizations receive some systematic changes --- this allows us to infer some general features of the data by comparing the initial amplitude shapes. Specifically, we find that the light quark data needs a slower BK evolution and a larger target size. We interpret this as the presence of a substantial non-perturbative hadronic contribution in the light-quark-only data. 

	To summarize, we performed the first NLO accuracy DIS fits to HERA data to determine the initial condition for the BK evolution. We found a good agreement with the theory and data for both the HERA and light-quark-only datasets. To assess the theoretical uncertainty, we used alternative prescriptions of key pieces of the computation and found the fit results to be quite robust. Thanks to the universality of the dipole amplitude, discussed in \se\ref{sec:dipole}, the dipole amplitudes determined in these fits can be used in other phenomenological calculations at NLO accuracy, as has already been done for exclusive heavy vector meson production~\cite{Mantysaari:2021ryb}.

\section{Assessing the theory uncertainty}
	\label{sec:fits-uncert}
	Any theory calculation of an observable has an intrinsic theoretical uncertainty --- perfect information of the physical system cannot be had, and various assumptions or approximations are made to make theory calculations possible. In the case of the NLO DIS cross sections evaluated in this work, some of this uncertainty arises due to choices in modeling that cause finite differences that are of higher order in the perturbative expansion. Examples of effects like these are the differences between the running coupling prescriptions and the differences between the resummation techniques of the enhanced BK equations. There is also uncertainty whether the BK evolution prescription should be done in the probe or target rapidity. While the latter seems to be a more natural prescription of the evolution one still has to perform a translation between the evolution prescriptions when computing the DIS impact factors that are derived in the probe rapidity prescription. Lastly, selecting the functional form and rapidity scale for the initial condition of the BK evolution introduces uncertainty.
	
	To gauge the magnitude of these effects we performed the data comparisons with a number of prescription combinations.
	The fits done in \cite{oma3} compare BK resummation prescriptions, running coupling schemes, BK initial condition rapidity scales, and HERA and light-quark-only datasets. As discussed above, the BK equation and running coupling prescription effects should be somewhat small, and in \cite{oma3} we demonstrated that all the used BK equation and running coupling choices were able to describe the data well.
	
	We ran two series of fits to test the sensitivity of the cross sections to the BK evolution initial rapidity scale. In the first set the initial scale was set to $\Yobk = \ln 1/0.01$, which has traditionally been considered as reasonably small $x$ to begin the evolution. However, in the cross section computation one needs to evaluate the dipole amplitude in the range $[0, \Yobk]$, which is before the evolution begins. We chose to freeze the dipole amplitudes at the initial scale in this region. In the second set the scale was taken to be $\Yobk = \Yoif = 0$, which forgoes the need for freezing, but now the initial scale of the evolution is at unnaturally large $x$. We interpret this as the usage a dipole amplitude initial form which is that of an evolved dipole amplitude of $\ln 1/0.01$ units of rapidity evolution. More details and discussion about the results are found in \cite{oma3}.
	
	Lastly, some uncertainty is introduced by the fact that we are making comparisons between inclusive cross sections for massless quarks and HERA data that has notable contributions from charm and bottom quarks especially at $Q^2 \gg m_c^2$. To address this we generated a light-quark-only reduced cross section dataset from HERA data by subtracting the charm and bottom contributions manually. Since the original inclusive and heavy quark datasets are not binned in the same way, we had to perform interpolation using a separate leading order IPsat parametrized fit~\cite{Mantysaari:2018nng}. The experimental uncertainties of the original HERA data were left unchanged, since the proper calculation of the uncertainties would need to be done by the experimentalists. We expect that this mostly affects the $\chi^2/N$ values of the light-quark-only data fits, and not the fit parametrizations that are found which are the key interest in this analysis. The NLO cross sections were found to describe both datasets well, which was briefly discussed in the previous section, and at more detail in~\cite{oma3}.
	
	Effects that we could not look into extensively include the functional form of the BK initial condition, the choice of rapidity scale in the NLO dipole contribution, the mismatch between Balitsky and smallest dipole running couplings to be discussed in \se\ref{sec:improvements}, and the choice of the $\Yoif$ scale in the impact factors \eqref{eq:NLO_qg_unsub_oma3}, which in principle could be taken as a fit parameter.

	\subsection{Assessing the impact of the NLO BK equation}
	\label{sec:assess-nlobk}
	
		\begin{table*}[ht]
			\centering
			\caption{NLO cross section comparisons to HERA data using NLO BK evolution starting from the initial conditions determined in \cite{oma3}. The change in $\chi^2/N$ is a measure of the difference between the NLO BK and the enhanced BK equations used in the fits. The notation (acc.$\downarrow$) refers to the relaxed numerical accuracy of the calculation done here, that matches the accuracy of the numerical NLO BK evolution. These lower accuracy values are provided as a reference to gauge the magnitude of the numerical uncertainty of the NLO BK evolution.}
			\begin{tabular}{|L{1.8cm}|L{1.1cm}|R{1.15cm}||R{2.1cm}|R{2.3cm}|R{3.0cm}|}
				\hline
				Fit BK & $\as$ & \multicolumn{1}{l||}{$\Yobk$} & \multicolumn{1}{l|}{$\chi^2/N$(\cite{oma3})} & \multicolumn{1}{l|}{$\chi^2/N$ (acc.$\downarrow$)} & \multicolumn{1}{l|}{$\chi^2/N$ (NLO BK)} \\
				\hline
				ResumBK & parent & {\small $\ln \frac{1}{0.01}$} & 2.24 & 2.3 & 2.9  \\
				KCBK 	& parent & {\small $\ln \frac{1}{0.01}$} & 1.85 & 1.9 & 4.0  \\
				TBK 	& parent & {\small $\ln \frac{1}{0.01}$} & 2.76 & 2.8 & 2.9  \\
				\hline
				ResumBK & parent & $0$ 					& 1.12 & 1.2 & 1.4  \\
				KCBK 	& parent & $0$ 					& 1.24 & 1.3 & 5.5  \\
				TBK 	& parent & $0$ 					& 1.03 & 1.1 & 1.4  \\
				\hline
			\end{tabular}
			\label{tab:fits-heradata-nlobk-evol}
		\end{table*}
	
		\begin{figure}
			\centering
			\includegraphics[width=0.75\textwidth]{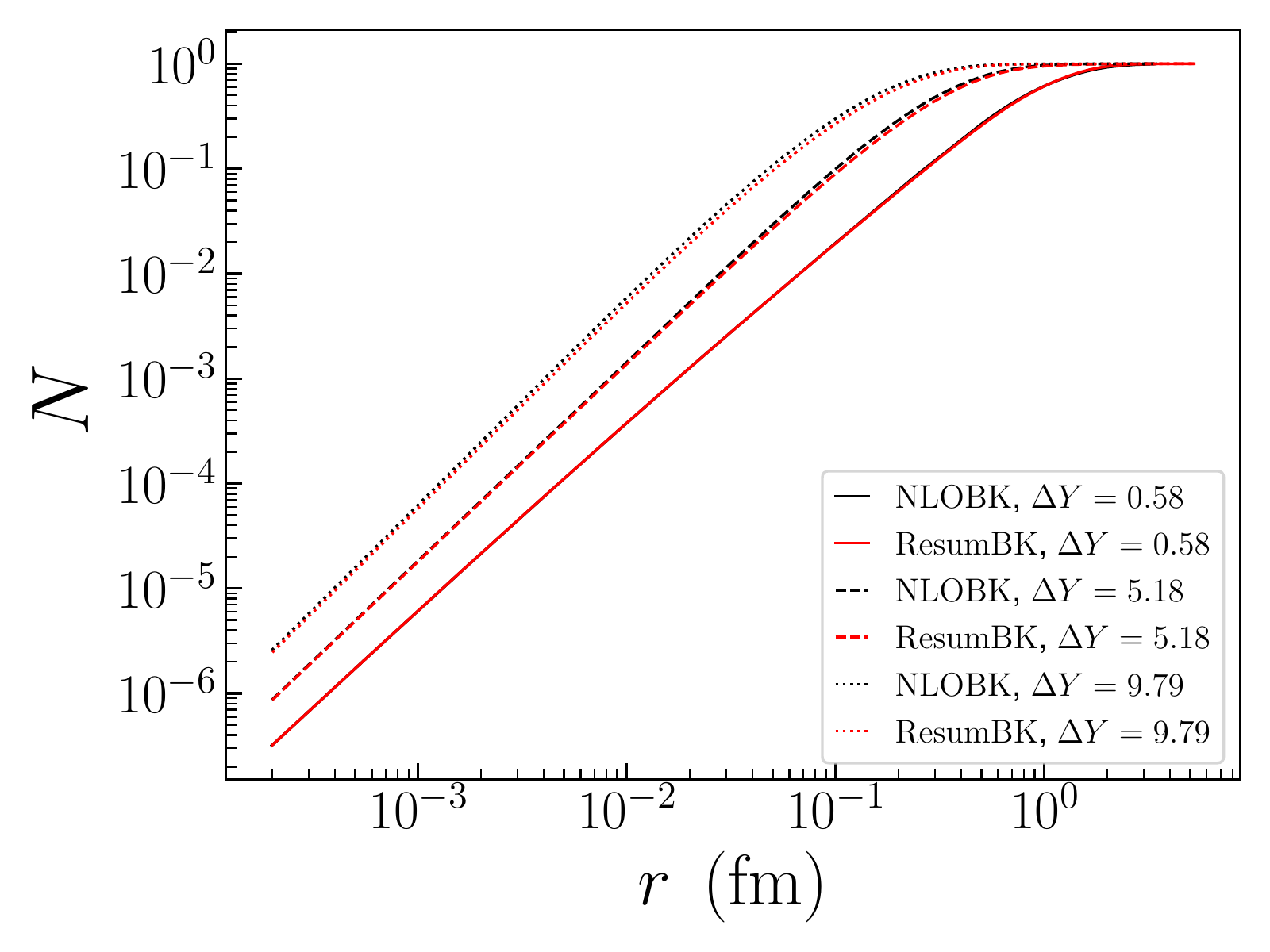}
			\caption{The same initial amplitude shape evolved using the ResumBK and NLO BK equations. The initial condition is from the ResumBK, parent dipole, $\Yobk = \ln \frac{1}{0.01}$ fit shown in Table~\ref{tab:fits-heradata-nlobk-evol}, determined in \cite{oma3}.}
			\label{fig:amplitudes-nlobk-resumbk}
		\end{figure}

		Aside from the above discussion is the question of how big an impact will the full next-to-leading order BK evolution \cite{Balitsky:2008zza, Lappi:2015fma, Lappi:2016fmu, Lappi:2020srm} have. Inclusion the full NLO BK would push the accuracy of the calculation to NLO+NLL order. To assess this, new comparisons between the used enhanced BK equations and the NLO BK including resummations~\cite{Lappi:2015fma,Lappi:2016fmu} were done for this thesis. Some of the initial shapes of the dipole amplitudes determined in \cite{oma3} were evolved using the full NLO BK equation~\cite{Lappi:2016fmu} and then used for computation of the cross sections for data comparison. Even though the different BK equations used in \cite{oma3} are quite different\footnote{ResumBK equation resums both single and double large transverse logs, KCBK equation resums double logs, and TBK equation resums the double logs, and it is formulated in the target momentum fraction picture instead of the projectile mom. fraction picture. See Article~\cite{oma3} and references therein for more details.} theoretically, they take the same functional shape as the initial condition. Thus this comparison gauges
		the importance of $\as^2$ contributions in the NLO BK equation that are not enhanced by large transverse logarithms
		based on how suitable the initial conditions determined in the fits are for the NLO BK evolution. This suitability is quantified by the change in the goodness of the fit $\chi^2/N$.
		
		As is shown in Table~\ref{tab:fits-heradata-nlobk-evol}, the full NLO BK equation causes a fairly small changes in the $\chi^2/N$ values of the fits. The smallest changes are seen with the initial conditions of the ResumBK and TBK equations, whereas the KCBK fit initial conditions see slightly larger changes. The ResumBK evolution is expected to approximate the NLO BK evolution the closest since it resums single transverse logs as well as the double logs resummed by the other two prescriptions. In addition to this, the specific prescription of the ResumBK equation used is numerically optimized to match the NLO BK equation with resummations as well as it can~\cite{Lappi:2016fmu}.
		On the other hand, there is little reason to expect that the TBK evolution initial conditions would be good for this formulation of the NLO BK evolution. The TBK evolution is derived in the target momentum fraction picture, whereas the NLO BK evolution is based on the projectile momentum fraction picture. Furthermore, the TBK equation used only resums the large transverse double logs, whereas the NLO BK resums the single transverse logs as well. Thus without further study, we can only conclude that the small changes seen with the TBK equation initial conditions are unexpected.
		
		In \fig\ref{fig:amplitudes-nlobk-resumbk} is shown the same initial condition evolved using the ResumBK and NLO BK equations, and the close approximation of the NLO BK result by ResumBK is evident. We see that the NLO BK evolution is slightly faster than the ResumBK evolution, as has been seen previously~\cite{Lappi:2016fmu}.
		
		\begin{figure}
			\centering
			\includegraphics[width=0.75\textwidth]{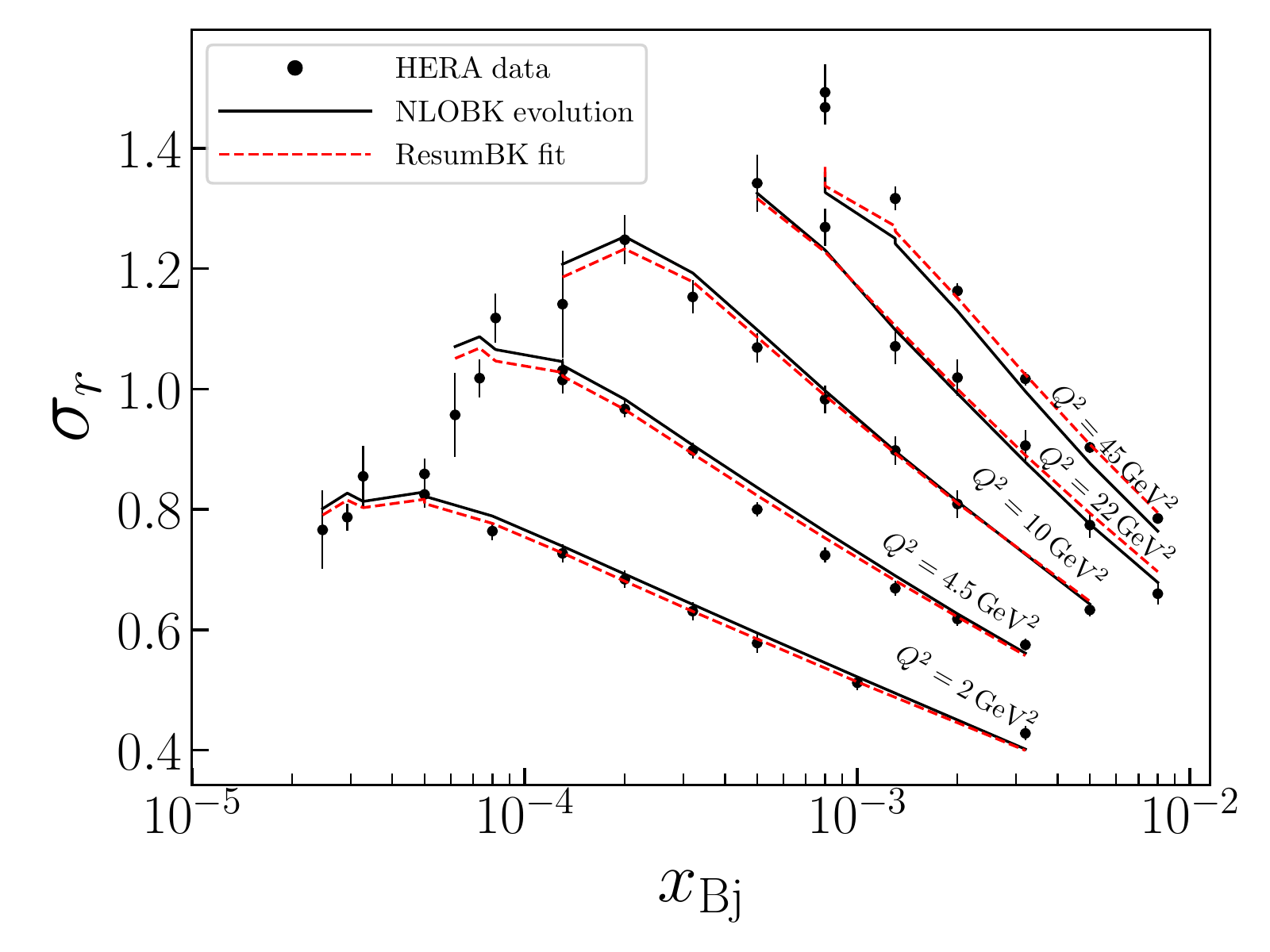}
			\caption{Reduced cross section computed using the NLO BK evolved dipole amplitude compared to HERA data. The initial shape is from the fit using the ResumBK equation, $\Yobk = \ln \frac{1}{0.01}$ shown in Table~\ref{tab:fits-heradata-nlobk-evol}, from \cite{oma3}.}
			\label{fig:nlobk-dipole-hera-data}
		\end{figure}
	
		This NLO BK evolved dipole amplitude is used to compute reduced cross sections for data comparison, which is shown in \fig\ref{fig:nlobk-dipole-hera-data}, together with the ResumBK evolved dipole amplitude that was fitted to the data. We see that the difference between the evolution equations is larger both at larger $Q^2$ and at smaller $x$, both of which correspond to a later stage in the evolution. This difference seen between the NLO BK and ResumBK results is of the same magnitude as between the enhanced BK equations found in~\cite{oma3}. It is unclear if the available data would discern between the NLO BK and the enhanced BK equations.

		The interpretation of this simple comparison is unfortunately muddled by the fact that the unsub scheme used in the computation cannot be consistently upgraded to NLL accuracy with a simple replacement of the evolution equation~\cite{Beuf:2017bpd}. The issue is that when one replaces a LL evolved scattering amplitude with an NLL evolved amplitude in the unsub form of the cross section, one is introducing next-to-next-to-leading order (NNLO) corrections together with the NLL contribution from the NLO BK. In the sub scheme these NNLO corrections are canceled between the $qg$ contribution and the subtraction term, producing a strict NLO result. Whereas in unsub scheme the NNLO contribution persists without further work. We note however that while this use of unsub scheme with NLO BK goes beyond strict next-to-leading order, it is the natural result of applying perturbative QCD. To elaborate, calculating all the relevant Feynman diagrams at this order produces this subset of NNLO corrections due to the nested nature of the problem \cite{Iancu:2016vyg} --- both the BK evolution and the DIS impact factors get separate $O(\as)$ contributions which compound naturally, and this higher order contribution would need to be removed manually. In principle it is possible to amend unsub scheme to strict NLO+NLL order by extracting the NNLO contribution with an additional term accounting for the mismatch between the LL evolution contribution contained in the unsub $qg$ term, and the NLL contribution arising from the NLO BK evolution. All this means that the calculations shown in Table~\ref{tab:fits-heradata-nlobk-evol} cannot claim strict NLO+NLL accuracy.

\section{An outlook for theory improvements to the fits}
	\label{sec:improvements}

	During the work on Article~\cite{oma3} some theory improvement opportunities for the fits were recognized, however they were out of scope for the project. This section discusses these improvements.
	
	\begin{itemize}
		\item \textbf{Quark and anti-quark momentum fraction constraint}.

			In the expressions \eqref{eq:NLO_dip}, \eqref{eq:NLO_qg_unsub_oma3} the quark and antiquark momentum fractions $z_0$ and $z_1$ can attain all values in their defined regime $[0,1]$. This implies that the invariant masses of the included $q \bar{q}$ and $q \bar{q} g$ states can become arbitrarily large:
			\vspace{-0.2cm}
			\begin{align*}
				M_{q \bar{q}}^2 & = (k_0 + k_1)^2 = \frac{\kt_0^2}{z_0} + \frac{\kt_1^2}{z_1} = \frac{\kt_1^2}{z_1(1-z_1)},
				\\
				M_{q \bar{q} g}^2 & = (k_0 + k_1 + k_2)^2 = \frac{\kt_0^2}{z_0} + \frac{\kt_1^2}{z_1} + \frac{\kt_2^2}{z_2},
			\end{align*}
			where $k_0, k_1, k_2$ are the four-momenta of the quark, antiquark and gluon, and $z_i$ are their respective longitudinal momentum fractions. This means that the limits $z_i \to 0$ are problematic. The production of $q \bar{q}$ and $q \bar{q} g$ states with an invariant mass larger than the c.m.s. energy of the scattering is forbidden, so a limit of the type:
			\begin{equation*}
				\frac{\kt_1^2}{z_1(1-z_1)} < W^2
			\end{equation*}
			is needed in the case of $q \bar{q}$ production. However, the consistent implementation of a limit of this type in mixed space of transverse positions and longitudinal momentum fractions is non-trivial and left for further work. Similar analysis needs to be done for the $q \bar{q} g$ contribution as well.
			Some suggestions for how to do this are outlined in~\cite{Beuf:2017bpd}.
			Traditionally cut-offs like this are not done for regular non-divergent integrals. Such a cut-off for the longitudinal momenta $\plusk > \plusk_{min}$ creates effects that are power suppressed in $\frac{\plusk_{min}}{W}$~\cite{Beuf:2017bpd}. These effects are in principle beyond the precision of the calculation.
			
		\item \textbf{NLO dipole term rapidity scale}.
		
			As is discussed in \cite{oma1} and \se\ref{sec:dip-z2}, at NLO the calculation of the photon splitting wavefunctions do not give guidance on what should be the rapidity scale of the dipole amplitude in the loop corrections to the $q \bar{q}$-amplitude. The most natural thing to do would be to take both the loop and the tree-level NLO corrections at the same rapidity scale, which needs to depend on the momentum fraction of the gluon, as discussed in \se\ref{sec:dis-nlo}. However, this was not possible in \cite{oma1,oma3}, since the NLO dipole contribution~\eqref{eq:NLO_dip} had been integrated over the gluon momentum fraction~\cite{Beuf:2016wdz}. In \se\ref{sec:dip-z2} a form is derived for the NLO dipole contribution, \eq\eqref{eq:NLO_dip_z2}, where the momentum fraction integration has been undone and the usage of a consistent rapidity scale with the $qg$-contribution is possible.
			
		\item \textbf{Exact running coupling matching in the BK and impact factors}.
		
			With the Balitsky~+~smallest dipole running coupling prescription used in \cite{oma3}, there is a subtle mismatch between the evolution equation and the NLO impact factor. Since this prescription uses the Balitsky coupling in the leading order BK kernel, and the smallest dipole in the BK resummation kernel and DIS impact factors, a sub-optimal finite leftover of order NNLO is produced in the factorization of the soft gluon large logarithm. Note, however, that the Balitsky prescription reduces to the smallest dipole coupling in the limit that one of the dipoles is much smaller than others, which suggests that the finite leftover is not quite as large as one would get without this limit-agreement of the couplings.
			
			As an NNLO effect beyond our precision, the importance of this mismatch is largely unknown. Some light could be shone on this by a comparison to fits using only the smallest dipole coupling. The proposed improvement here is to perform fits using some other realistic running coupling prescriptions that can be implemented consistently in both the BK equation and the impact factors, such as the one proposed in \cite{Beuf:2017bpd}. See also Ref.~\cite{Ducloue:2017dit}, where an improved running coupling is developed for NLO calculations of forward hadron production in pA-collisions. However, this prescription might only be optimal for observables computed in momentum space.
			
		\item \textbf{Quark masses}.
		
			A substantial improvement to the theory would be the inclusion of the quark masses in the calculation of the impact factors. Recently the NLO DIS impact factor with massive quarks for the longitudinal photon has been made available \cite{Beuf:2021qqa}.
			
			The inclusion of the quark masses has the largest effect on the impact factors at small $Q^2$. A comparison between LO cross sections with massive and massless quarks has been done where the effect was largest at $Q^2 \lesssim 1 \gev^2$ \cite{GolecBiernat:1998js}. In this region we are not fitting extensively; the lower limit in our fits was $Q^2 \geq 0.75 \gev^2$~\cite{oma3}. Based on this, it would seem feasible that the fit qualities of the inclusive data fits are mostly affected at small $Q^2$. However, the fit regime in $Q$ contains the mass scales of the charm and bottom quarks, and especially the contribution of the charm quark to the total cross sections is considerable~\cite{H1:2018flt} up to the upper limit of the fit regime in $Q$. This suggests that there is notable uncertainty regarding the magnitude of the change to the inclusive data fits brought on by the inclusion of the quark masses in the calculation.
			
			In contrast, a novel possibility with the massive quark impact factors is the ability to fit heavy quark cross sections separately. In \cite{oma3} we found evidence that the inclusive DIS cross sections contain a substantial non-perturbative contribution from the light quarks. This suggests that the charm and bottom quark cross sections should be better perturbative quantities.
			The inclusion of heavy quarks in the theory calculations is an important improvement, and will hopefully shed light on the tension between the inclusive and heavy quark data fits seen at leading order~\cite{Ducloue:2019jmy}.
		
		\item \textbf{Next-to-Leading Order BK evolution}.
		
			The remaining improvement for the cross section calculation to reach full NLO+NLL accuracy is to use the NLO BK equation \cite{Balitsky:2008zza, Lappi:2015fma, Lappi:2016fmu, Lappi:2020srm}. This, however, is a non-trivial undertaking due to the computational cost of the NLO BK equation. Even with relaxed numerical precision, the evaluation of the NLO BK evolution is slower by more than two orders of magnitude in comparison to the implementation of the ResumBK equation, which is caused by the larger phase space of the NLO BK equation. This means that a fit is mostly feasible in a small region. It might be possible to use the resummed BK equations as guidance in the determination of the fitting region, as they approximate the NLO BK fairly well \cite{Lappi:2016fmu}, which was also verified with some of the fits in \se\ref{sec:fits-uncert}.
	\end{itemize}
	\chapter{Diffractive Deep Inelastic Scattering}
	\label{ch:ddis}

\section{Diffraction in particle collisions}
	\label{sec:ddis-general}
	
	In the 1990s, a striking discovery was made when the high-energy electron-proton collisions began at DESY-HERA. As the electrons struck the target protons with immense energy, in roughly 1 in 10 collisions the proton remained intact~\cite{Albacete:2014fwa}, and instead the virtual photon emitted by the electron would create a shower of hadrons. These hadron showers were seen to be separated by a substantial angle from the beam axis. Diffractive events had been seen in the deep inelastic scattering (DIS) experiment.
	
	The observations of the diffractive DIS (DDIS) events at HERA are nigh on tantalizing for two features. First, it was surprising to see such a large proportion of large angular separation --- or large rapidity\footnote{Pseudorapidity is used to parameterize the  angle away from the beam axis: $\eta \coloneqq - \ln \left[ \tan \left(\frac{\theta}{2}\right) \right]$. A gap in pseudorapidity is equivalent with a gap in the angular distribution.} gap --- events. The simple expectation from QCD is that such large rapidity gap events would be exponentially suppressed \cite{Bjorken:1992er}.
	The second interesting feature of the HERA data is that the ratio of the diffractive to all events ($\sigma^\textrm{diff}/\sigma^\textrm{tot}$) is almost constant with varying center-of-mass energy of the virtual photon-proton system, $W^2$~\cite{Gelis:2010nm}. Here the expectation was that the ratio would grow rapidly with increasing energy, which was not seen in the experiment~\cite{Gelis:2010nm}.
	
	For theoretical calculations of high-energy diffractive scattering, it is useful to state an equivalent definition of diffraction~\cite{Bjorken:1992er,Barone:2002cv}: a scattering at \textit{high-energy} in which no quantum numbers are exchanged between the colliding particles is a diffractive scattering. The requirement on high-energy is due to the existence of the exponentially suppressed non-diffractive reactions, which asymptotically become negligible at high-energy. In practice this means that after the virtual photon scatters off the proton, both the target and the parton shower produced by the photon must be in color-singlet states. If they had net color-charge after the scattering, the rapidity gap would be filled by the gluon-bremsstrahlung as the target and produced system would color neutralize during hadronization. 
	
	A number of theoretical mechanisms have been constructed for the color-neutral formation of the target and produced system. They can roughly be grouped in three categories:
	pomeron exchange models, soft color interaction models, and dipole models.
	For reviews on the topic see Refs.~\cite{Hebecker:1999ej,Boonekamp:2009yd}, and for a more pedagogical discussion Ref.~\cite{Barone:2002cv} --- note however that these do not cover some of the important saturation framework prescriptions of DDIS, which are covered in some detail in Refs.~\cite{Gelis:2010nm,Albacete:2014fwa}.

	The first theoretical description of diffractive DIS was postulated by Ingelman and Schlein (IS)~\cite{Ingelman:1984ns} already before it was observed in experiment. In the IS model the target wavefunction contains a hadronic color-singlet component, a so-called pomeron, whose parton structure the virtual photon would probe. This picture would imply that the pomeron structure functions would be universal, which however was falsified by $p \bar p$ collision experiments, where hard-diffractive jet pair events were found to only constitute $1-2\%$ of all jet events, in contrast with the $10\%$ of the events in DDIS~\cite{Brodsky:2004hi}. Later pomeron exchange models consider the pomeron as a dynamically emergent object in DDIS, which carries the quantum numbers of the vacuum, and has an internal structure of quarks and gluons~\cite{Barone:2002cv,Martin:2004xw}.
	
	In the soft color interaction (SCI) framework~\cite{Edin:1995gi,Edin:1996mw,Ingelman:1996mq}, the view is that the hard part of the process --- the virtual photon scattering off a parton --- is the same between DIS and DDIS. The color-neutralization then takes place via soft gluon interactions between the target remnants and the scattered off system. While SCI is considered phenomenologically successful, it has also been seen to be limited due to the ad hoc nature of the color exchange and not accounting for perturbative effects like color transparency~\cite{Barone:2002cv}. Later work has connected the SCI model with a similar soft interaction framework, providing it with a theoretical basis that includes rescattering effects~\cite{Brodsky:2004hi}.

	The dipole model family of DDIS descriptions has two principal branches that have connections to present day CGC framework saturation phenomenology. One is the semiclassical approach to DIS and DDIS by Hebecker et al.~\cite{Hebecker:1997gp,Hebecker:1999ej}, which can be considered an early precursor of the CGC framework. As such, the physical picture of the scattering from the semiclassical color-fields of the target is the same as in the case of dipole picture DIS, discussed in Sec.~\ref{sec:dipole}. Diffraction is introduced by projecting the scattered state into a color-singlet state. In the dipole picture a Fock state with a soft gluon becomes important when the final state has large invariant mass, and a leading twist approximation for this --- power counting-wise NLO --- contribution is derived in Ref.~\cite{Hebecker:1997gp}. A missing piece from the framework in comparison to present day CGC is the unitarization of the dipole model, which introduces the non-linear effects that lead to saturation.
	
	The second branch is the two-gluon exchange approximation of DDIS by Wusthoff et al.~\cite{Bartels:1994jj,Wusthoff:1995hd,Wusthoff:1997fz,Wusthoff:1999cr}. In this model the diffractive scattering proceeds by the virtual photon fluctuating into a $q \bar q$ or $q \bar q g$ dipole, which then scatters off the target by the exchange of two soft gluons. The $q \bar q g$-contribution to the cross sections is derived in a leading twist, i.e. high-$Q^2$, approximation. This model was combined with a phenomenological description of saturation, which described the HERA data well~\cite{GolecBiernat:1999qd}. Another early approximative saturation approach was calculated by Munier and Shoshi~\cite{Munier:2003zb} in the limit that the invariant mass of the diffractive system is large. The Wusthoff and Munier--Shoshi approaches were later connected by Marquet~\cite{Marquet:2007nf} in a work that gave the first CGC framework description of diffraction that builds on the virtual photon wavefunction derived by Wusthoff et al~\cite{Wusthoff:1995hd,Wusthoff:1997fz,GolecBiernat:1999qd}. A good description of HERA data was found using the CGC prescription~\cite{Kowalski:2008sa}. In Sec.~\ref{sec:ddis-nlo} the $q \bar q g$-contribution to the diffractive structure functions is calculated in full NLO accuracy without kinematic approximations in the CGC framework for the first time for both longitudinal and transverse virtual photons.

	\begin{figure}
		\centering
		\includegraphics[width=0.5\textwidth]{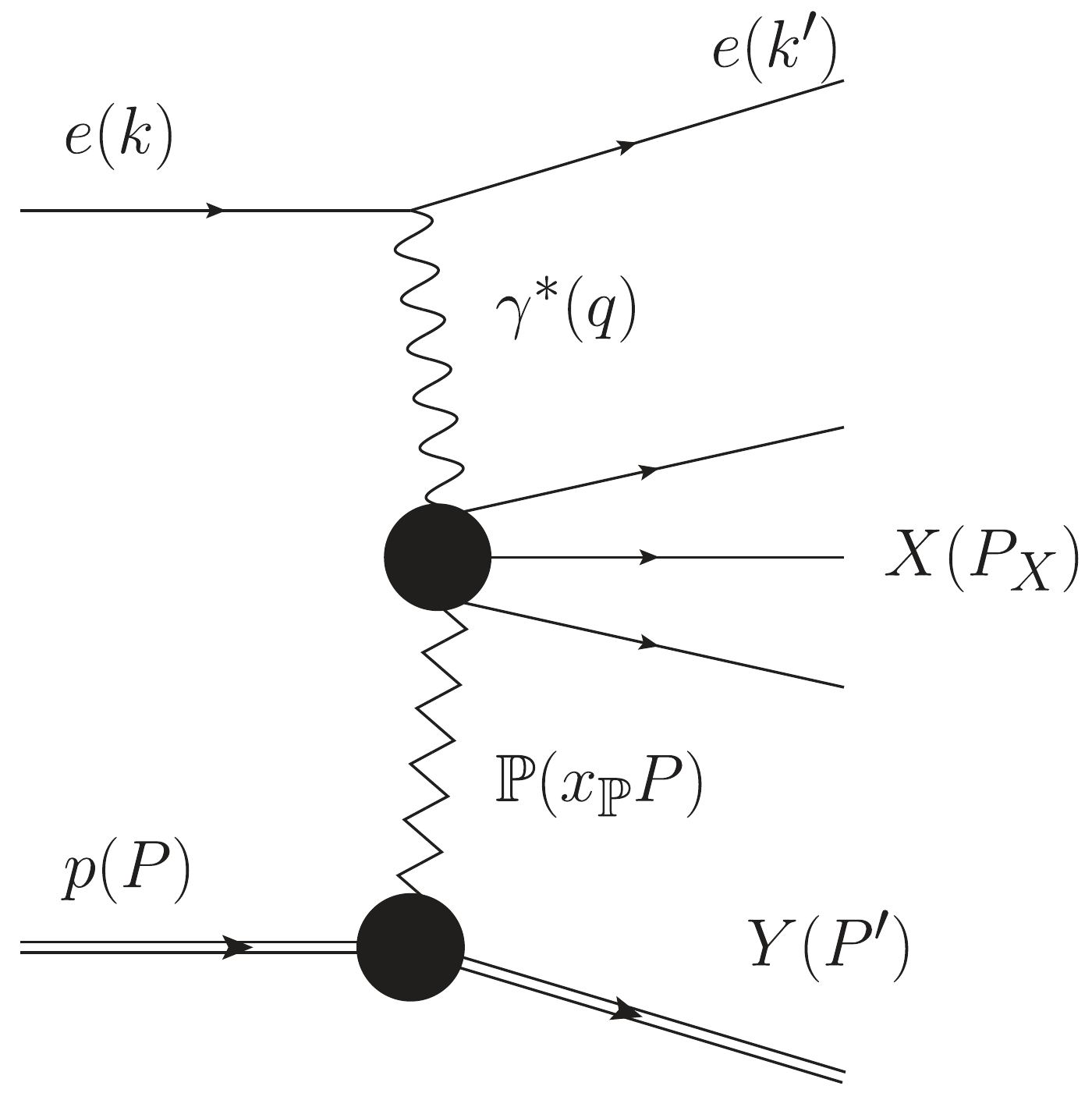}
		\caption{Diffractive electron-proton deep inelastic scattering. The proton can either stay intact and $Y=p$ or it can dissociate into the forward direction.}
		\label{fig:ddis}
	\end{figure}

	To discuss the general kinematics of diffractive DIS, let us consider diffraction to take place as a pomeron exchange without reference to any specific mechanism discussed above. In Fig.~\ref{fig:ddis} is depicted a general diffractive scattering mediated by the virtual photon scattering off the pomeron, which produces the diffractive color-singlet system $X$, and the target evolves into the color-singlet system $Y$. In elastic diffraction the target proton remains intact, or in the inelastic case it dissociates into a hadron shower into the forward direction, either vanishing into the beam pipe or showing up in a forward detector. 
	Analogously to the inclusive DIS discussed in Sec.~\ref{sec:dis-general}, diffractive DIS of a virtual photon off a proton is parametrized by four quantities. We need $Q^2$ and $\xbj$ like with inclusive DIS, the invariant momentum transfer $t$, and the invariant mass of the diffractive system $\mx$:
	\begin{align}
		t & \coloneqq -(P-P')^2
		\\
		\mx^2 & \coloneqq P_X^2
		\\
		\xpom & \coloneqq \frac{(P-P') \cdot q}{P \cdot q}
				= \frac{\mx^2 + Q^2 - t}{W^2 + Q^2 - M^2}
				\approx \frac{\mx^2 + Q^2}{W^2 + Q^2}
		\\
		\beta & \coloneqq \frac{Q^2}{2 q \cdot \left(P-P'\right)}
				\equiv \frac{\xbj}{\xpom}
				= \frac{Q^2}{Q^2 + M_X^2 - t} 
				\approx \frac{Q^2}{Q^2 + M_X^2},
	\end{align}
	where $\xpom$ and $\beta$ are alternative ways to parameterize the $\mx$ dependence, and the momenta are as shown in Fig.~\ref{fig:ddis}. They can be given interpretation in the IS model: $\xpom$ is the fraction of proton longitudinal momentum the pomeron carries, and $\beta$ is the fraction of the pomeron momentum the parton struck by the photon carries. The mass of the proton is $M$ and $W^2 = (P+q)^2$ is the energy of the $\gamma^*p$ system.
	
	Due to the distinct physics in the diffractive scattering, specific structure functions are defined for DDIS by writing the total diffractive cross section as~\cite{Barone:2002cv}:
	\begin{align}
		\sigma^{D(4)}_{\textrm{tot}}
		& = \frac{(2\pi)^2 \aem}{Q^2} F_{2}^{D(4)}(\xbj, Q^2, \mx, t) 
		\\
		& = \frac{(2\pi)^2 \aem}{Q^2} \left( F_{T}^{D(4)}(\xbj, Q^2, \mx, t) + F_{L}^{D(4)}(\xbj, Q^2, \mx, t)\right)  ,
		\label{eq:diffractive-FL-FT}
	\end{align}
	where the superscript $(4)$ refers to the dependence of the diffractive structure functions on four kinematic parameters. As experimental quantities the diffractive structure functions are $t$-integrated in HERA data\footnote{The Electron-Ion Collider experiment~\cite{Accardi:2012qut,Aschenauer:2017jsk,AbdulKhalek:2021gbh} is considering measuring $|t|$-differential cross sections, which could be made possible thanks to the higher luminosity of the experiment, in comparison to HERA.}, and in this case the $t$-independent structure functions are defined as $F_{2,L,T}^{D(3)} \coloneqq F_{2,L,T}^{D(3)} (\xbj, Q^2, \mx)$, or using equivalent parameter choices such as $ F_{2,L,T}^{D(3)} (\xpom, Q^2, \beta)$.

	The study of diffraction in high-energy collisions gives us invaluable information on the hadronic structure of the proton or nucleus. As discussed above, there is still much to learn about the mechanism that is taking place, and about the relation between the pomeron exchange and the saturation pictures. Understanding of these details gives insight into the appropriate high-energy degrees-of-freedom of QCD. Diffraction is more sensitive to the non-linear effects of saturation than fully inclusive DIS~\cite{GolecBiernat:1999qd,Marquet:2007nf}, which is seen explicitly at the cross section level as the dependence on the dipole amplitude squared $\sigma^{\textrm{diff}} \sim (1-S)^2$ in comparison to the inclusive case $\sigma^{\textrm{tot}} \sim \Re (1-S)$. This makes diffraction an invaluable tool in the study of the Color Glass Condensate framework.

\section{DDIS in the dipole picture at leading order}
	\label{sec:ddis-lo}
	
\begin{figure}
		\centering
		\includegraphics[width=0.75\textwidth]{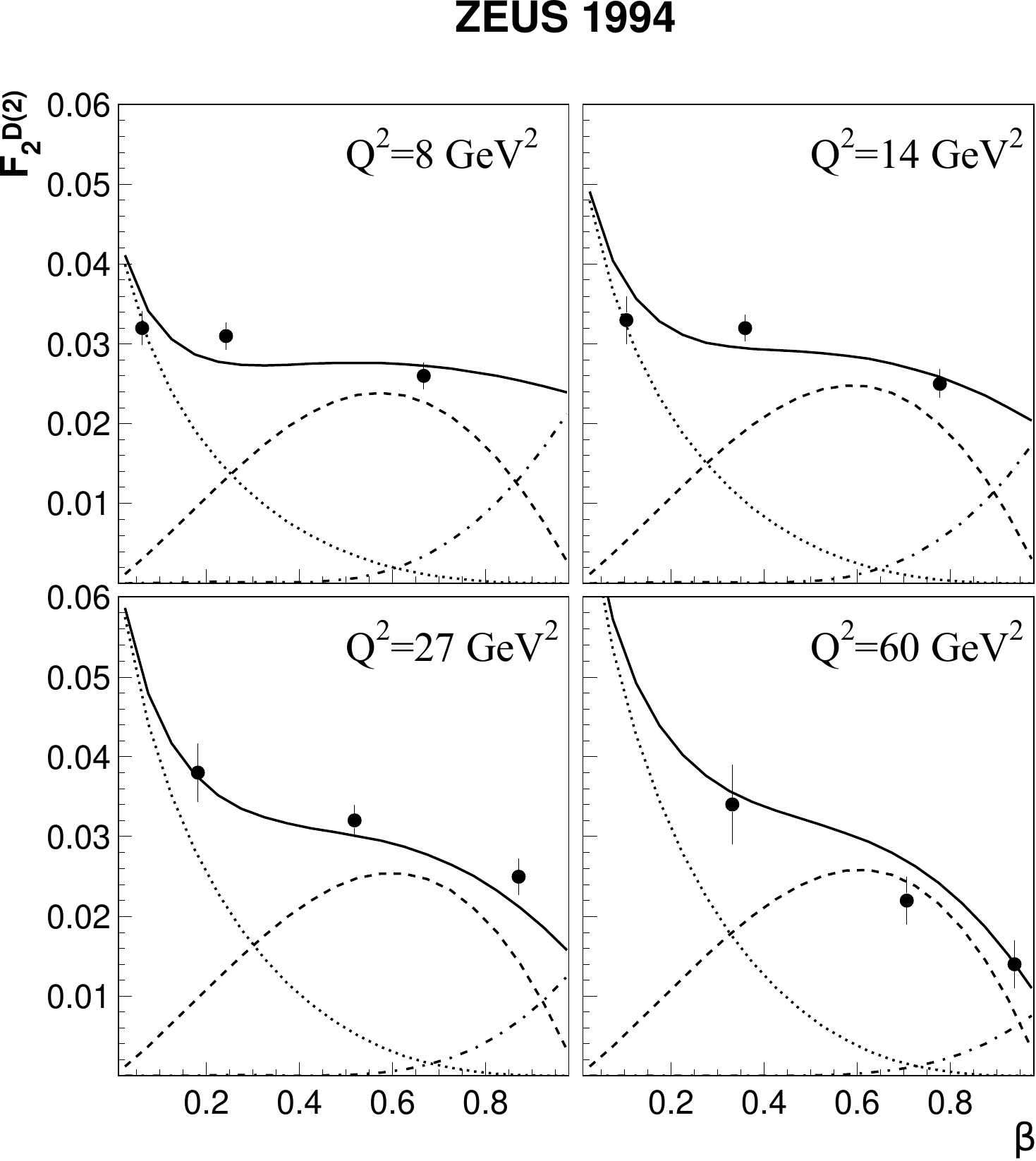}
		\caption{
			The leading contributions to the diffractive structure function $F_2^D$ compared to HERA data~\cite{GolecBiernat:1999qd}.
			\textbf{Dashed} line: $q \bar q$ contribution to $F_T^D$ \textbf{dot-dashed} line: $q \bar q$ contribution to $F_L^D$ 
			\textbf{dotted} line: $q \bar q g$ contribution to $F_T^D$.
			Here it is explicitly seen that the power counting-wise strictly LO $q \bar q$-contribution completely fails to describe the data at small-$\beta$, or equivalently large-$\mx$. In this regime the formally NLO $q \bar q g$-contribution is enhanced by a large logarithm $\log(Q^2)$ or $\log(1/\beta)$, and so becomes of order $\as \log(Q^2) \sim 1$ and the main contribution to the structure function at small-$\beta$.
			Reprinted figure with permission from
			K. Golec-Biernat, and M. Wusthoff,
			Phys. Rev. D, 60,
			114023, 1999.
			Copyright (1999) by the American Physical Society.
			}
		\label{fig:ddis-gbw-components}
	\end{figure}

	At leading order in the dipole picture the process of diffractive scattering is remarkably similar to the one of the fully inclusive deep inelastic scattering discussed in Sec.~\ref{sec:dis-lo}. The incoming virtual photon fluctuates into the $q \bar q$ dipole that then scatters eikonally from the target. Diffraction is introduced into the process by requiring that in the dipole-target scattering outgoing $q \bar q$-state is a color-singlet. Early LO calculations of DDIS in the dipole picture were done by Ryskin~\cite{Ryskin:1990fb}, and Nikolaev and Zakharov~\cite{Nikolaev:1991et,Nikolaev:1993th}. It however turns out, that the LO $q \bar q$ contribution alone was utterly incapable of describing the diffraction observations seen at HERA --- the contribution vanishes when the invariant mass of the produced system $\mx$ is large, or equivalently when $\beta \ll 1$. This is seen in Fig.~\ref{fig:ddis-gbw-components} where the $q \bar q$ contributions substantially fall short of the data.
	
	The question then is, how are large invariant mass systems created in the dipole picture. To see this, we write the invariant masses of the leading $q_0 \bar q_1$, and the next-to-leading $q_0 \bar q_1 g_2$ Fock states:
	\begin{align*}
		M_{q \bar{q}}^2 & = (k_0 + k_1)^2 = \frac{\kt_0^2}{z_0} + \frac{\kt_1^2}{z_1},
		\\
		M_{q \bar{q} g}^2 & = (k_0 + k_1 + k_2)^2 = \frac{\kt_0^2}{z_0} + \frac{\kt_1^2}{z_1} + \frac{\kt_2^2}{z_2},
	\end{align*}
	where the indices $0, 1, 2$ refer to the quark, antiquark and gluon, and the frame is set so that the transverse momentum of the photon is zero: $\qt = 0$. We see that large $\mx$ arises either if some of the partons have large transverse momentum, or if they have a low longitudinal momentum fraction $z_i \coloneqq \kplus / \qplus$. In fact, the production of soft gluons is enhanced over the production of soft quarks. This is seen by looking at the schematic $\kplus_i$-dependence of the dipole formation vertex for a transversely polarized photon~\cite{Beuf:2011xd}:
	\begin{equation*}
		\mathcal{V}_{\gamma^*_T(\qplus, \qt) \to q(k_0^+, \kt_0) \bar q(k_1^+, \kt_1)} 
		\!\sim\!
		\sqrt{\!\kplus_0 \! \kplus_1} \varepsilon_\lambda \!\cdot\! \left[
			\frac{\kt_0 \!+\! \kt_1}{\kplus_0 \!+\! \kplus_1} \!-\! \frac{1\!-\!2h_0\lambda}{2} \frac{\kt_0}{\kplus_0} \!-\! \frac{1\!+\!2h_0\lambda}{2} \frac{\kt_1}{\kplus_1}
		\right]
	\end{equation*}
	and the gluon emission from a quark vertex~\cite{Beuf:2011xd}:
	\begin{equation*}
		\mathcal{V}_{q(k_0^+ + k_2^+
		) \to q(k_0^+, \kt_0) g(k_2^+, \kt_2)} 
		\!
		\sim
		\!
		\sqrt{\!(\kplus_0 \!+\! \kplus_2) \kplus_0} \varepsilon^*_\lambda \!\cdot\! \left[
		\frac{\kt_2}{\kplus_2} \!-\! \frac{1\!+\!2h_0\lambda}{2} \frac{\kt_0}{\kplus_0} \!-\! \frac{1\!-\!2h_0\lambda}{2} \frac{\kt_0 \!+\! \kt_2}{k_0^+ \!+\! k_2^+}
		\right]\!,
	\end{equation*}
	where $h_0$ is the helicity of the quark, and $\lambda$ that of the photon or gluon. Out of the two vertices above, we see that the gluon emission grows the fastest as $\kplus_i \to 0$. This suggests that the leading process that produces large-$\mx$ diffractive systems is the emission of a gluon from the $q \bar q$ Fock state, a formally NLO contribution.
	
	Early attempts to calculate the $q \bar q g$ contribution to DDIS were done by Mueller, Ryskin, and Nikolaev\&Zakharov~\cite{Mueller:1989st,Ryskin:1990fb,Nikolaev:1991et,Nikolaev:1993th}. A breakthrough was the calculation of the $q \bar q g$ contribution at large-$Q^2$ by Wusthoff et al.~\cite{Levin:1992bz,Bartels:1994jj,Wusthoff:1995hd,Wusthoff:1997fz,Wusthoff:1999cr}. In the large-$Q^2$ regime the contribution is enhanced by a large logarithm $\log(Q^2)$, making the contribution formally of the order $\as \log(Q^2) \sim 1$, i.e. perturbatively leading order.
	Specifically, the calculation exclusively considers the contributions where the transverse virtual photon fluctuates into a quark-antiquark pair, after which a gluon is emitted. The $q \bar q g$-state then scatters from the proton as an effective $gg$-dipole by the exchange of two gluons, allowing the diffractive system and proton remnant remain color-neutral. The contribution from the longitudinal photon is not included since it is higher-twist, i.e. suppressed by one power in the $Q^2$-expansion~\cite{Wusthoff:1997fz}. This result together with a phenomenological model --- the GBW model --- was used with tremendous success to describe HERA data~\cite{GolecBiernat:1999qd}. The analysis used the dipole amplitude as determined from inclusive DIS fit~\cite{GolecBiernat:1998js} to HERA data, which is seen as the success of saturation phenomenology.
	
	A full NLO accuracy calculation without approximations of the $q \bar q g$-contribu\-tion has not been performed before, let alone the implied loop contribu\-tions. These will be discussed in more detail in Sec.~\ref{sec:ddis-nlo}. Due to this, the large-$Q^2$ $q \bar q g$-contribution derived by Wusthoff has remained among the most accurate descriptions\footnote{Other studies of the $q \bar q g$ contribution include the calculation by Hebecker et al.~\cite{Hebecker:1997gp,Buchmuller:1998jv}, and in the small-$\beta$ approximation the analyses~\cite{Kopeliovich:1999am,Bartels:1999tn,Kovchegov:2001ni,Marquet:2004xa,Golec-Biernat:2005prq}.} of the $q \bar q g$-contribution, one other being the small-$\beta$ leading $\log(1/\beta)$ contribution~\cite{Munier:2003zb}. The latter being less general, since the structure function result is to be taken at the limit $\beta \equiv 0$. In the following section we will discuss the analysis of Marquet and Kowalski et al.~\cite{Marquet:2007nf,Kowalski:2008sa}, who combined the large-$Q^2$ and small-$\beta$ results to have the most accurate dipole picture prescription of DDIS and HERA data to date.

	\subsection{Leading contributions to the diffractive structure functions in the dipole picture}

	\begin{figure}
		\centering
		\includegraphics[width=0.6\textwidth]{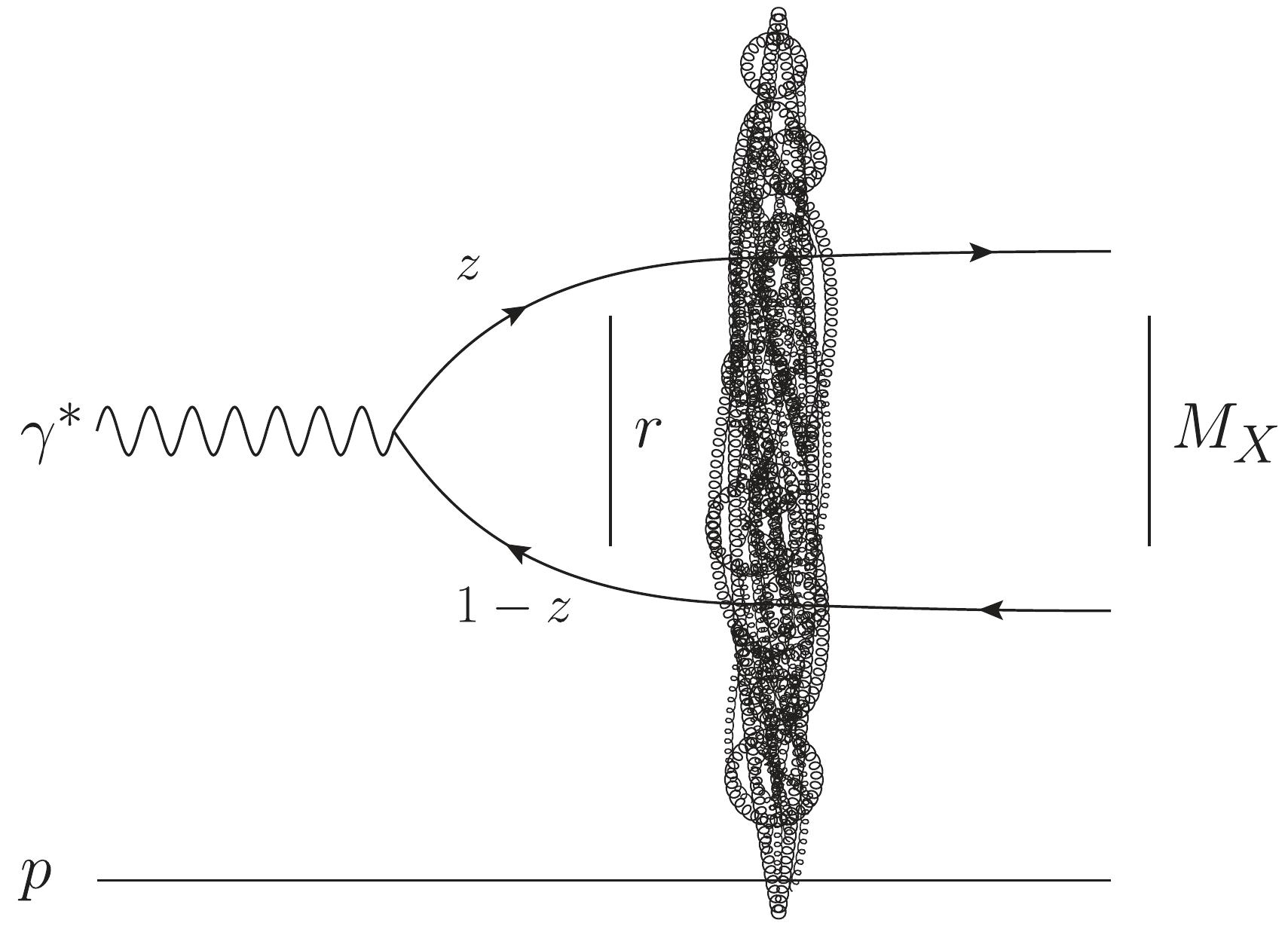}
		\caption{Diffractive $\gamma^*p \to Xp$ deep inelastic scattering at leading order in the dipole picture. The diffractive system with invariant mass $\mx$ is formed by the $q \bar q$ dipole in the color-singlet state. The figure depicts the scattering amplitude for the process.}
		\label{fig:ddis-lo}
	\end{figure}

	The leading order contributions of the $q \bar q$ dipole, depicted in Fig.~\ref{fig:ddis-lo}, to the diffractive structure functions are~\cite{Marquet:2007nf,Kowalski:2008sa}:
	\begin{align}
		\xpom F_{T, q \bar q}^{D} (\xpom, \beta, Q^2)
		& = 
		\frac{\nc Q^4}{16 \pi^3 \beta} \sum_f e_f^2 \int_{z_\textrm{min}}^{\frac{1}{2}} \ud z \, z(1-z)
		\nonumber
		\\
		& \hspace{3cm} \times
		\left[ \varepsilon^2 \left( z^2 + (1-z)^2 \right) \mathcal{J}_1 + m_f^2 \mathcal{J}_0 \right] ,
		\label{eq:ddis-lo-qqbar-ft}
		\\
		\xpom F_{L, q \bar q}^{D} (\xpom, \beta, Q^2)
		& =
		\frac{\nc Q^6}{4 \pi^3 \beta} \sum_f e_f^2 \int_{z_\textrm{min}}^{\frac{1}{2}} \ud z \, z^3(1-z)^3 \mathcal{J}_0,
		\label{eq:ddis-lo-qqbar-fl}
	\end{align}
	where the auxiliary function is defined as
	\begin{equation}
		\mathcal{J}_n \coloneqq \int \ud^2 \bt \left[ \int_0^\infty dr r \besk_n(\varepsilon r) \besj_n(\kappa r) \left(2 \mathcal{N}(\bt, r, \xpom)\right) \right]^2,
	\end{equation}
	and $\varepsilon^2 \coloneqq z(1-z)Q^2 + m_f^2$, $\kappa^2 \coloneqq z(1-z)\mx^2-m_f^2$, $z_\textrm{min} \!\coloneqq\! \left(1\!-\!\sqrt{1\!-\!4 m_f^2/\mx^2}\right)\!/2$. The notation was altered from the one in~\cite{Marquet:2007nf,Kowalski:2008sa} intentionally to clarify two details. First, the name of $\kappa$ was changed to distinguish it from the variable $k$ present in the $q\bar q g$ contribution to be discussed next, since they are not the same. Second, the substitution $\frac{\ud \sigma_{\textrm{dip}}}{d^2 \bt}(\bt,r,\xpom) \equiv 2 \mathcal{N}(\bt,r,\xpom)$ was made to make explicit the normalization used for the dipole amplitude.
	
	Some assumptions or approximations have been made in the expressions~\eqref{eq:ddis-lo-qqbar-ft} and~\eqref{eq:ddis-lo-qqbar-fl}. First, it has been assumed that the dipole amplitude $\mathcal{N}(\bt, \rt, \xpom)$ does not depend on the angle $\angle(\bt,\rt)$, which frees one to perform the angular integration of $\rt$ in the direct and complex conjugate amplitudes. This leads to the polarization dependent forms with Bessel functions $\besj_0$ and $\besj_1$.
	Secondly, the $t$-integration has been performed under the assumption that a term in the final state transverse momentum integration can be neglected. Specifically, if one assumes that $\exp(i \Deltat \cdot (\overline \rt-\rt)) \sim 1$, one gets the identity: 
	\begin{equation}
		\label{eq:delta-bt-cbt}
		\int_{-\infty}^0 \ud t
		\int \frac{\ud^2 \Deltat}{(2\pi)^2}  \delta(\Deltat^2 - \abs{t}) e^{i \Deltat \cdot \left(\overline \bt - \bt
		\right)}
		= \delta^{(2)}\left(\overline \bt - \bt
		\right),
	\end{equation}
	where $\Deltat \coloneqq \pt_0 + \pt_1 - \qt$ is the momentum transfer in the scattering. This assumption has been justified by stating $| \rt | \approx 1/Q$, which is taken to be small.
	This leads to the elimination of one of the two impact parameter integrals over $\bt$ and $\overline \bt$ that are associated with the calculation of the squared scattering amplitude, which explains how the diffractive structure functions~\eqref{eq:ddis-lo-qqbar-ft} and~\eqref{eq:ddis-lo-qqbar-fl} have two dipole size $\rt$ integrations, but only one impact parameter integral.

	\begin{figure}
	\centering
	\includegraphics[width=0.6\textwidth]{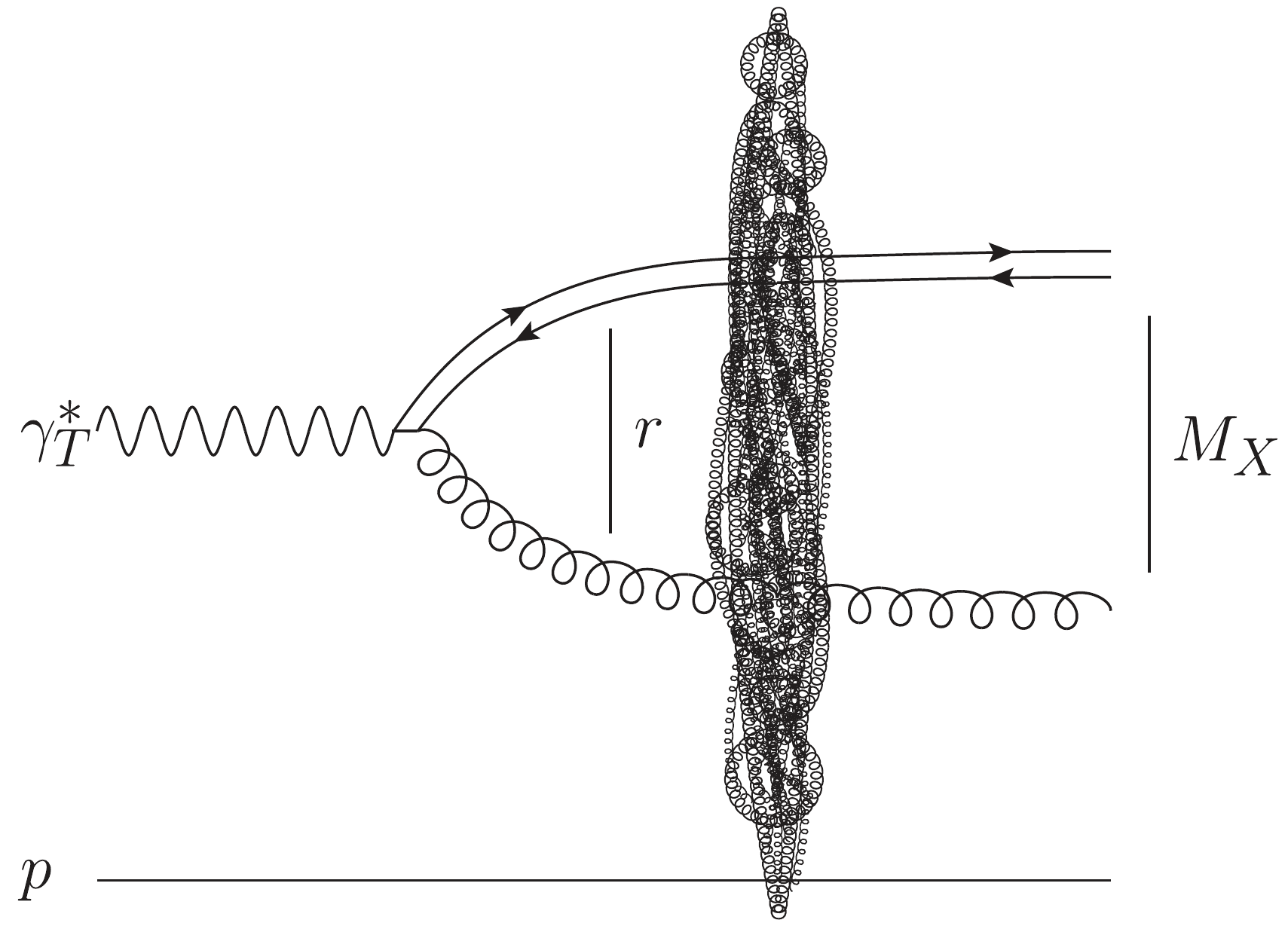}
	\caption{Schematic depiction of the leading logarithm large-$Q^2$ contribution of the transverse virtual photon splitting function to the diffractive $q \bar q g$ production. This becomes dominant over the LO $q \bar q$ production at small but finite $\beta$, or equivalently at large $\mx$.}
	\label{fig:ddis-wust}
	\end{figure}

	Moving onto the leading $\log(Q^2)$ description of the $q \bar q g$-contribution, illustrated schematically in Fig.~\ref{fig:ddis-wust}. It is at leading order only relevant for the transverse structure function, as discussed in the previous section. The original $q \bar q g$-contribution in question~\cite{Wusthoff:1997fz,GolecBiernat:1999qd} was written in the following form in Refs.~\cite{Marquet:2007nf,Kowalski:2008sa}:
	\begin{multline}
		\label{eq:ddis-ft-wusthoff}
		\xpom F_{T, q \bar q g}^{D ~(\textrm{GBW})} (\xpom, \beta, Q^2)
		\\
		=
		\frac{\as \beta}{8 \pi^4} \sum_f e_f^2 \int \ud^2 \bt \int_0^{Q^2} \ud k^2 \int_\beta^1 \ud z
		\Bigg\lbrace
			k^4 \ln \frac{Q^2}{k^2}	
			\left[\left(1 - \frac{\beta}{z}\right)^2 + \left(\frac{\beta}{z}\right)^2\right]
		\\
			\times
			\left[\int_0^\infty \ud r r \besk_2\left(\sqrt{z}kr\right) \besj_2\left(\sqrt{1-z}kr\right) \left(2 \tilde{\mathcal{N}}(\bt, r, \xpom) \right) \right]^2
		\Bigg\rbrace
	\end{multline}
	where $r$ is the size of the effective $gg$-dipole, $k^2$ is the virtuality of the final state gluon, and $z$ is the minus-momentum fraction of the $t$-channel gluon with respect to the incoming gluon. The $t$-channel gluon is present in the two-gluon exchange and it is exchanged between the $q \bar q$ dipole and gluon coming from the target, which is not shown in Fig.~\ref{fig:ddis-wust}. In this large-$Q^2$ leading logarithm approximation the scattering $q \bar q g$ state is an adjoint representation $gg$ dipole, and so the dipole amplitude $\tilde{\mathcal{N}}(\bt, r, \xpom)$ in the adjoint representation is to be used, which in the large-$\nc$ limit can be written as:
	\begin{equation}
		\tilde{\mathcal{N}}(\bt, r, \xpom) = 1 - \left(1 - \mathcal{N}(\bt, r, \xpom)\right)^2.
	\end{equation}

	The leading $\log(1/\beta)$ $q \bar q g$-contribution to the structure function~\cite{Munier:2003zb} is written as~\cite{Marquet:2007nf,Kowalski:2008sa}:
	\begin{multline}
		\label{eq:ddis-ft-ms}
		\xpom F_{T, q \bar q g}^{D ~(\textrm{MS})} (\xpom, \beta \equiv 0, Q^2)
		\\
		=
		\frac{\cf \as Q^2}{4 \pi^4 \aem}
		\int \ud^2 \rt \int_0^1 \ud z
		\frac{1}{4\pi} \left| \Psi_{\gamma_{T}^{*} \to q \bar{q}} \right|^2
		\int \ud^2 \bt A(\rt,\xpom,\bt),
	\end{multline}
	where the wavefunction is the same as for the inclusive LO DIS \eqref{eq:lfwf-sq-lo-dis-t}, with the normalization $1/4\pi$ added to account for the different normalization used in Eq.~\eqref{eq:lfwf-sq-lo-dis-t} and Refs.~\cite{Marquet:2007nf,Kowalski:2008sa}. In this limit the dipole amplitude of the $q \bar q g$ tripole $\mathcal{N}^{(2)}$ factorizes into the dipole amplitudes of two dipoles of sizes $\rt'$ and $\rt-\rt'$ and so the amplitude used is in the auxiliary function:
	\begin{multline}
		A(\rt,\xpom,\bt) =
			\int \ud^2 \rt' \frac{\rt^2}{\rt'^2(\rt-\rt')^2}
			\Big[
				\mathcal{N}(\rt') + \mathcal{N}(\rt - \rt')
				\\
				- \mathcal{N}(\rt) - \mathcal{N}(\rt')\mathcal{N}(\rt - \rt')
			\Big]^2.
	\end{multline}
	
	It was shown in Ref.~\cite{Marquet:2007nf} that the large-$Q^2$ result \eqref{eq:ddis-ft-wusthoff} does not correctly coincide with the small-$\beta$ result \eqref{eq:ddis-ft-ms} in the limit $\beta \to 0$ at moderate $Q^2$, where the small-$\beta$ result is correct by definition. However, the results were shown to agree in the validity regime $Q^2 \gg Q_s^2$ of the leading $\log(Q^2)$ result Eq.~\eqref{eq:ddis-ft-wusthoff}. To remedy this incompatibility, an interpolation model for $F_{T,q \bar q g}^D$ was constructed:
	\begin{equation}
		\label{eq:ddis-ft-interp}
		\xpom F_{T, q \bar q g}^{D ~\textrm{(interp)}} (\xpom, \beta, Q^2)
			=
			\frac{\xpom F_{T, q \bar q}^{D ~(\textrm{GBW})} (\xpom, \beta, Q^2) \times \xpom F_{T, q \bar q}^{D ~(\textrm{MS})} (\xpom, Q^2)}{\xpom F_{T, q \bar q}^{D ~(\textrm{GBW})} (\xpom, \beta = 0, Q^2)},
	\end{equation}
	which leverages both results to get a better description of $F_{T,q \bar q g}^D$ at moderate $Q^2$ and small $\beta$. With this the complete model for $F_2^D$ is 
	\begin{multline}
		\xpom F_{2, ~\mathrm{LO}+\mathrm{LL}(Q^2)+\mathrm{LL}(1/\beta)}^{D} (\xpom, \beta, Q^2)
		=
		\xpom F_{L, q \bar q}^{D} (\xpom, \beta, Q^2)
		+ \xpom F_{T, q \bar q}^{D} (\xpom, \beta, Q^2)
		\\
		+ \xpom F_{T, q \bar q g}^{D ~\textrm{(interp)}} (\xpom, \beta, Q^2).
	\end{multline}
	This formulation has been the most precise description of DDIS in the dipole picture, and was used to describe HERA data very well~\cite{Kowalski:2008sa}. After a brief aside in the next section, in Sec.~\ref{sec:ddis-nlo} we will discuss how this is superseded by NLO calculations.

	\subsection{Impact parameter dependence of the \texorpdfstring{$q \bar q g$}{qq̅g}-contribu\-tion in the large-\texorpdfstring{$Q^2$}{Q²} limit}
		
		Some work is needed to get the form of the GBW structure function~\eqref{eq:ddis-ft-wusthoff} starting from the original result in Ref.~\cite{GolecBiernat:1999qd}. The intermediate steps of this calculation are not shown in the literature, which is amended here with a calculation connecting the two results.
		
		Our starting point is the $q \bar q g$ contribution to $F_T^D$ from Ref.~\cite{GolecBiernat:1999qd}\footnote{Italicized variables $r, r', k_t$ are the lengths of corresponding vector quantities.}:
		\begin{multline}
			\label{eq:ddis-ft-wusthoff-orig}
			\xpom F_{T, q \bar q g}^{D ~(\textrm{GBW})}
			=
			\frac{81 \beta}{512 \pi^5 B_D} \sum_f e_f^2 \frac{\alpha_s}{2\pi}
			\int_\beta^1 \frac{\ud z}{z} \left[ \left( 1 - \frac{\beta}{z} \right)^2 +\left( \frac{\beta}{z} \right)^2 \right]
			\frac{z}{(1-z)^3}
			\\
			\times
			\int \frac{\ud^2 \ktt}{(2\pi)^2}
			\ktt^4 \ln\left( \frac{(1-z)Q^2}{\ktt^2} \right) \Theta((1-z)Q^2 - \ktt^2)
			\\
			\times
			\int \ud^2 \rt \int \ud^2 \rt' e^{i \ktt \cdot (\rt - \rt')}
			\hat \sigma(r,\xpom) \hat \sigma(r',\xpom)
			\left( \delta^{mn} - 2 \frac{\rt^{m} \rt^{n}}{r^{2}} \right)
			\\
			\times
			\left( \delta^{mn} - 2 \frac{\rt'^{m} \rt'^{n}}{r'^{2}} \right)
			\besk_2\left(\sqrt{\frac{z}{1-z}\ktt^2 r^2}\right)
			\besk_2\left(\sqrt{\frac{z}{1-z}\ktt^2 r'^2}\right).
		\end{multline}
		The angular integrals in the  coordinate space can be performed, so we separate the terms with angular dependence:
		\begin{align}
			\mathcal{I} 
			&
			\coloneqq
			\int \ud^2 \rt \int \ud^2 \rt' e^{i \ktt \cdot (\rt - \rt')}
			\left( \delta^{mn} - 2 \frac{\rt^{m} \rt^{n}}{r^{2}} \right)
			\left( \delta^{mn} - 2 \frac{\rt'^{m} \rt'^{n}}{r'^{2}} \right)
			\nonumber
			\\
			&
			=
			\int \ud^2 \rt \int \ud^2 \rt' 
			e^{i \ktt \cdot (\rt - \rt')} \left[ 4 \frac{(\rt \cdot \rt')^2}{r^2 r'^2} - 2 \right]
			\nonumber
			\\
			&
			\eqqcolon 2 \mathcal{I}_1 - 2 \mathcal{I}_2.
		\end{align}
		We will need the identity
		\begin{equation}
			\label{eq:Jn}
			\besj_n(z) = \frac{1}{2\pi i^n} \int_0^{2\pi} \ud \theta \cos(n\theta) e^{i z \cos\theta}.
		\end{equation}
		The second integral simply yields
		\begin{equation}
			\mathcal{I}_2 = (2\pi)^2 \int r \ud r \besj_0(k_t r) \int r' \ud r' \besj_0(k_t r').
		\end{equation}
		For $\mathcal{I}_1$ we write, parameterizing the angles as $\angle(\rt,\ktt) \eqqcolon \theta$,  $\angle(\rt',\ktt) \eqqcolon \phi$:
		\begin{equation}
			\rt \cdot \rt' = r r' \cos(\phi - \theta) = r r' (\cos \phi \cos \theta + \sin \phi \sin \theta).
		\end{equation}
		This can be simplified with some trigonometric algebra:
		\begin{align*}
			(\cos \phi \cos \theta + \sin \phi \sin \theta)^2
			& =
			\cos^2 \phi \cos^2 \theta + \sin^2 \phi \sin^2 \theta
			\\
			& \quad + 2 \cos \phi \cos \theta \cancelto{0}{\sin \phi \sin \theta}
			\\
			& = \frac{1}{2} (\cos 2\theta \cos 2\phi + 1),
		\end{align*}
		where the cross-term linear in $\sin \phi$ vanishes in the integration.
		Finally computing the first integral, we find
		\begin{align}
			\mathcal{I}_1
			& =
			\int r \ud r \ud \theta \int r' \ud r' \ud \phi \, e^{i k_t r \cos \theta} e^{-i k_t r' \cos \phi}
			\,
			2 \frac{(\rt \cdot \rt')^2}{r^2 r'^2}
			\nonumber
			\\
			& =
			\int r \ud r \ud \theta \int r' \ud r' \ud \phi \, e^{i k_t r \cos \theta} e^{-i k_t r' \cos \phi}
			(\cos 2\theta \cos 2\phi + 1)
			\nonumber
			\\
			& = (2\pi)^2 \int r \ud r \besj_2(k_t r) \int r' \ud r' \besj_2(k_t r') + \mathcal{I}_2.
		\end{align}
		Thus we have found
		\begin{align}
			\mathcal{I} = 2 (2\pi)^2 \int r \ud r \besj_2(k_t r) \int r' \ud r' \besj_2(k_t r').
		\end{align} 
		
		Integrating over the angle of $\ktt$ and imposing\footnote{The variable $k^2$ is the mean virtuality of the exchanged t-channel gluon~\cite{Wusthoff:1997fz,GolecBiernat:1999qd}, defined as $k^2 \coloneqq \frac{\ktt^2}{1-z}$.} $k_t^2 \equiv (1-z) k^2$, we can write the GBW result~\eqref{eq:ddis-ft-wusthoff-orig} in the form:
		\begin{multline}
			\xpom F_{T, q \bar q g}^{D ~(\textrm{GBW})}
			= 
			\frac{81 \as \beta}{512 \pi^5 B_D} \sum_f e_f^2 \int_\beta^1 dz \left[ \left( 1 - \frac{\beta}{z} \right)^2 +\left( \frac{\beta}{z} \right)^2 \right]
			\\
			\times
			\int_0^\infty \ud k^2 k^4 \ln\left( \frac{Q^2}{k^2} \right) \Theta(Q^2 - k^2)
			\\
			\times
			\left[ \int r \ud r \hat \sigma(r, x_{\mathbb{P}}) \besk_2 \left( \sqrt{ z } k r \right) \besj_2 \left( \sqrt{ 1-z} k r \right) \right]^2.
		\end{multline}
		It remains to upgrade the model for the adjoint dipole amplitude to be impact parameter dependent. This begins by recognizing that in~\cite{GolecBiernat:1999qd} the model used is~\cite{Marquet:2007nf}:
		\begin{equation}
			\tilde \sigma (r , \xpom) \approx \frac{\nc}{\cf} \hat \sigma(r , \xpom) = \frac{9}{4} \hat \sigma(r , \xpom),
		\end{equation}
		where $\tilde \sigma$ is the adjoint amplitude of the $gg$ dipole. In the large-$Q^2$ limit the $q\bar q$ dipole size is much smaller than the quark-gluon distance, and so the $q \bar q g$-tripole scattering is represented as an effective $gg$-dipole scattering~\cite{GolecBiernat:1999qd}. Secondly, based on the correct replacement of the diffractive slope model introduced in~\cite{Marquet:2007nf}, a factor of $1/(4\pi B_D)$ must be absorbed into the normalization of the dipole amplitudes.
		This is seen explicitly by assuming that the impact parameter dependence factorizes from the dipole amplitude and taking the normalized Gaussian proton impact parameter profile used in~\cite{Kowalski:2008sa} $T_p(\bt) \coloneqq \frac{1}{2 \pi B_D} \exp(-\frac{\bt^2}{2 B_D})$ and integrating:
		\begin{equation}
			\int \ud \bt^2 \left(T_p (\bt)\right)^2
			= \int \ud \bt^2 \frac{1}{(2 \pi B_D)^2} \exp(-\frac{\bt^2}{B_D})
			= \frac{1}{4 \pi B_D},
		\end{equation}
		where the fact is used that the $t$-integration done in~\cite{Kowalski:2008sa} over the full range $[-\infty,0]$ has forced the impact parameters in the direct and conjugate amplitude to be the same $\bt' \equiv \bt$.
		Thus we  essentially undo the impact parameter integration to incorporate the impact parameter dependence back into the dipole amplitude, and make the replacement
		\begin{align}
			\frac{81}{16} \frac{\hat \sigma(r, x_{\mathbb{P}}) \hat \sigma(r', x_{\mathbb{P}})}{4\pi B_D}
			& =
			\frac{\tilde \sigma(r, x_{\mathbb{P}}) \tilde \sigma(r', x_{\mathbb{P}})}{4\pi B_D}
			\nonumber
			\\
			&
			=
			\int \ud^2 \bt 
			\frac{\ud \tilde \sigma}{\ud^2 \bt}(\bt, \rt, x_{\mathbb{P}})
			\frac{\ud \tilde \sigma}{\ud^2 \bt}(\bt, \rt', x_{\mathbb{P}}).
		\end{align}
		With this we finally have
		\begin{multline}
			\xpom F_{T, q \bar q g}^{D ~(\textrm{GBW})}
			= 
			\frac{\as \beta}{8 \pi^4} \sum_f e_f^2 \int_\beta^1 dz \left[ \left( 1 - \frac{\beta}{z} \right)^2 +\left( \frac{\beta}{z} \right)^2 \right]
			\int_0^{Q^2} \ud k^2 k^4 \ln\left( \frac{Q^2}{k^2} \right)
			\\
			\times
			\int \ud^2 \bt
			\left[ \int r \ud r \frac{\ud \tilde \sigma}{\ud^2 \bt}(\bt, \rt', x_{\xpom}) \besk_2 \left( \sqrt{ z } k r \right) \besj_2 \left( \sqrt{ 1-z} k r \right) \right]^2,
		\end{multline}
		which is the result derived in~\cite{Kowalski:2008sa} and shown in Eq.~\eqref{eq:ddis-ft-wusthoff}.

\section{DDIS in the dipole picture at next-to-lead\-ing order}
	\label{sec:ddis-nlo}

	\begin{figure}
		\centering
		\includegraphics[width=0.6\textwidth]{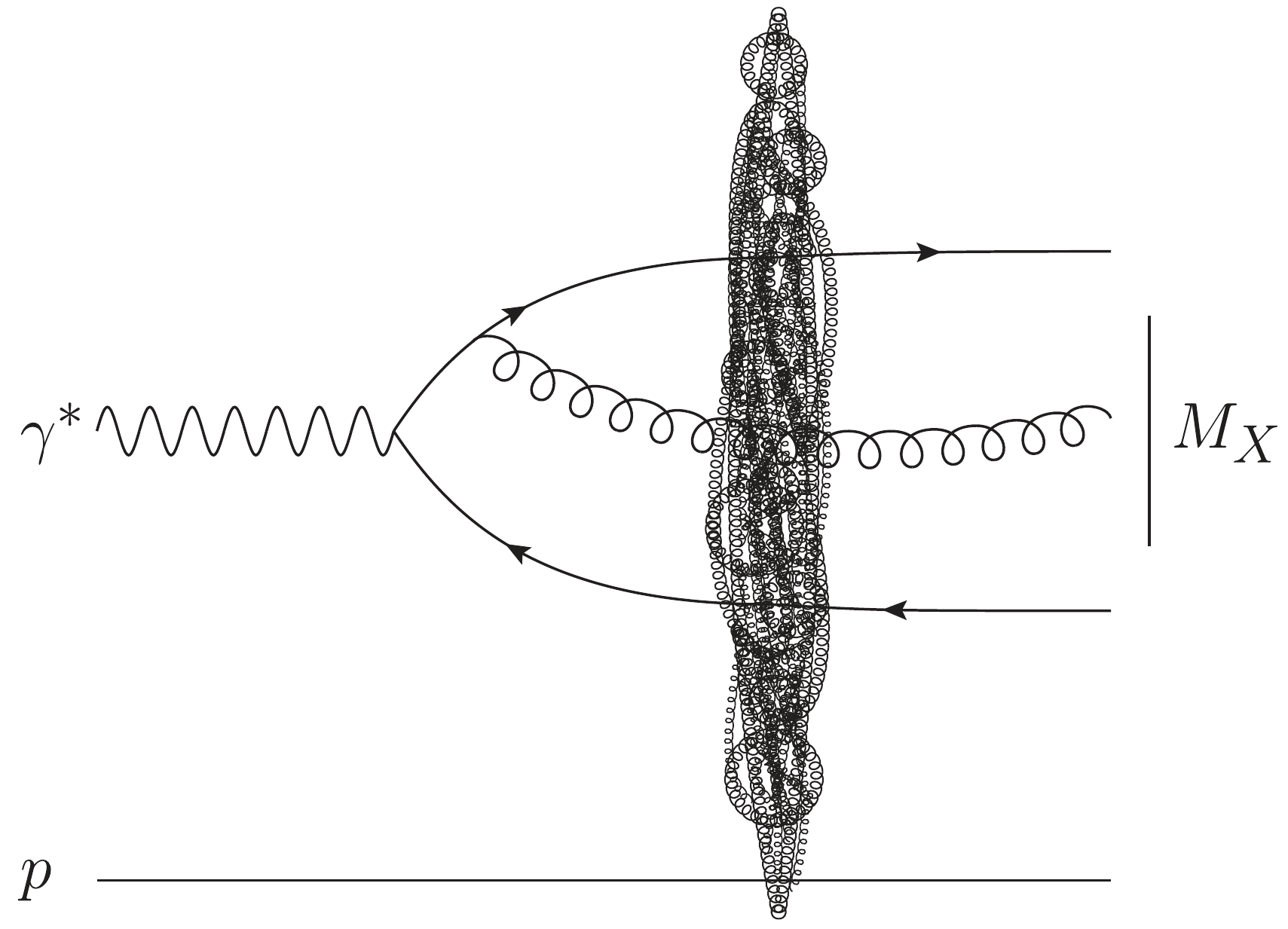}
		\caption{An example of a tree-level NLO diagram that contributes to the $q \bar q g$-contribution to $F_T^D$ and $F_L^D$. A gluon is emitted before the scattering by one of the quarks, and all three partons scatter from the color-field of the target into the diffractive color-singlet final state.}
		\label{fig:ddis-nlo-real}
	\end{figure}

	\begin{figure*}
		\centering
		\begin{subfigure}[]{0.48\textwidth}
			\centering
			\includegraphics[width=\textwidth]{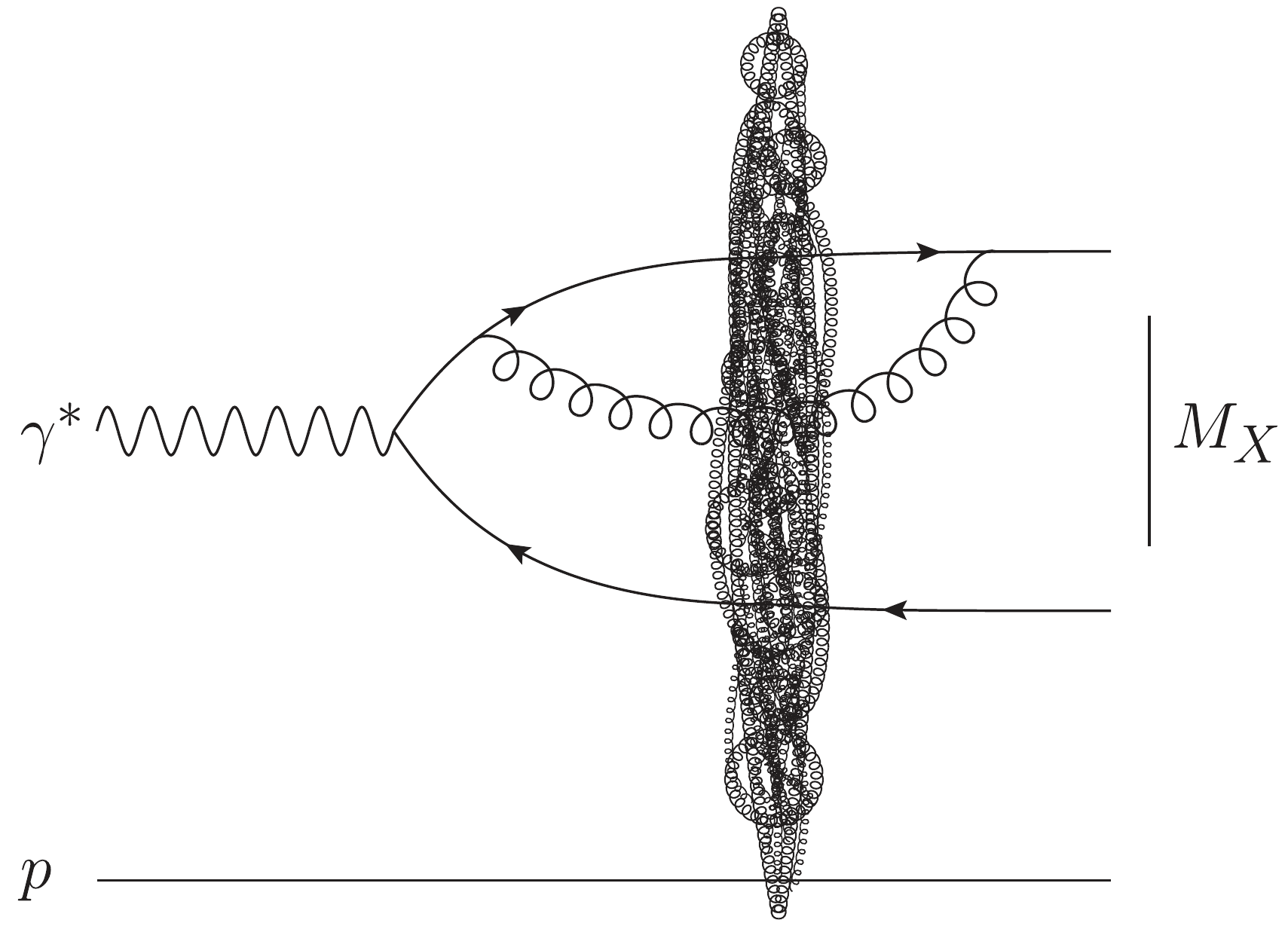}
							\caption[]%
							{{Propagator loop, gluon scatters}}    
							\label{fig:ddis-nlo-prop-loop}
		\end{subfigure}
		\hfill
		\begin{subfigure}[]{0.48\textwidth}  
			\centering 
			\includegraphics[width=\textwidth]{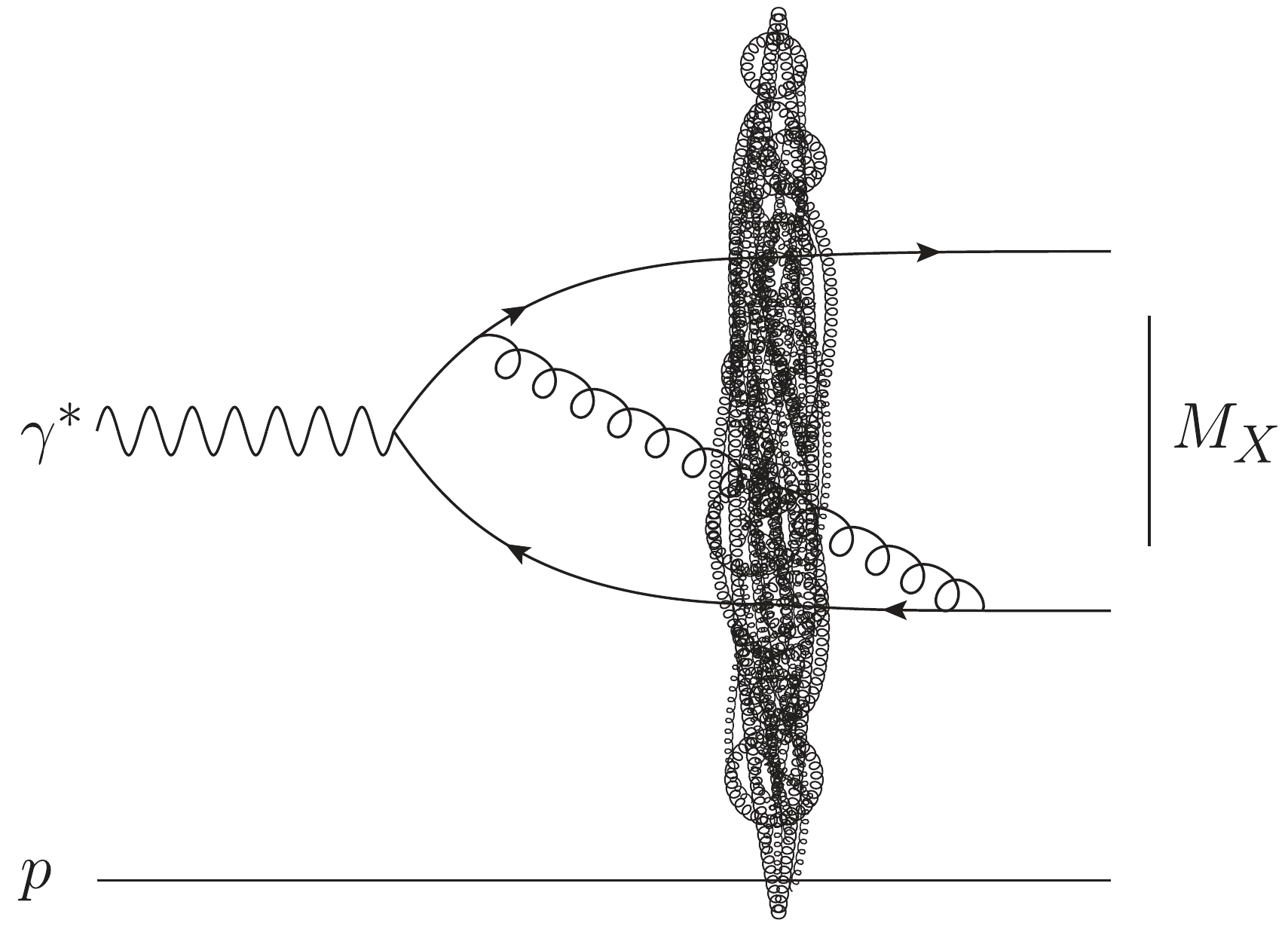}
							\caption[]%
							{{Vertex loop, gluon scatters}}   
							\label{fig:ddis-nlo-vertex-loop}
		\end{subfigure}
		\vskip\baselineskip
		\begin{subfigure}[]{0.48\textwidth}   
			\centering 
			\includegraphics[width=\textwidth]{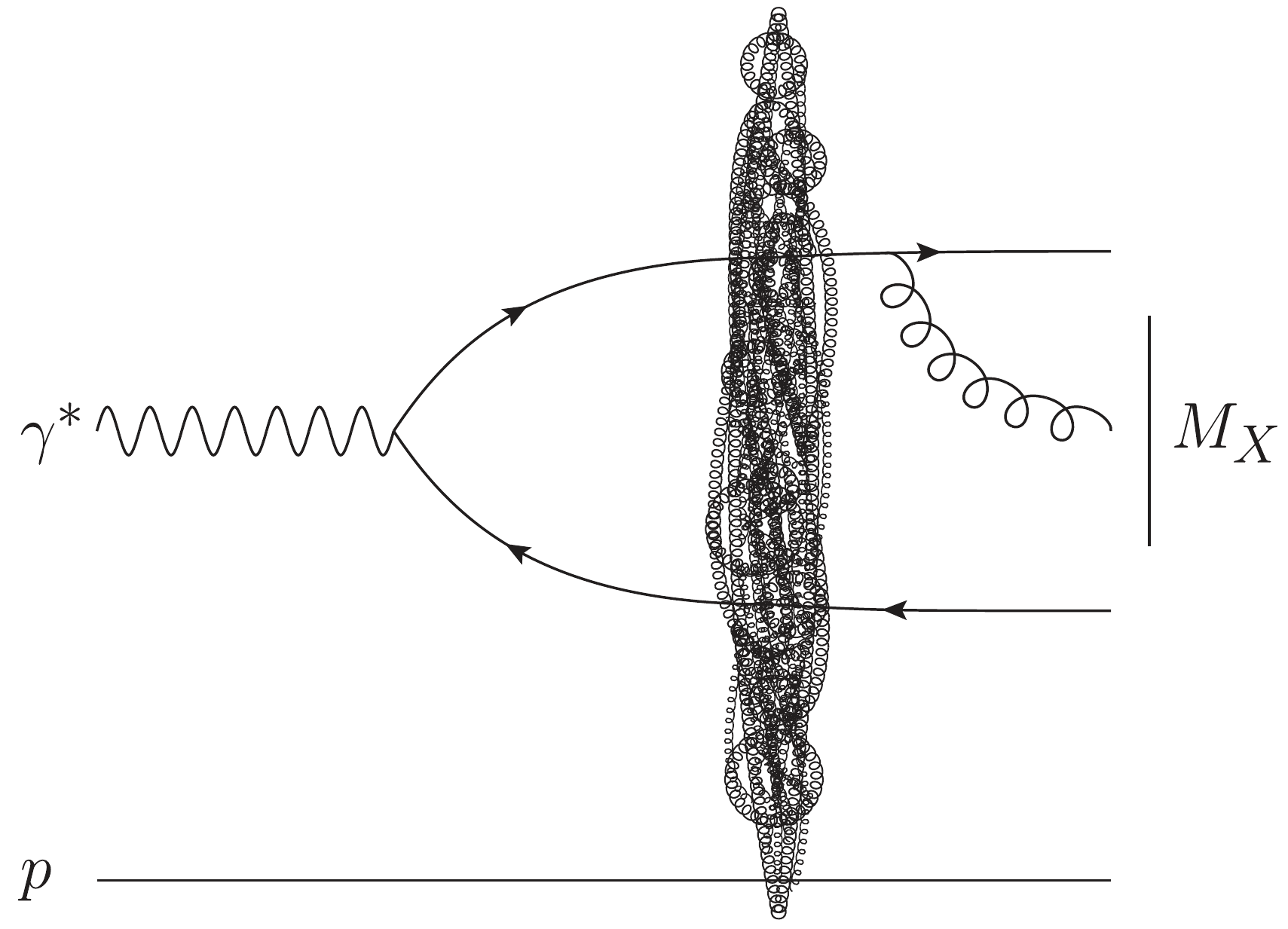}
			\caption[]%
			{{Final state emission}}    
			\label{fig:ddis-nlo-fs-emission}
		\end{subfigure}
		\hfill
		\begin{subfigure}[]{0.48\textwidth}   
			\centering 
			\includegraphics[width=\textwidth]{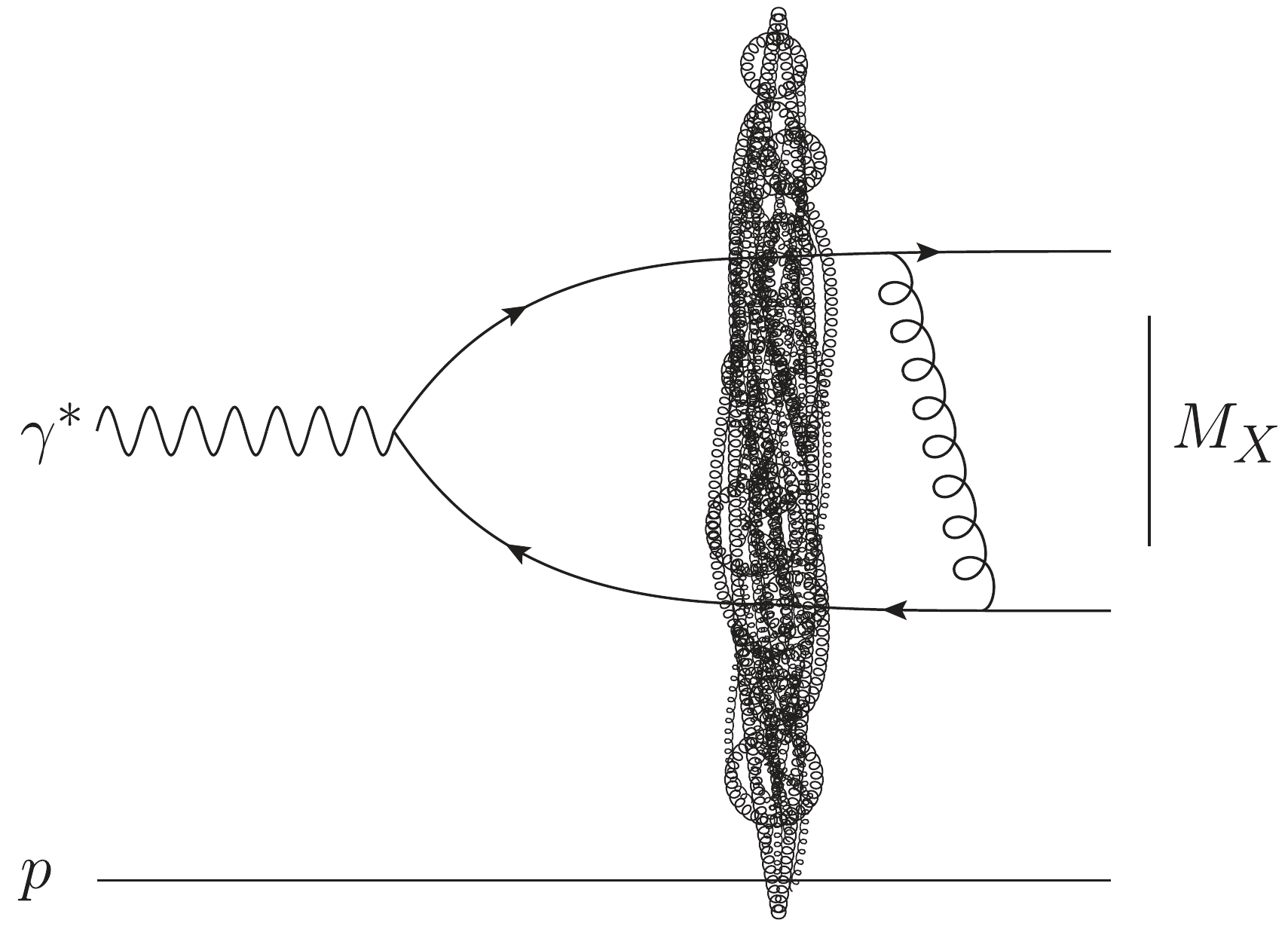}
			\caption[]%
			{{Final state interaction}}    
			\label{fig:ddis-nlo-fs-interaction}
		\end{subfigure}
		\vskip\baselineskip
		\begin{subfigure}[]{0.48\textwidth}   
			\centering 
			\includegraphics[width=\textwidth]{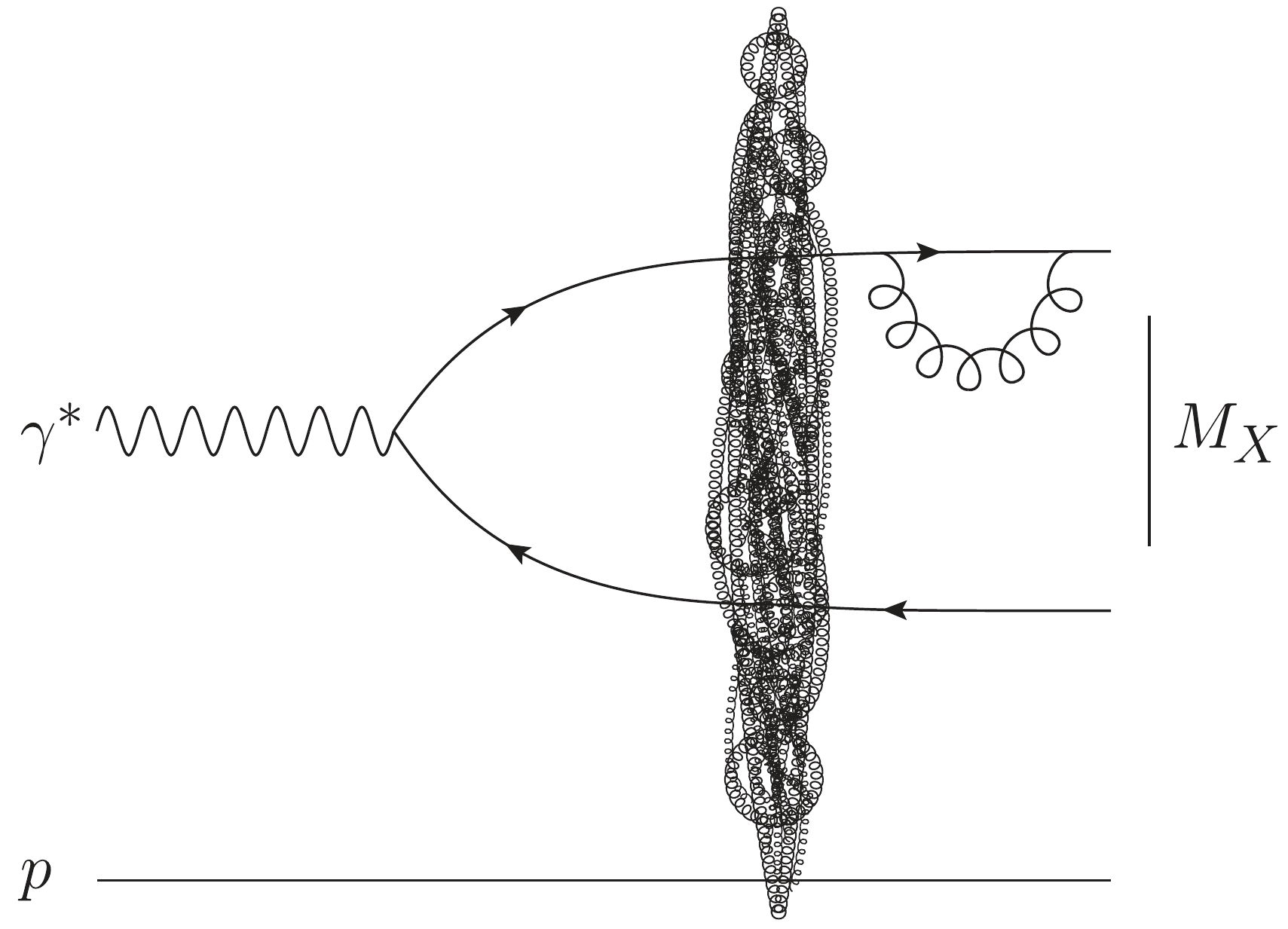}
							\caption[]%
							{{Final state propagator loop}}    
							\label{fig:ddis-nlo-fs-prop-loop}
		\end{subfigure}
		\caption{\small Five types of NLO corrections to DDIS that are novel in comparison to NLO DIS~\cite{Beuf:2016wdz,Beuf:2017bpd}. This is not an exhaustive presentation of relevant diagrams, only one out of multiple diagrams is shown for each type.} 
		\label{fig:ddis-nlo-loops-and-finalstate}
	\end{figure*}
	
	The next-to-leading order diffractive cross sections receive numerous contributions from both tree-level and loop diagrams, some of which are novel in comparison to NLO DIS~\cite{Beuf:2017bpd}. The large-$Q^2$ limit $q \bar q g$-contribution discussed in Sec.~\ref{sec:ddis-lo} includes a subset of the NLO tree-level $q \bar q g$-contributions under the assumption of strongly ordered kinematics illustrated in Fig.~\ref{fig:ddis-wust}.
	In contrast to this, in Fig.~\ref{fig:ddis-nlo-real} is shown the corresponding $q \bar q g$-contribution in general kinematics at NLO, where there is no strong ordering of the dipole sizes.
	At full NLO accuracy both the transverse and longitudinal virtual photon contribute to the $q \bar q g$-production, and additionally the transversely polarized photon can instantaneously fluctuate into the $q \bar q g$ Fock state, introducing new diagrams shown in Fig.~\ref{fig:gammaT_qqbarg}. Then in the CGC framework the formed $q \bar q g$-state scatters eikonally from the color-field of the target so that the scattered state is a color-singlet.
	At NLO the emission of a gluon in the final state is possible, shown in Fig.~\ref{fig:ddis-nlo-fs-emission}, but these contributions are not included in the large-$Q^2$ calculation~\cite{Wusthoff:1995hd,Wusthoff:1997fz}. The full NLO dipole picture result corresponding to the large-$Q^2$ diffractive $q \bar q g$-structure function Eq.~\eqref{eq:ddis-ft-wusthoff} is discussed in more detail and calculated in the next section.

	On top of these tree-level contributions, the diffractive cross sections receive contributions from a variety of loop diagrams. Completely analogously to NLO DIS~\cite{Beuf:2016wdz}, gluon loop corrections to the quark and antiquark propagators before the scattering contribute, as well as gluon loop corrections to the $\gamma_{T,L}^* q \bar q$- and $\gamma_T^* q \bar q g$-vertices. These contributions to the virtual photon LFWFs have been calculated in Ref.~\cite{Beuf:2016wdz} and should be usable as-is for the calculation of the respective corrections to the diffractive cross sections.
	
	NLO DDIS however receives new types of loop contributions as well, shown in Fig.~\ref{fig:ddis-nlo-loops-and-finalstate}. These include propagator and vertex loops where the emission happens before and the absorption after the scattering from the target, shown in Figs.~\ref{fig:ddis-nlo-prop-loop} and~\ref{fig:ddis-nlo-vertex-loop} respectively. Completely new type of contributions come from the final state, where the quark and antiquark can: emit a gluon, interact by a gluon exchange, or either can receive a propagator loop correction, shown in Figs.~\ref{fig:ddis-nlo-fs-emission},~\ref{fig:ddis-nlo-fs-interaction} and~\ref{fig:ddis-nlo-fs-prop-loop}. These types of diagrams, and their interferences with the above tree-level and loop diagrams familiar from NLO DIS, are novel and will need to be calculated from scratch. These contributions will also include UV divergences which must be regularized and canceled out between the diagrams. The standard method is to use dimensional regularization, which means all tree-level and loop contributions will need to be consistently calculated in $D$-dimensions, analogously to NLO DIS~\cite{Beuf:2016wdz,Beuf:2017bpd}.

	\subsection{The \texorpdfstring{$q \bar q g$}{qq̅g}-contribution to the DDIS structure functions at NLO}
	\label{sec:ddis-nlo-qqbarg}
		
		As discussed above, the final state emissions are not included in the original large-$Q^2$ limit calculation of the $q \bar q g$-contribution~\cite{Wusthoff:1995hd,Wusthoff:1997fz}. This means that the derivation of the corresponding full NLO accuracy result without approximations is fairly straightforward: only the tree-level NLO diagrams where the emitted gluon crosses both the shockwaves --- in the direct and complex conjugate amplitude --- and the cut contribute. An example diagram is shown in Fig.~\ref{fig:ddis-nlo-real}, and all at the stage of the transverse photon splitting are shown in Fig.~\ref{fig:gammaT_qqbarg}. For a longitudinal photon only the diagrams $(a)$ and $(b)$ are relevant. A further simplification is that since the final state has a definite invariant mass $\mx$, all of the relevant contributions --- which are tree-level, and as such have no loops --- become UV finite and the calculation may simply be done in $D=4$.
		
		In this section this calculation is done for both the longitudinally and transversely polarized virtual photon, taking advantage of the NLO virtual photon splitting functions calculated in Ref.~\cite{Beuf:2017bpd}.		
		The calculation is structured as follows. First we write down the relevant definitions to calculate the diffractive cross section for the $\gamma^* p$ scattering, after which the photon splitting wavefunction contributions are calculated in the case of DDIS. Finally we write the results for $F_L^D$ and $F_T^D$ in NLO accuracy, in somewhat preliminary form.
	
		\subsection{Definitions: the \texorpdfstring{$q \bar q g$}{qq̅g} diffractive cross section}
			The differential cross section for the virtual photon-proton scattering via the $q \bar q g$ Fock state is defined as:
			\begin{multline}
			    \label{eq:diff-ddis-cs-nlo-qqg}
				\ud \sigma_{\gamma^*_\lambda \rightarrow q \bar{q} g ~ \textrm{singlet}}^{\textrm{NLO}}
				\coloneqq
				(2 \qplus) 2\pi \delta(\pplus_0 + \pplus_1 + \pplus_2 - \qplus)
				\\
				\times
				\frac{\ud^3 \vecp{0}}{(2\pi)^3} \frac{\theta(\pplus_0)}{2 \pplus_0} \frac{\ud^3 \vecp{1}}{(2\pi)^3} \frac{\theta(\pplus_1)}{2 \pplus_1} \frac{\ud^3 \vecp{2}}{(2\pi)^3} \frac{\theta(\pplus_2)}{2 \pplus_2}
				\sum_{q_0 \bar q_1 g_2 ~ \textrm{F. states} } \left| \mathcal{M}_{\gamma^* \rightarrow q \bar q g ~ \textrm{singlet}}^{\textrm{NLO}} \right|^2,
			\end{multline}
			where the scattering amplitude is defined with the relation:
			\begin{multline}
			    \label{eq:ddis-amplitude-nlo-qbarqg}
				\bra{( g_2 \bar q_1 q_0)_H \vphantom{\opse}} \left( \opse - \bf 1 \right) \ket{\gamma^*_\lambda(\qplus, \qt; Q^2)_H \vphantom{\opse}}
				\\
				=
				(2 \qplus) 2\pi \delta(\pplus_0 + \pplus_1 + \pplus_2 - \qplus) i \mathcal{M}_{\gamma^* \rightarrow q \bar q g}^{\textrm{NLO}},
			\end{multline}
			where $H$ subscript refers to the Heisenberg picture dressed states~\cite{Beuf:2016wdz}. The outgoing state is
			\begin{multline}
			    \label{eq:fock-qqbarg}
				\bra{(g_2 \bar q_1 q_0)_H \vphantom{\se}}
				=
				\sqrt{Z_{g}} \sqrt{Z_{q}}^2
				\left\lbrace
				\bra{0} a(\pplus_2, \pt_2) d(\pplus_1, \pt_1) b(\pplus_0, \pt_0) + \dots
				\right\rbrace
				\\
				=
				\sqrt{Z_{g}} \sqrt{Z_{q}}^2
				\bigg\lbrace
				\int \ud^2 \xt_0 \ud^2 \xt_1 \ud^2 \xt_2 e^{-i \pt_0 \cdot \xt_0} e^{-i \pt_1 \cdot \xt_1} e^{-i \pt_2 \cdot \xt_2}
				\\
				\times
				\bra{0} \tilde{a}(\pplus_2, \xt_2) \tilde{d}(\pplus_1, \xt_1) \tilde{b}(\pplus_0, \xt_0) + \dots
				\bigg\rbrace,
			\end{multline}
			where the normalization is of the order $Z_g Z_q^2 = 1 + \ocal(\as)$, and the correction is related to the self-energy loop corrections of the $q \bar q$ final state, which do not contribute to the tree-level $q \bar q g$-contribution and thus can be dropped in what follows.
			Using this, the expression~\eqref{eq:fock-gamma} for the dressed photon wavefunction and the commutation~\eqref{eq:op-com-anticom-b}--\eqref{eq:op-com-a} and eikonal scattering operator~\eqref{com_a_dag_SE}--\eqref{com_d_dag_SE} relations from Sec.~\ref{sec:lcpt} we find for the overlap
			\begin{multline}
				\bra{0} a(\pplus_2, \cdot) d(\pplus_1, \cdot) b(\pplus_0, \cdot) (\opse - \textbf{1} ) b^\dagger(\kprimplus_0, \cdot) d^\dagger(\kprimplus_1, \cdot) a^\dagger(\kprimplus_2, \cdot) \ket{0}
				\\
				=
				(2\pplus_0) (2 \pi) (2\pplus_1) (2 \pi) (2\pplus_2) (2 \pi)
				\delta(\kprimplus_0 - \pplus_0) \delta(\kprimplus_1 - \pplus_1) \delta(\kprimplus_2 - \pplus_2)
				\\
				\times
				\delta^{(2)}(\xt'_0 - \xt_0) \delta^{(2)}(\xt'_1 - \xt_1) \delta^{(2)}(\xt'_2 - \xt_2)
				\delta_{h_0', h_0} \delta_{h_1', h_1} \delta_{\lambda_2', \lambda_2}
				\\
				\times
				\left[
				U_F(\xt_{0}')_{\beta_0' \alpha_0'}
				U_F^\dagger(\xt_1')_{\alpha_1' \beta_1'}
				U_A(\xt_2')_{b' a'}
				\delta_{\alpha_0, \beta_0'} \delta_{\alpha_1, \beta_1'} \delta_{a, b'}
				- 
				\delta_{\alpha_0, \alpha_0'} \delta_{\alpha_1, \alpha_1'} \delta_{a, a'}
				\right],
			\end{multline}
			where the shorthand is defined as $b^\dagger(\kprimplus_0, \cdot) \coloneqq b^\dagger(\kprimplus_0, \xt_0', h_0', \alpha_0')$ in the ket-state and $b(\pplus_0, \cdot) \coloneqq b(\pplus_0, \xt_0, h_0, \alpha_0)$ in the bra-state. Analogously for the antiquark operators with subscript $1$. For the gluon they are $a^\dagger(\kprimplus_2, \cdot) \coloneqq a^\dagger(\kprimplus_2, \xt_2', \lambda_2', a_0')$ and $a(\pplus_2, \cdot) \coloneqq a(\pplus_2, \xt_2, \lambda_2, a_0)$.
			With this we find for the scattering amplitude
			\begin{align}
				\label{eq:qqg-M-amplitude}
				(2 \qplus) & 2\pi \delta(\pplus_0 + \pplus_1 + \pplus_2 - \qplus) i \mathcal{M}_{\gamma^* \rightarrow q \bar q g
				}^{\tree}
				\nonumber\\
				= &
				\int \ud^2 \xt_0 \int \ud^2 \xt_1 \int \ud^2 \xt_2
				\int \ud^2 \xt_0' \int \ud^2 \xt_1' \int \ud^2 \xt_2'
				\nonumber\\
				& \int \frac{\ud \kprimplus_0}{(2\pi)2\kprimplus_0}
				\int \frac{\ud \kprimplus_1}{(2\pi)2\kprimplus_1}
				\int \frac{\ud \kprimplus_2}{(2\pi)2\kprimplus_2}
				e^{-i(\pt_0 \cdot \xt_0  + \pt_1 \cdot \xt_1 + \pt_2 \cdot \xt_2)}
				\nonumber\\
				& \times
				(2\qplus) 2\pi \delta(\kprimplus_0 + \kprimplus_1 + \kprimplus_2 - \qplus)
				e^{i \frac{\qt}{\qplus} \cdot (\kprimplus_0 \xt_0' + \kprimplus_1 \xt_1' + \kprimplus_2 \xt_2')}
				{t}^{a_2}_{\alpha_0 \alpha_1} \widetilde{\psi}_{\gamma^{*}_\lambda \rightarrow q_0 \bar{q}_1 g_2}
				\nonumber\\
				& \times
				\bra{0} a(\pplus_2, \cdot) d(\pplus_1, \cdot) b(\pplus_0, \cdot) (\opse - \textbf{1} ) b^\dagger(\kprimplus_0, \cdot) d^\dagger(\kprimplus_1, \cdot) a^\dagger(\kprimplus_2, \cdot) \ket{0}
				\nonumber\\
				= &
				(2\qplus) 2\pi \delta(\pplus_0 + \pplus_1 + \pplus_2 - \qplus)
				\int \ud^2 \xt_0 \int \ud^2 \xt_1 \int \ud^2 \xt_2
				\nonumber\\
				& e^{-i(\pt_0 \cdot \xt_0 + \pt_1 \cdot \xt_1 + \pt_2 \cdot \xt_2)}
				e^{i \frac{\qt}{\qplus} \cdot (\pplus_0 \xt_0 + \pplus_1 \xt_1 + \pplus_2 \xt_2)}
				\nonumber\\
				& \times
				\widetilde{\psi}_{\gamma^{*}_\lambda \rightarrow q_0 \bar{q}_1 g_2}
				\left[
				U_F(\xt_{0})_{\alpha_0 \alpha_0'}
				{t}^{a'}_{\alpha_0' \alpha_1'}
				U_F^\dagger(\xt_1)_{\alpha_1' \alpha_1}
				U_A(\xt_2)_{a a'}
				- 
				{t}^{a}_{\alpha_0 \alpha_1}
				\right],
			\end{align}
		
			The diffractive system must be in a color-singlet state, which is enforced with a color projection operator:
			\begin{equation}
				P_{q \bar q g}^{\textrm{singlet}} \coloneqq \frac{ \left(t^{a'}\right)_{\alpha'_1 \alpha'_0} \left(t^{a} \vphantom{t^{a'}} \right)_{\alpha_0 \alpha_1} }{d(F) C_F},
			\end{equation}
			where $d(F) \equiv \nc$.
			The singlet projection acts on the color factor seen above as
			\begin{multline}
				\frac{ \left(t^{a}\right)_{\alpha_1 \alpha_0} \left(t^{b} \right)_{\beta_0 \beta_1} }{d(F) C_F}
				\left[
				U_F(\xt_{0})_{\alpha_0 \alpha_0'}
				{t}^{a'}_{\alpha_0' \alpha_1'}
				U_F^\dagger(\xt_1)_{\alpha_1' \alpha_1}
				U_A(\xt_2)_{a a'}
				- 
				{t}^{a}_{\alpha_0 \alpha_1}
				\right]
				\\
				=
				\left(t^{b} \right)_{\beta_0 \beta_1}
				\left[
				\frac{1}{N_c C_F} \tr \left[ U_F(\xt_0) t^{a'} U_F^\dagger(\xt_1) t^a \right] U_A(\xt_2)_{a a'}
				-
				1
				\right]
				\\
				=
				\left(t^{b} \right)_{\beta_0 \beta_1}
				\left[
				S_{012}^{(3)}
				-
				1
				\right],
			\end{multline}
			where the tripole scattering amplitude is defined
			\begin{equation}
				S_{012}^{(3)} \coloneqq \frac{1}{N_c C_F} \tr \left(t^a U_F(\xt_0) t^{a'} U_F^\dagger(\xt_1) \right) U_A(\xt_2)_{aa'}.
			\end{equation}
			Thus the squared amplitude of the $q \bar q g$ singlet production is
			\begin{multline}
				\sum_{q_0 \bar q_1 g_2 ~ \textrm{F. states} } \left| \mathcal{M}_{\gamma^* \rightarrow q \bar q g ~ \textrm{singlet}}^{\tree} \right|^2
				\\
				=
				\nc \cf
				\int \ud^2 \xt_0 \int \ud^2 \xt_1 \int \ud^2 \xt_2
				\int \ud^2 \cxt_{0} \int \ud^2 \cxt_{1} \int \ud^2 \cxt_{2}
				\\
				\times
				e^{i \xt_{\conz0}(\pt_0 - \frac{\pplus_0}{\qplus} \qt)}
				e^{i \xt_{\cono1}(\pt_1 - \frac{\pplus_1}{\qplus} \qt)}
				e^{i \xt_{\cont2}(\pt_2 - \frac{\pplus_2}{\qplus} \qt)}
				\\
				\times
				\sum_{h_0, h_1, \lambda_2}
				\left(\tilde{\psi}_{\gamma^{*}_\lambda \rightarrow q_{\conz} \Bar{q}_{\cono} g_{\cont}}\right)^\dagger
				\left(\tilde{\psi}_{\gamma^{*}_\lambda \rightarrow q_0 \Bar{q}_1 g_2}\right)
				\left[ 1 - S_{\contrip}^{(3)\dagger}\right] \left[ 1 - S_{012}^{(3)}\right],
			\end{multline}
			where we used $\big(\!\! \left(t^{b} \right)_{\beta_0 \beta_1} \! \big)^{\! \dagger} \left(t^{b} \right)_{\beta_0 \beta_1} = \tr (t^b t^b) = \nc \cf$, and the imaginary part of $\mathcal{M}$ was chosen to have a positive sign, i.e. $-(S-1) = (1-S)$.
		
			In order to specify the invariant mass $\mx$ of the diffractive system and the invariant momentum transfer $t$ in the scattering, it will be convenient to define a change of variables to the final state transverse momenta:
			\begin{equation}
				\pp_i \coloneqq \pt_i - z_i \qt.
			\end{equation}
			Defining $\mx^2 \coloneqq (p_0 + p_1 + p_2)^2$ and $\Deltat \coloneqq \pt_0 + \pt_1 + \pt_2 - \qt$, the shifted momenta satisfy the relations
			\begin{align}
				\Deltat & = \pp_0 + \pp_1 + \pp_2,
				\\
				\mx^2 & = \frac{\pp_0^2}{z_0} + \frac{\pp_1^2}{z_1} + \frac{\pp_2^2}{z_2} - \Deltat^2.
			\end{align}
			Now we are able to write the cross sections for the $q \bar q g$ contribution to the diffractive $\gamma^*p$ scattering, where the gluon is emitted before the interaction:
			\begin{align}
				\notag
				\frac{\ud \sigma^{\text{diff} ~ q \bar q g}_{\lambda}}{\ud M_{X}^{2} \ud \abs{t}}
				=
				&
				\frac{\nc \cf}{(4\pi)^2}
				\int \frac{\ud^2 \pp_0}{(2\pi)^2}
				\int \frac{\ud^2 \pp_1}{(2\pi)^2}
				\int \frac{\ud^2 \pp_2}{(2\pi)^2}
				\int_0^1 \frac{\ud z_0}{z_0}
				\int_0^1 \frac{\ud z_1}{z_1}
				\int_0^1 \frac{\ud z_2}{z_2}
				\nonumber
				\\
				& \times
				\delta(z_0 \! + \! z_1 \! + \! z_2 \! - \! 1)
				\delta(\Deltat^2 \! - \! \abs{t})
				\delta \left(\frac{\pp_0^2}{z_0} \! + \! \frac{\pp_1^2}{z_1} \! + \! \frac{\pp_2^2}{z_2} \! - \! \Deltat^2 \! - \! M_X^2 \right)
				\nonumber
				\\
				&
				\times
				\int_{\xt_0} \int_{\xt_1} \int_{\xt_2} \int_{\cxt_0} \int_{\cxt_1} \int_{\cxt_2} (2\pi)^6
				e^{i \xt_{\conz0} \pp_0}
				e^{i \xt_{\cono1} \pp_1}
				e^{i \xt_{\cont2} \pp_2}
				\nonumber
				\\
				&
				\times
				\sum_{h_0, h_1, \lambda_2}
				\left(\tilde{\psi}_{\gamma^{*}_\lambda \rightarrow q_{\conz} \Bar{q}_{\cono} g_{\cont}}\right)^\dagger
				\left(\tilde{\psi}_{\gamma^{*}_\lambda \rightarrow q_0 \Bar{q}_1 g_2}\right)
				\left[ 1 - S_{\contrip}^{(3)\dagger} \right] \left[ 1 - S_{012}^{(3)} \right],
				\label{eq:ddis-qqbarg-cs-nlo}
			\end{align}
			where the shorthand $\int_{\xt} \coloneqq \int \frac{\ud^2 \xt}{2\pi}$ was introduced. The last big missing pieces are the squared wavefunctions.

		\subsection{Squaring the wavefunctions}
	
		\begin{figure*}
			\centering
			\begin{subfigure}[b]{0.475\textwidth}
				\centering
				\includegraphics[width=\textwidth]{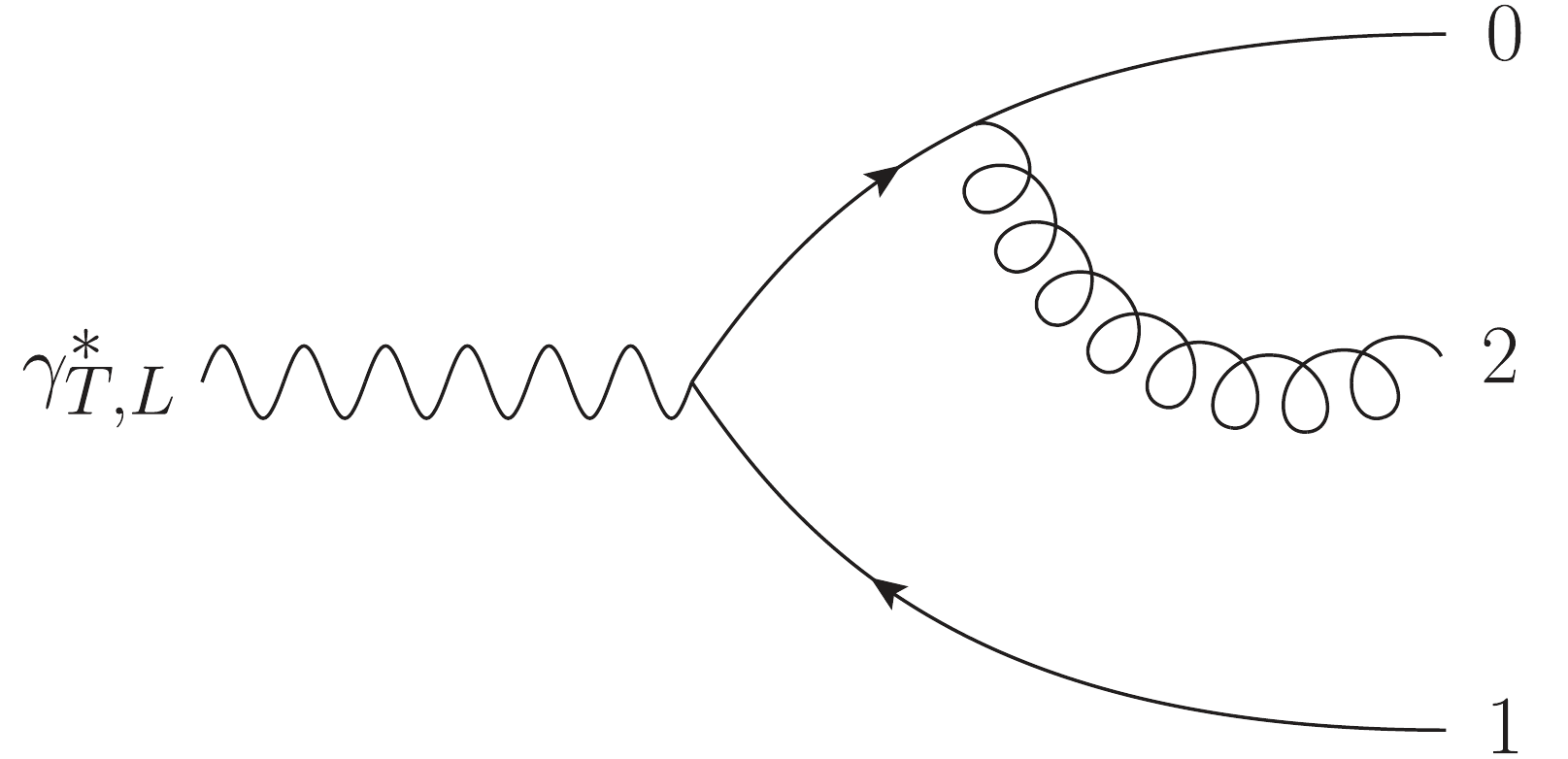}
				\caption*{Diagram $(a)$}
			\end{subfigure}
			\hfill
			\begin{subfigure}[b]{0.475\textwidth}  
				\centering 
				\includegraphics[width=\textwidth]{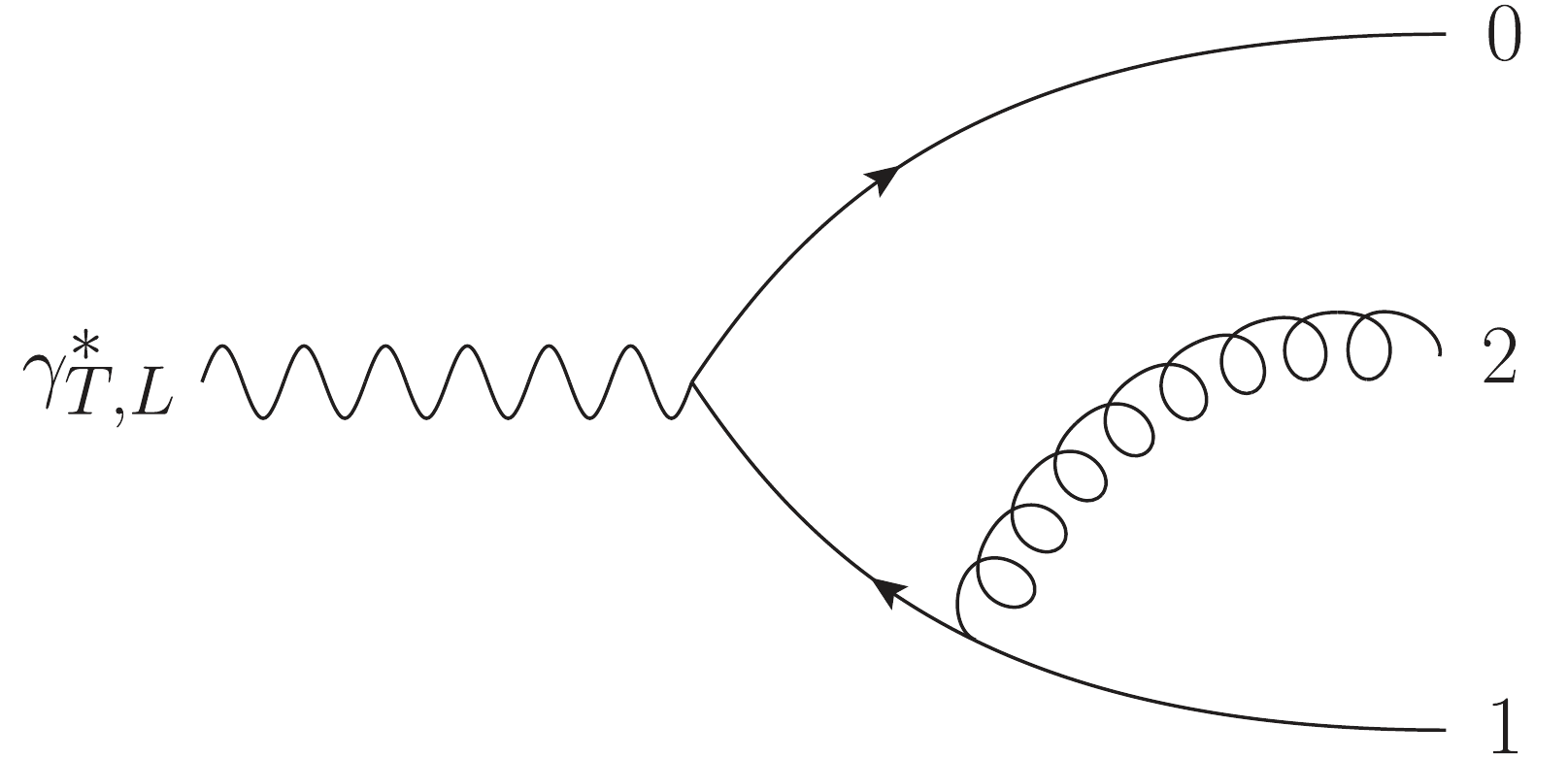}
				\caption*{Diagram $(b)$}   
			\end{subfigure}
			\vskip\baselineskip
			\begin{subfigure}[b]{0.475\textwidth}   
				\centering 
				\includegraphics[width=\textwidth]{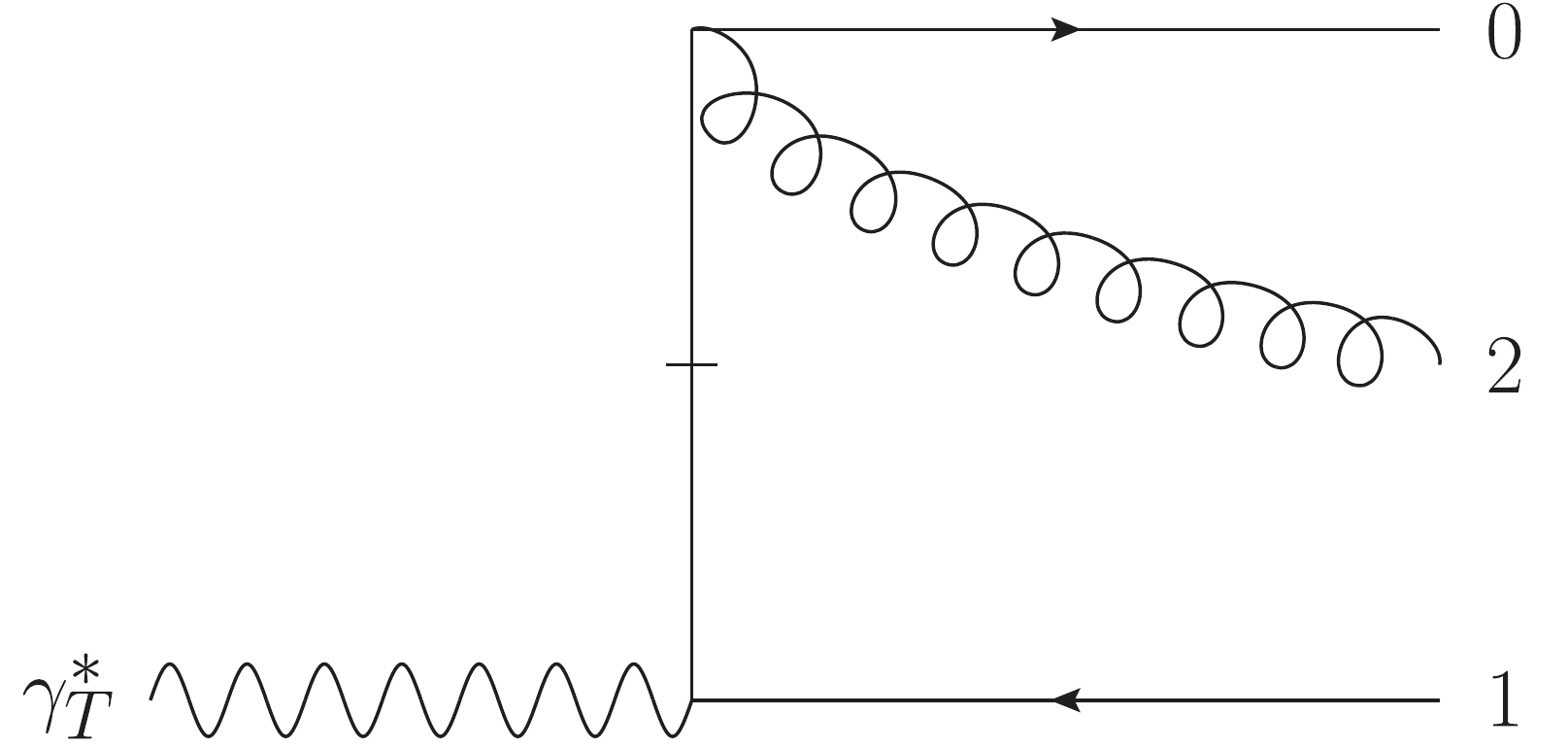}
				\caption*{Diagram $(a')$} 
			\end{subfigure}
			\hfill
			\begin{subfigure}[b]{0.475\textwidth}   
				\centering 
				\includegraphics[width=\textwidth]{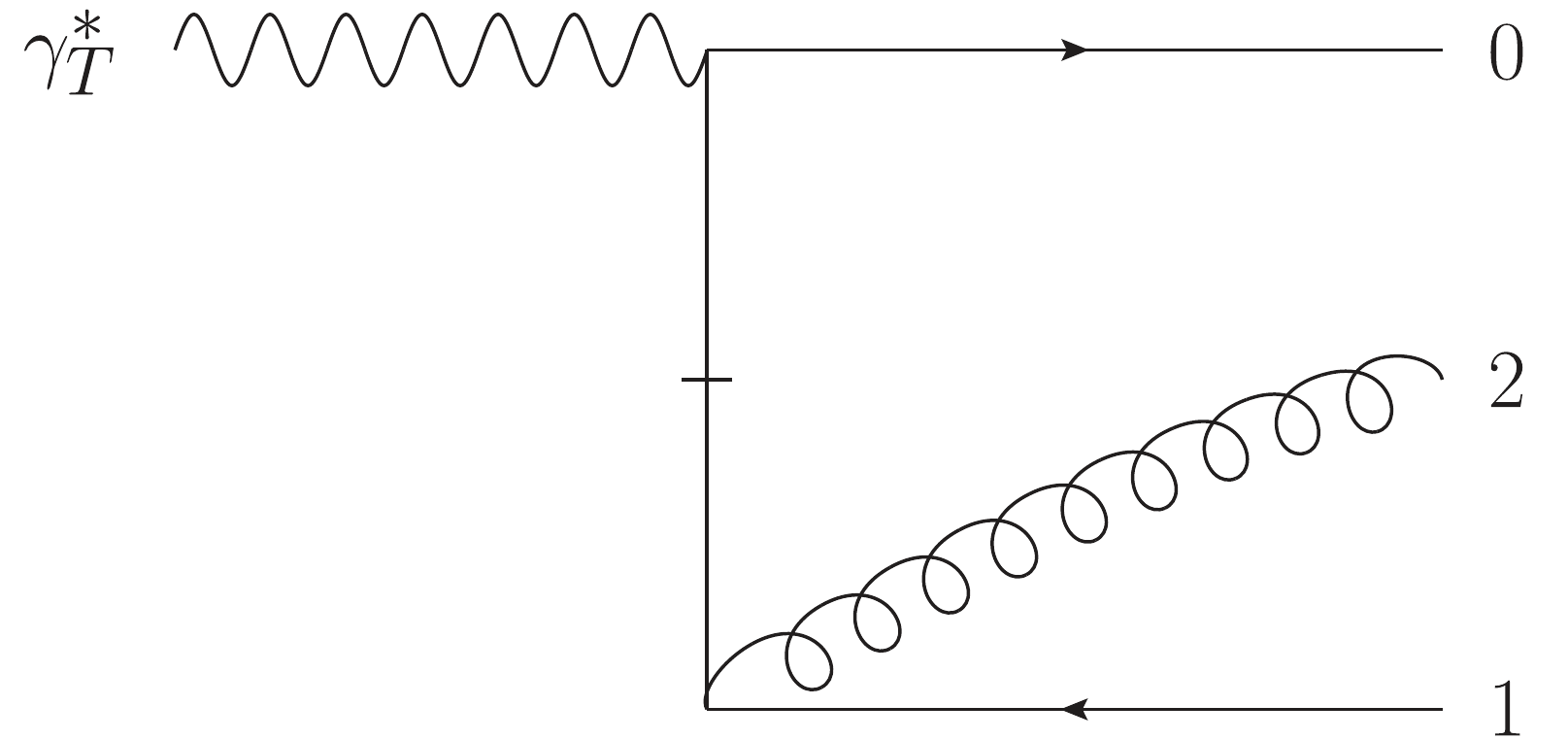}
				\caption*{Diagram $(b')$}  
			\end{subfigure}
			\caption{\small The tree-level gluon emission diagrams that contribute at NLO to the $\gamma^* \to q \bar q g$ wavefunction~\eqref{eq:qqbarg_WF_T_mixed_4D}. Diagram labels follow the convention of Ref.~\cite{Beuf:2017bpd}.} 
			\label{fig:gammaT_qqbarg}
		\end{figure*}

		The $q \bar q g$-contribution to the NLO DDIS cross sections we are after only gets a contribution from the $q \bar q g$ Fock state of the virtual photon. This means that it is relatively straightforward to derive this contribution from the virtual photon light-front wavefunctions for the $q \bar q g$ splitting derived in Ref.~\cite{Beuf:2017bpd}. In this section the calculation of the DDIS impact factor that goes into the cross sections is shown.
		
		We begin by writing the $D=4$ wavefunctions for the $\gamma^* \to q \bar q g$ splitting~\cite{Beuf:2017bpd} in the normalization scheme for the reduced wavefunctions~\eqref{eq:gamma-lfwf-qqbarg} introduced in Sec.~\ref{sec:lcpt}. For the longitudinal photon the LFWF becomes
		\begin{multline}
			\widetilde{\psi}_{\gamma_L^{*}\rightarrow q_0\bar{q}_1g_2}^{\textrm{Tree}}
			=      e\, e_f\, g\; \frac{i}{(2\pi)^2} \varepsilon_{\lambda_2}^{j *}\,
			2Q\; \textrm{K}_0\!\left(Q X_{012}\right)
			\sqrt{z_0}\, \sqrt{z_1}\; \delta_{h_1,-h_0}
			\\
			\times
			\Bigg\{ z_{1} 
			\Big[(2 z_{0} \!+\! z_{2}) \delta^{jm}
			-i(2h_0)\, z_{2}\,  \epsilon^{jm}\Big]\:
			\left(\frac{\xt_{20}^m}{\xt_{20}^2}\right)
			\\
			-z_{0}
			\Big[(2 z_{1} \!+\! z_{2} ) \delta^{jm}
			+i(2h_0)\, z_{2} \,  \epsilon^{jm}\Big]\:
			\left(\frac{\xt_{21}^m}{\xt_{21}^2}\right)
			\Bigg\}
			\, ,
			\label{eq:qqbarg_WF_L_mixed_4D}
		\end{multline}
		and for the transverse photon:
		\begin{align}
			&\widetilde{\psi}_{\gamma_{\lambda}^{*}\rightarrow q_0\bar{q}_1g_2}^{\textrm{Tree}}
			=    \frac{ e\, e_f\, g}{(2\pi)^2}\;
			\varepsilon_{\lambda}^{i}\, \varepsilon_{\lambda_2}^{j *}\;
			\sqrt{z_0}\, \sqrt{z_1}\; \delta_{h_1,-h_0}\;
			\frac{Q}{X_{012}}\; \textrm{K}_1\!\left(Q X_{012}\right)
			\nonumber\\
			&
			\times
			\Bigg\{ z_1
			\Big[(2 z_0 \! + \! z_{2}) \delta^{jm}
			-i\, (2h_0)\, z_{2} \,  \epsilon^{jm} \Big]\!
			\Big[(2 z_{1} \! - \! 1) \delta^{il}
			-i\, (2h_0) \, \epsilon^{il} \Big]\;
			\xt_{0+2;1}^l\, \left(\frac{\xt_{20}^m}{\xt_{20}^2}\right)
			\nonumber\\
			&
			\hphantom{\Bigg\lbrace}
			+z_0
			\Big[(2z_{1} \! + \! z_{2}) \delta^{jm}
			+i(2h_0)\, z_{2} \,  \epsilon^{jm}\Big]\! \Big[(2 z_{0} \!-\! 1) \delta^{il}
			+ i\, (2h_0) \, \epsilon^{il}\Big] \;
			\xt_{0;1+2}^l\, \left(\frac{\xt_{21}^m}{\xt_{21}^2}\right)
			\nonumber\\
			&
			\hphantom{\Bigg\lbrace}
			- \frac{z_{0} z_{1} z_{2}}{z_{0} \! + \! z_{2}}\;
			\Big[\delta^{ij}  - i\, (2h_0)\, \epsilon^{ij}\Big]
			+ \frac{z_{0} z_{1} z_{2}}{z_{1} \!+\! z_{2}}\;
			\Big[\delta^{ij} + i\, (2h_0)\, \epsilon^{ij}\Big]
			\Bigg\}
			\, ,
			\label{eq:qqbarg_WF_T_mixed_4D}
		\end{align}
		where $X_{012}$, $\xt_{0+2;1}$ and $\xt_{0;1+2}$ are defined
		\begin{align}
			X_{012}^2 
			& \coloneqq
			z_{0} z_{1} \xt_{01}^2 + z_{0} z_{2} \xt_{02}^2 + z_{1} z_{2} \xt_{12}^2
			\\
			\xt_{0+2;1}
			& \coloneqq
			- \frac{z_0}{z_0+z_2} \xt_{20} + \xt_{21}
			=
			\xt_{01} + \frac{z_2}{z_0 + z_2} \xt_{{2} {0}}
			\\
			\xt_{0;1+2}
			& \coloneqq
			-\xt_{{2} {0}} +\frac{z_1}{z_1+z_2} \xt_{{2} {1}}
			=
			\xt_{01} - \frac{z_2}{z_1+z_2} \xt_{{2} {1}}
			\, .
		\end{align}
		Following the convention of Ref.~\cite{Beuf:2017bpd}, in the discussion that follows the contributions of the four terms in the curly braces of Eq.~\eqref{eq:qqbarg_WF_T_mixed_4D} are denoted by $(a)$, $(b)$, $(a')$ and $(b')$ --- visualized in Fig.~\ref{fig:gammaT_qqbarg}.

		The DDIS cross sections~\eqref{eq:ddis-qqbarg-cs-nlo} depend on the squares of the above wavefunctions summed over the quantum numbers of the quark, antiquark and gluon. Specifically, the momentum fractions of the partons are conserved and therefore the same in the direct and complex conjugate amplitude, but the transverse coordinates of the particles are different: $\xt_0, \xt_1, \xt_2$ in the direct amplitude, and $\cxt_{{0}}, \cxt_{{1}}, \cxt_{{2}}$ in the c.c. amplitude, respectively for the quark, antiquark and gluon. Squaring the longitudinal splitting wavefunction is a straightforward affair:
		\begin{multline}
			\label{eq:ddis-qqbarg-psiL^2}
			\sum_{h_0, h_1, \lambda_2}
			\left( \widetilde{\psi}_{\gamma_L^{*}\rightarrow q_{\conj{0}} \bar{q}_{\conj{1}} g_{\conj{2}}}^{\textrm{Tree}} \right)^*
			\widetilde{\psi}_{\gamma_L^{*}\rightarrow q_0\bar{q}_1g_2}^{\textrm{Tree}}
			=
			\\
			\frac{e^2 e_f^2 g^2}{(2\pi)^4}  \left(\sum_{\lambda_2} \varepsilon_{\lambda_2}^{j '} \varepsilon_{\lambda_2}^{j *} \right)
			4 z_0 z_1 Q^2 \besk_0 \left( Q X_{012}\right) \besk_0 \left( Q X_{\conj{0} \conj{1} \conj{2}}\right)
			\sum_{h_0, h_1} \delta_{h_1,-h_0}
			\\
			\times \! 
				\Bigg\{ \! z_{1} \! 
				\Big[\! (2 z_{0} \!+\! z_{2}) \delta^{j'm'}
				\! + i(2h_0) z_{2} \epsilon^{j'm'} \! \Big] \! \! 
				\left(\frac{\xt_{\conj{2} \conj{0}}^{m'}}{\xt_{\conj{2} \conj{0}}^2}\right)
				\! - z_{0} \! 
				\Big[\! (2 z_{1} \!+\! z_{2} ) \delta^{j'm'}
				\! -i(2h_0) z_{2} \epsilon^{j'm'} \! \Big] \! \! 
				\left(\frac{\xt_{ \conj{2} \conj{1}}^{m'}}{\xt_{\conj{2} \conj{1}}^2}\right)
				\! \! \Bigg\}
			\\
			\times \! \! 
				\Bigg\{ \! z_{1} \! 
				\Big[(2 z_{0} \!+\! z_{2}) \delta^{jm}
				\! -i(2h_0) z_{2} \epsilon^{jm} \! \Big] \! \! 
				\left(\frac{\xt_{20}^m}{\xt_{20}^2}\right)
				\! -z_{0} \! 
				\Big[(2 z_{1} \!+\! z_{2} ) \delta^{jm}
				+i(2h_0) z_{2} \epsilon^{jm} \! \Big] \! \! 
				\left(\frac{\xt_{21}^m}{\xt_{21}^2}\right)
				\! \! \Bigg\}
			\\ 
			=
			2
			\frac{\aem \as e_f^2 }{\pi^2}
			4 z_0 z_1 Q^2 \besk_0 \left( Q X_{012}\right) \besk_0 \left( Q X_{\conj{0} \conj{1} \conj{2}}\right)
			\Bigg\{
				2 z_1^2 \left( 2z_0 (z_0+z_2) + z_2^2 \right)
				\frac{\xt_{\conj{2} \conj{0}} \cdot \xt_{20}}{\xt_{\conj{2} \conj{0}}^2 \xt_{20}^2}
				\\
				-
				z_0 z_1 \left( 2z_0 (z_1 + z_2) + 2z_1(z_0+z_2) \right)
				\left( 
					\frac{\xt_{\conj{2} \conj{0}} \cdot \xt_{21}}{\xt_{\conj{2} \conj{0}}^2 \xt_{21}^2}
					+
					\frac{\xt_{{2} {0}} \cdot \xt_{\conj{2} \conj{1}}}{\xt_{{2} {0}}^2 \xt_{\conj{2} \conj{1}}^2}
				\right)
				\\
				+
				2 z_0^2 \left( 2 z_1 (z_1 + z_2) + z_2^2 \right) \frac{\xt_{\conj{2} \conj{1}} \cdot \xt_{{2} {1}}}{\xt_{\conj{2} \conj{1}}^2 \xt_{{2} {1}}^2}
			\Bigg\}
			\\ 
			\hspace{-4cm}
			=
			4
			\frac{\aem \as e_f^2 }{\pi^2}
			4 z_0 z_1 Q^2 \besk_0 \left( Q X_{012}\right) \besk_0 \left( Q X_{\conj{0} \conj{1} \conj{2}}\right)
			\Bigg\{
			\\
				z_1^2
				\Bigg[
					\left(2 z_0 (z_0 + z_2) + z_2^2\right)
					\left(
						\frac{\xt_{20}}{\xt_{20}^2} \cdot
							\left( \frac{\xt_{\conj{2} \conj{0}}}{\xt_{\conj{2} \conj{0}}^2}
							- \half \frac{\xt_{\conj{2} \conj{1}}}{\xt_{\conj{2} \conj{1}}^2} \right)
						- \half \frac{\xt_{\conj{2} \conj{0}} \cdot \xt_{21}}{\xt_{\conj{2} \conj{0}}^2 \xt_{21}^2} \right)
					\\
					+ \frac{z_2^2}{2}
					\left(
						\frac{\xt_{\conj{2} \conj{0}} \cdot \xt_{21}}{\xt_{\conj{2} \conj{0}}^2 \xt_{21}^2}
						+
						\frac{\xt_{{2} {0}} \cdot \xt_{\conj{2} \conj{1}}}{\xt_{{2} {0}}^2 \xt_{\conj{2} \conj{1}}^2}
					\right)
				\Bigg]
			\\
			+	
			z_0^2
			\Bigg[
				\left(2 z_1 (z_1 + z_2) + z_2^2\right)
				\left(
					\frac{\xt_{21}}{\xt_{21}^2} \cdot
						\left( \frac{\xt_{\conj{2} \conj{1}}}{\xt_{\conj{2} \conj{1}}^2}
						- \half \frac{\xt_{\conj{2} \conj{0}}}{\xt_{\conj{2} \conj{0}}^2} \right)
					- \half \frac{\xt_{{2} {0}} \cdot \xt_{\conj{2} \conj{1}}}{\xt_{{2} {0}}^2 \xt_{\conj{2} \conj{1}}^2} \right)
			\\
			+ \frac{z_2^2}{2}
			\left(
				\frac{\xt_{\conj{2} \conj{0}} \cdot \xt_{21}}{\xt_{\conj{2} \conj{0}}^2 \xt_{21}^2}
				+
				\frac{\xt_{{2} {0}} \cdot \xt_{\conj{2} \conj{1}}}{\xt_{{2} {0}}^2 \xt_{\conj{2} \conj{1}}^2}
			\right)
			\Bigg]
			\Bigg\}.
		\end{multline}
		In the second equality --- analogously to Ref.~\cite{Beuf:2017bpd} ---  the result is rearranged to be symmetric in the exchanges of the quark and antiquark: $(z_0, \xt_0) \leftrightarrow (z_1, \xt_1)$ and $(z_0, \cxt_{{0}}) \leftrightarrow (z_1, \cxt_{{1}})$. In the limit $\cxt_{i} \to \xt_i$, the first and second term match the corresponding results in Eqs.~(84) and (85) of the NLO DIS calculation~\cite{Beuf:2017bpd}. However, the same simplifications as in Ref.~\cite{Beuf:2017bpd} are not seen since the particle coordinates are not the same in the direct and c.c. amplitude.

		The calculation is more involved for the transversely polarized virtual photon. To begin the discussion, let us write down the different terms to be considered by referring to them by their respective diagrams $(a)$, $(b)$, $(a')$ and $(b')$ and coordinates. Specifically, let $\conj{(a)}$ denote the contribution of the first term in the complex conjugate of the wavefunction~\eqref{eq:qqbarg_WF_T_mixed_4D}, which implies the usage of the transverse coordinates $\cxt_i$ of the complex conjugate. Thus the square consists of the terms:
		\begin{align*}
			\left( \widetilde{\psi}_{\gamma_\lambda^{*}\rightarrow q_{\conj{0}} \bar{q}_{\conj{1}} g_{\conj{2}}}^{\textrm{Tree}} \right)^*
			\widetilde{\psi}_{\gamma_\lambda^{*}\rightarrow q_0\bar{q}_1g_2}^{\textrm{Tree}}
			= &
			\conj{(a)} (a) + \conj{(b)} (b)
			+ \conj{(a)} (b) + \conj{(b)} (a)
			\\
			& + \conj{(a')} (a') + 	\conj{(a')}(a) + \conj{(a)} (a') + \conj{(a')}(b) + \conj{(b)} (a')
			\\
			& + \conj{(b')} (b') + 	\conj{(b')}(a) + \conj{(a)} (b') + \conj{(b')}(b) + \conj{(b)} (b')
			\\
			& + \conj{(a')}(b') + \conj{(b')}(a')
			.
		\end{align*}
		The interference terms of the instantaneous contributions $\conj{(a')}(b')$ and $\conj{(b')}(a')$ vanish exactly since both are proportional to $\delta^{ij} \delta^{ij} - \epsilon^{ij} \epsilon^{ij} \equiv 0$, similarly as happens in the case of NLO DIS~\cite{Beuf:2017bpd}. Terms containing either instantaneous contribution could have feasibly been read from the intermediate results of Ref.~\cite{Beuf:2017bpd} and then replacing $\xt_i \to \cxt_i$ as would be appropriate. However this could not have been done for the contributions of the regular emissions $(a)$ and $(b)$, where the transverse structures interact non-trivially. Furthermore the squared contributions $\conj{(a)} (a)$, $\conj{(b)} (b)$ are not computed in $D=4$ at all due to the UV regularization that is required~\cite{Beuf:2017bpd}. As stated previously, UV regularization is not necessary for the real 3-parton contribution to DDIS in question here, since the invariant mass constraint does it for us.
		
		Beginning with the contributions of the regular emissions, we have for $(a)^2$:
		\begin{multline}
			\sum_{T \, \textrm{pol.}\, \lambda, \lambda_2}
			\sum_{h_0, h_1}
			\left( \widetilde{\psi}_{\gamma_\lambda^{*}\rightarrow q_{\conj{0}} \bar{q}_{\conj{1}} g_{\conj{2}}}^{(a)} \right)^*
			\widetilde{\psi}_{\gamma_\lambda^{*}\rightarrow q_0\bar{q}_1g_2}^{(a)}
			=
			\frac{e^2 g^2 e_f^2}{(2\pi)^4} \frac{z_0 z_1 Q^2}{X_{012} X_{\contrip}} \besk_1 (QX_{012}) \besk_1 (QX_{\contrip})
			\\
			\times
			z_1^2 \Bigg[
			(4z_0(z_0 + z_2) + 2z_2^2) (2 - 4z_1 (1-z_1))
			\left(\xt_{\conj{0} + \conj{2}; \conj{1}} \cdot \xt_{0+2;1} \right)
			\frac{(\xt_{\conj{2} \conj{0}} \cdot \xt_{{2} {0}})}{\xt_{\conj{2} \conj{0}}^2 \xt_{{2} {0}}^2}
			\\
			\hphantom{z_1^2 \Bigg[}
			- 4 z_2 (2z_0 + z_2)(2z_1 - 1)
			\left( \xt_{\conj{0} + \conj{2}; \conj{1}} \wedge \xt_{0+2;1} \right)
			\frac{(\xt_{\conj{2} \conj{0}} \wedge \xt_{{2} {0}})}{\xt_{\conj{2} \conj{0}}^2 \xt_{{2} {0}}^2}
			\Bigg]
			,
		\end{multline}
		and for $(b)^2$:
		\begin{multline}
			\sum_{T \, \textrm{pol.}\, \lambda, \lambda_2}
			\sum_{h_0, h_1}
			\left( \widetilde{\psi}_{\gamma_\lambda^{*}\rightarrow q_{\conj{0}} \bar{q}_{\conj{1}} g_{\conj{2}}}^{(b)} \right)^*
			\widetilde{\psi}_{\gamma_\lambda^{*}\rightarrow q_0\bar{q}_1g_2}^{(b)}
			=
			\frac{e^2 g^2 e_f^2}{(2\pi)^4} \frac{z_0 z_1 Q^2}{X_{012} X_{\contrip}} \besk_1 (QX_{012}) \besk_1 (QX_{\contrip})
			\\
			\times
			z_0^2 \Bigg[
			(4z_1(z_1 + z_2) + 2z_2^2) (2 - 4z_0 (1-z_0))
			\left(\xt_{\conj{0}; \conj{1} + \conj{2}} \cdot \xt_{0;1+2} \right)
			\frac{(\xt_{\conj{2} \conj{1}} \cdot \xt_{{2} {1}})}{\xt_{\conj{2} \conj{1}}^2 \xt_{{2} {1}}^2}
			\\
			\hphantom{z_0^2 \Bigg[}
			- 4 z_2 (2z_1 + z_2)(2z_0 - 1)
			\left( \xt_{\conj{0}; \conj{1} + \conj{2}} \wedge \xt_{0;1+2} \right)
			\frac{(\xt_{\conj{2} \conj{1}} \wedge \xt_{{2} {1}})}{\xt_{\conj{2} \conj{1}}^2 \xt_{{2} {1}}^2}
			\Bigg]
			,
		\end{multline}
		and for the interference of $(a)$ and $(b)$:
		\begin{multline}
			\sum_{T \, \textrm{pol.}\, \lambda, \lambda_2}
			\sum_{h_0, h_1}
			\bigg\lbrace
			\left( \widetilde{\psi}_{\gamma_\lambda^{*}\rightarrow q_{\conj{0}} \bar{q}_{\conj{1}} g_{\conj{2}}}^{(a)} \right)^*
			\widetilde{\psi}_{\gamma_\lambda^{*}\rightarrow q_0\bar{q}_1g_2}^{(b)}
			+
			\left( \widetilde{\psi}_{\gamma_\lambda^{*}\rightarrow q_{\conj{0}} \bar{q}_{\conj{1}} g_{\conj{2}}}^{(b)} \right)^*
			\widetilde{\psi}_{\gamma_\lambda^{*}\rightarrow q_0\bar{q}_1g_2}^{(a)}
			\bigg\rbrace
			\\
			=
			\frac{e^2 g^2 e_f^2}{(2\pi)^4} \frac{z_0 z_1 Q^2}{X_{012} X_{\contrip}} \besk_1 (QX_{012}) \besk_1 (QX_{\contrip})
			\\
			\times
			\Bigg\lbrace
			- z_0 z_1
			\left[ 2z_1(z_0 + z_2) + 2z_0(z_1 + z_2) \right]
			\left[ 2z_0(z_0 + z_2) + 2z_1(z_1 + z_2) \right]
			\\
			\times \!
			\left[
			\left( \xt_{\conj{0} + \conj{2}; \conj{1}} \cdot \xt_{0;1+2} \right)
			\frac{(\xt_{\conj{2} \conj{0}} \cdot \xt_{{2} {1}})}{\xt_{\conj{2} \conj{0}}^2 \xt_{{2} {1}}^2}
			+
			\left( \xt_{\conj{0}; \conj{1} + \conj{2}} \cdot \xt_{0+2;1} \right)
			\frac{(\xt_{\conj{2} \conj{1}} \cdot \xt_{{2} {0}})}{\xt_{\conj{2} \conj{1}}^2 \xt_{{2} {0}}^2}
			\right]
			+ 4 z_0 z_1 z_2 (z_0-z_1)^2
			\\
			\times
			\left[
			\left( \xt_{\conj{0} + \conj{2}; \conj{1}} \! \wedge \! \xt_{0;1+2} \right) \! 
			\frac{(\xt_{\conj{2} \conj{0}} \! \wedge \! \xt_{{2} {1}})}{\xt_{\conj{2} \conj{0}}^2 \xt_{{2} {1}}^2}
			\! + \! 
			\left( \xt_{\conj{0}; \conj{1} + \conj{2}} \! \wedge \! \xt_{0+2;1} \right) \! 
			\frac{(\xt_{\conj{2} \conj{1}} \! \wedge \! \xt_{{2} {0}})}{\xt_{\conj{2} \conj{1}}^2 \xt_{{2} {0}}^2}
			\right] \! \!
			\Bigg\rbrace
			.
		\end{multline}
		Next, the contributions involving instantaneous diagram $(a')$ are
		\begin{multline}
			\sum_{T \, \textrm{pol.}\, \lambda, \lambda_2}
			\sum_{h_0, h_1}
			\bigg\lbrace
			\left( \widetilde{\psi}_{\gamma_\lambda^{*}\rightarrow q_{\conj{0}} \bar{q}_{\conj{1}} g_{\conj{2}}}^{(a')} \right)^*
			\widetilde{\psi}_{\gamma_\lambda^{*}\rightarrow q_0\bar{q}_1g_2}^{(a')}
			+
			\left( \widetilde{\psi}_{\gamma_\lambda^{*}\rightarrow q_{\conj{0}} \bar{q}_{\conj{1}} g_{\conj{2}}}^{(a')} \right)^*
			\widetilde{\psi}_{\gamma_\lambda^{*}\rightarrow q_0\bar{q}_1g_2}^{(a)}
			\\
			+
			\left( \widetilde{\psi}_{\gamma_\lambda^{*}\rightarrow q_{\conj{0}} \bar{q}_{\conj{1}} g_{\conj{2}}}^{(a)} \right)^*
			\widetilde{\psi}_{\gamma_\lambda^{*}\rightarrow q_0\bar{q}_1g_2}^{(a')}
			+
			\left( \widetilde{\psi}_{\gamma_\lambda^{*}\rightarrow q_{\conj{0}} \bar{q}_{\conj{1}} g_{\conj{2}}}^{(a')} \right)^*
			\widetilde{\psi}_{\gamma_\lambda^{*}\rightarrow q_0\bar{q}_1g_2}^{(b)}
			+
			\left( \widetilde{\psi}_{\gamma_\lambda^{*}\rightarrow q_{\conj{0}} \bar{q}_{\conj{1}} g_{\conj{2}}}^{(b)} \right)^*
			\widetilde{\psi}_{\gamma_\lambda^{*}\rightarrow q_0\bar{q}_1g_2}^{(a')}
			\bigg\rbrace
			\\
			=
			\frac{e^2 g^2 e_f^2}{(2\pi)^4} \frac{z_0 z_1 Q^2}{X_{012} X_{\contrip}} \besk_1 (QX_{012}) \besk_1 (QX_{\contrip})
			\Bigg\lbrace
			4\frac{z_0^2 z_1^2 z_2^2}{(z_0 + z_2)^2}
			\\
			- 4\frac{z_0^2 z_1^3 z_2}{z_0 + z_2}
			\left( 
			\frac{\xt_{{0} + {2}; {1}} \cdot \xt_{{2} {0}}}{\xt_{{2} {0}}^2}
			+
			\frac{\xt_{\conj{0} + \conj{2}; \conj{1}} \cdot \xt_{\conj{2} \conj{0}}}{\xt_{\conj{2} \conj{0}}^2}
			\right)
			\\
			+ 4\frac{z_0^2 z_1 (z_1 + z_2)^2 z_2}{z_0 + z_2}
			\left( 
			\frac{\xt_{{0}; {1}+{2}} \cdot \xt_{{2} {1}}}{\xt_{{2} {1}}^2}
			+
			\frac{\xt_{\conj{0}; \conj{1}+ \conj{2}} \cdot \xt_{\conj{2} \conj{1}}}{\xt_{\conj{2} \conj{1}}^2}
			\right)
			\Bigg\rbrace
			.
		\end{multline}
		The contributions involving $(b')$ are
		\begin{multline}
			\sum_{T \, \textrm{pol.}\, \lambda, \lambda_2}
			\sum_{h_0, h_1}
			\bigg\lbrace
			\left( \widetilde{\psi}_{\gamma_\lambda^{*}\rightarrow q_{\conj{0}} \bar{q}_{\conj{1}} g_{\conj{2}}}^{(b')} \right)^*
			\widetilde{\psi}_{\gamma_\lambda^{*}\rightarrow q_0\bar{q}_1g_2}^{(b')}
			+
			\left( \widetilde{\psi}_{\gamma_\lambda^{*}\rightarrow q_{\conj{0}} \bar{q}_{\conj{1}} g_{\conj{2}}}^{(b')} \right)^*
			\widetilde{\psi}_{\gamma_\lambda^{*}\rightarrow q_0\bar{q}_1g_2}^{(a)}
			\\
			+
			\left( \widetilde{\psi}_{\gamma_\lambda^{*}\rightarrow q_{\conj{0}} \bar{q}_{\conj{1}} g_{\conj{2}}}^{(a)} \right)^*
			\widetilde{\psi}_{\gamma_\lambda^{*}\rightarrow q_0\bar{q}_1g_2}^{(b')}
			+
			\left( \widetilde{\psi}_{\gamma_\lambda^{*}\rightarrow q_{\conj{0}} \bar{q}_{\conj{1}} g_{\conj{2}}}^{(b')} \right)^*
			\widetilde{\psi}_{\gamma_\lambda^{*}\rightarrow q_0\bar{q}_1g_2}^{(b)}
			+
			\left( \widetilde{\psi}_{\gamma_\lambda^{*}\rightarrow q_{\conj{0}} \bar{q}_{\conj{1}} g_{\conj{2}}}^{(b)} \right)^*
			\widetilde{\psi}_{\gamma_\lambda^{*}\rightarrow q_0\bar{q}_1g_2}^{(b')}
			\bigg\rbrace
			\\
			=
			\frac{e^2 g^2 e_f^2}{(2\pi)^4} \frac{z_0 z_1 Q^2}{X_{012} X_{\contrip}} \besk_1 (QX_{012}) \besk_1 (QX_{\contrip})
			\Bigg\lbrace
			4\frac{z_0^2 z_1^2 z_2^2}{(z_1 + z_2)^2}
			\\
			- 4\frac{z_0 z_1^2 (z_0 + z_2)^2 z_2}{z_1 + z_2}
			\left( 
			\frac{\xt_{{0} + {2}; {1}} \cdot \xt_{{2} {0}}}{\xt_{{2} {0}}^2}
			+
			\frac{\xt_{\conj{0} + \conj{2}; \conj{1}} \cdot \xt_{\conj{2} \conj{0}}}{\xt_{\conj{2} \conj{0}}^2}
			\right)
			\\
			+ 4\frac{z_0^3 z_1^2 z_2}{z_1 + z_2}
			\left( 
			\frac{\xt_{{0}; {1}+{2}} \cdot \xt_{{2} {1}}}{\xt_{{2} {1}}^2}
			+
			\frac{\xt_{\conj{0}; \conj{1}+ \conj{2}} \cdot \xt_{\conj{2} \conj{1}}}{\xt_{\conj{2} \conj{1}}^2}
			\right)
			\Bigg\rbrace
			.
		\end{multline}
		
		In the above expressions, the exterior products are defined as
		\begin{equation}
			\xt \wedge \B{y} \coloneqq \epsilon^{ij} \xt^i \B{y}^j.
		\end{equation}
		The products of two exterior products can be simplified with the identity\footnote{The determinant form can be used as a mnemonic since in that case the indices have a row-column pattern. The identity generalizes to higher dimensions.} for the product of two Levi-Civita symbols:
		\begin{equation}
			\epsilon^{ij}\epsilon^{mn} = 
				\left|
					\begin{matrix}
						\delta^{im} & \delta^{in} \\
						\delta^{jm} & \delta^{jn}
					\end{matrix}
				\right|
				= \delta^{im}\delta^{jn} - \delta^{in}\delta^{jm}.	
		\end{equation}

		Collecting the results, the squared splitting function for the transverse photon becomes
		\begin{multline}
			\label{eq:ddis-qqbarg-psiT^2}
			\half
			\sum_{T \, \textrm{pol.}\, \lambda, \lambda_2}
			\sum_{h_0, h_1}
			\left( \widetilde{\psi}_{\gamma_\lambda^{*}\rightarrow q_{\conj{0}} \bar{q}_{\conj{1}} g_{\conj{2}}}^{\textrm{Tree}} \right)^*
			\widetilde{\psi}_{\gamma_\lambda^{*}\rightarrow q_0\bar{q}_1g_2}^{\textrm{Tree}}
			\\
			=
			\half \frac{\aem \as e_f^2}{\pi^2} z_0 z_1 \frac{Q^2}{X_{012} X_{\conj{0}\conj{1}\conj{2} }}
			\besk_1\left(QX_{012}\right) \besk_1\left(Q X_{\conj{0}\conj{1}\conj{2} }\right)
			\\
			\times
			4 \left\lbrace
				\Upsilon^{(|a|^2)}_{\textrm{reg.}} + \Upsilon^{(|b|^2)}_{\textrm{reg.}} + \Upsilon^{(a')}_{\textrm{inst.}} + \Upsilon^{(b')}_{\textrm{inst.}} + \Upsilon^{(ab)}_{\textrm{interf.}}
			\right\rbrace ,
		\end{multline}
		where
		\begin{align}
			\label{eq:ddis-qqbarg-nlo-upsilon-a^2}
			\Upsilon^{(|a|^2)}_{\textrm{reg.}}
			=&
			z_1^2 \Bigg[
				(2z_0(z_0 + z_2) + z_2^2) (1 - 2z_1 (1-z_1))
				\left(\xt_{\conj{0} + \conj{2}; \conj{1}} \cdot \xt_{0+2;1} \right)
				\frac{(\xt_{\conj{2} \conj{0}} \cdot \xt_{{2} {0}})}{\xt_{\conj{2} \conj{0}}^2 \xt_{{2} {0}}^2}
				\nonumber\\
				&
				\hphantom{z_1^2 \Bigg[}
				- z_2 (2z_0 + z_2)(2z_1 - 1)
				\left( \xt_{\conj{0} + \conj{2}; \conj{1}} \wedge \xt_{0+2;1} \right)
				\frac{(\xt_{\conj{2} \conj{0}} \wedge \xt_{{2} {0}})}{\xt_{\conj{2} \conj{0}}^2 \xt_{{2} {0}}^2}
			\Bigg]
			\\
			\label{eq:ddis-qqbarg-nlo-upsilon-b^2}
			\Upsilon^{(|b|^2)}_{\textrm{reg.}}
			=&
			z_0^2 \Bigg[
				(2z_1(z_1 + z_2) + z_2^2) (1 - 2z_0 (1-z_0))
				\left(\xt_{\conj{0}; \conj{1} + \conj{2}} \cdot \xt_{0;1+2} \right)
				\frac{(\xt_{\conj{2} \conj{1}} \cdot \xt_{{2} {1}})}{\xt_{\conj{2} \conj{1}}^2 \xt_{{2} {1}}^2}
				\nonumber\\
				&
				\hphantom{z_0^2 \Bigg[}
				- z_2 (2z_1 + z_2)(2z_0 - 1)
				\left( \xt_{\conj{0}; \conj{1} + \conj{2}} \wedge \xt_{0;1+2} \right)
				\frac{(\xt_{\conj{2} \conj{1}} \wedge \xt_{{2} {1}})}{\xt_{\conj{2} \conj{1}}^2 \xt_{{2} {1}}^2}
			\Bigg]
			\\
			\label{eq:ddis-qqbarg-nlo-upsilon-a'}
			\Upsilon^{(a')}_{\textrm{inst.}}
			=&
			\frac{z_0^2 z_1^2 z_2^2}{(z_0 + z_2)^2}
			- \frac{z_0^2 z_1^3 z_2}{z_0 + z_2}
			\left( 
				\frac{\xt_{{0} + {2}; {1}} \cdot \xt_{{2} {0}}}{\xt_{{2} {0}}^2}
				+
				\frac{\xt_{\conj{0} + \conj{2}; \conj{1}} \cdot \xt_{\conj{2} \conj{0}}}{\xt_{\conj{2} \conj{0}}^2}
				\right)
			\nonumber\\
			&
			+ \frac{z_0^2 z_1 (z_1 + z_2)^2 z_2}{z_0 + z_2}
			\left( 
				\frac{\xt_{{0}; {1}+{2}} \cdot \xt_{{2} {1}}}{\xt_{{2} {1}}^2}
				+
				\frac{\xt_{\conj{0}; \conj{1}+ \conj{2}} \cdot \xt_{\conj{2} \conj{1}}}{\xt_{\conj{2} \conj{1}}^2}
				\right)
			\\
			\label{eq:ddis-qqbarg-nlo-upsilon-b'}
			\Upsilon^{(b')}_{\textrm{inst.}}
			=&
			\frac{z_0^2 z_1^2 z_2^2}{(z_1 + z_2)^2}
			- \frac{z_0 z_1^2 (z_0 + z_2)^2 z_2}{z_1 + z_2}
			\left( 
				\frac{\xt_{{0} + {2}; {1}} \cdot \xt_{{2} {0}}}{\xt_{{2} {0}}^2}
				+
				\frac{\xt_{\conj{0} + \conj{2}; \conj{1}} \cdot \xt_{\conj{2} \conj{0}}}{\xt_{\conj{2} \conj{0}}^2}
				\right)
			\nonumber\\
			&
			+ \frac{z_0^3 z_1^2 z_2}{z_1 + z_2}
			\left( 
				\frac{\xt_{{0}; {1}+{2}} \cdot \xt_{{2} {1}}}{\xt_{{2} {1}}^2}
				+
				\frac{\xt_{\conj{0}; \conj{1}+ \conj{2}} \cdot \xt_{\conj{2} \conj{1}}}{\xt_{\conj{2} \conj{1}}^2}
				\right)
			\\
			\label{eq:ddis-qqbarg-nlo-upsilon-ab}
			\Upsilon^{(ab)}_{\textrm{interf.}}
			=&
			- z_0 z_1
			\left[ z_1(z_0 + z_2) + z_0(z_1 + z_2) \right]
			\left[ z_0(z_0 + z_2) + z_1(z_1 + z_2) \right]
			\nonumber\\
			&\times \!
			\left[
				\left( \xt_{\conj{0} + \conj{2}; \conj{1}} \cdot \xt_{0;1+2} \right)
				\frac{(\xt_{\conj{2} \conj{0}} \cdot \xt_{{2} {1}})}{\xt_{\conj{2} \conj{0}}^2 \xt_{{2} {1}}^2}
				+
				\left( \xt_{\conj{0}; \conj{1} + \conj{2}} \cdot \xt_{0+2;1} \right)
				\frac{(\xt_{\conj{2} \conj{1}} \cdot \xt_{{2} {0}})}{\xt_{\conj{2} \conj{1}}^2 \xt_{{2} {0}}^2}
			\right]
			\nonumber\\
			&
			+ z_0 z_1 z_2 (z_0-z_1)^2
			\nonumber\\
			& \times \!
			\left[
				\left( \xt_{\conj{0} + \conj{2}; \conj{1}} \! \wedge \! \xt_{0;1+2} \right)
				\frac{(\xt_{\conj{2} \conj{0}} \! \wedge \! \xt_{{2} {1}})}{\xt_{\conj{2} \conj{0}}^2 \xt_{{2} {1}}^2}
				+
				\left( \xt_{\conj{0}; \conj{1} + \conj{2}} \! \wedge \! \xt_{0+2;1} \right)
				\frac{(\xt_{\conj{2} \conj{1}} \! \wedge \! \xt_{{2} {0}})}{\xt_{\conj{2} \conj{1}}^2 \xt_{{2} {0}}^2}
			\right] \!.
		\end{align}
		The wedge products in the interference contribution can be expanded as
		\begin{align}
			\left( \xt_{\conj{0} + \conj{2}; \conj{1}} \right.&\left.\! \wedge \, \xt_{0;1+2} \right)
			\frac{(\xt_{\conj{2} \conj{0}} \! \wedge \! \xt_{{2} {1}})}{\xt_{\conj{2} \conj{0}}^2 \xt_{{2} {1}}^2}
			+
			\left( \xt_{\conj{0}; \conj{1} + \conj{2}} \! \wedge \! \xt_{0+2;1} \right)
			\frac{(\xt_{\conj{2} \conj{1}} \! \wedge \! \xt_{{2} {0}})}{\xt_{\conj{2} \conj{1}}^2 \xt_{{2} {0}}^2}
			\nonumber\\
			=&
			\left(
				\frac{z_0}{z_0+z_2} \xt_{\conj{2} \conj{0}} \! \wedge \! \xt_{{2} {0}}
				+ \frac{z_1}{z_1+z_2} \xt_{\conj{2} \conj{1}} \! \wedge \! \xt_{{2} {1}}
			\right)
			\left(
				\frac{(\xt_{\conj{2} \conj{0}} \! \wedge \! \xt_{{2} {1}})}{\xt_{\conj{2} \conj{0}}^2 \xt_{{2} {1}}^2}
				+
				\frac{(\xt_{\conj{2} \conj{1}} \! \wedge \! \xt_{{2} {0}})}{\xt_{\conj{2} \conj{1}}^2 \xt_{{2} {0}}^2}
			\right)
			\nonumber\\
			&-
			\left(
				\frac{z_0 z_1}{(z_0+z_2)(z_1+z_2)} (\xt_{\conj{2} \conj{0}} \! \wedge \! \xt_{{2} {1}})
				+ (\xt_{\conj{2} \conj{1}} \! \wedge \! \xt_{{2} {0}})
			\right) \frac{(\xt_{\conj{2} \conj{0}} \! \wedge \! \xt_{{2} {1}})}{\xt_{\conj{2} \conj{0}}^2 \xt_{{2} {1}}^2}
			\nonumber\\
			&-
			\left(
			\frac{z_0 z_1}{(z_0+z_2)(z_1+z_2)} (\xt_{\conj{2} \conj{1}} \! \wedge \! \xt_{{2} {0}})
			+ (\xt_{\conj{2} \conj{0}} \! \wedge \! \xt_{{2} {1}})
			\right) \frac{(\xt_{\conj{2} \conj{1}} \! \wedge \! \xt_{{2} {0}})}{\xt_{\conj{2} \conj{1}}^2 \xt_{{2} {0}}^2}
			\\
			&
			\mkern-18mu
			\xrightarrow{\conj{0},\conj{1},\conj{2} \to 0,1,2}
			\frac{2 z_2}{(z_0+z_2)(z_1+z_2)}
			\frac{(\xt_{{2} {0}} \! \wedge \! \xt_{{2} {1}})^2}{\xt_{{2} {0}}^2 \xt_{{2} {1}}^2},
		\end{align}
		where the correct correspondence to Ref.~\cite{Beuf:2017bpd} is seen in the DIS limit $\xt_{\conj{0}} \to \xt_0,\xt_{\conj{1}} \to \xt_1, \xt_{\conj{2}} \to \xt_2$.

		\subsection{The \texorpdfstring{$q \bar q g$}{qq̅g} diffractive structure functions at next-to-leading order}

			Now we have the key pieces to write the $q \bar q  g$ structure functions at NLO accuracy. Let us first define an auxiliary integral to encapsulate the integrations over the final state momenta:
			\begin{align}
				\label{eq:ddis-final-state-int}
				\ical_{\textrm{F.S.}} (\mx, t)
				\coloneqq
				&
				\int \frac{\ud^2 \pp_0}{(2\pi)^2}
				\int \frac{\ud^2 \pp_1}{(2\pi)^2}
				\int \frac{\ud^2 \pp_2}{(2\pi)^2}
				e^{i \xt_{\conz0} \pp_0}
				e^{i \xt_{\cono1} \pp_1}
				e^{i \xt_{\cont2} \pp_2}
				\nonumber
				\\*
				& \times
				\delta(\Deltat^2 \! - \! \abs{t})
				\delta \left(\frac{\pp_0^2}{z_0} \! + \! \frac{\pp_1^2}{z_1} \! + \! \frac{\pp_2^2}{z_2} \! - \! \Deltat^2 \! - \! M_X^2 \right).
			\end{align}
			With this and the squared wavefunctions~\eqref{eq:ddis-qqbarg-psiL^2}, \eqref{eq:ddis-qqbarg-psiT^2} we may write the NLO cross section~\eqref{eq:ddis-qqbarg-cs-nlo} into diffractive structure functions $F_L^D$ and $F_T^D$~\eqref{eq:diffractive-FL-FT}:
			\begin{align}
				\notag
				& F_{L, \, q \bar q  g}^{D(4) \, \textrm{NLO}} (\xbj, Q^2, \mx, t)
				=
				4\nc Q^2
				\as \cf
				\sum_f e_f^2
				\int_0^1 \frac{\ud z_0}{z_0}
				\int_0^1 \frac{\ud z_1}{z_1}
				\int_0^1 \frac{\ud z_2}{z_2}
				\nonumber
				\\
				& \qquad \times
				\delta(z_0 \! + \! z_1 \! + \! z_2 \! - \! 1)
				\int_{\xt_0} \int_{\xt_1} \int_{\xt_2} \int_{\cxt_0} \int_{\cxt_1} \int_{\cxt_2}
				\ical_{\textrm{F.S.}} (\mx, t)
				\nonumber
				\\
				& \qquad \times
				4 z_0 z_1 Q^2 \besk_0 \left( Q X_{012}\right) \besk_0 \left( Q X_{\conj{0} \conj{1} \conj{2}}\right)
				\Bigg\{
				\nonumber
				\\
				&\qquad 
				z_1^2
				\Bigg[
				\left(2 z_0 (z_0 + z_2) + z_2^2\right)
				\left(
				\frac{\xt_{20}}{\xt_{20}^2} \cdot
				\left( \frac{\xt_{\conj{2} \conj{0}}}{\xt_{\conj{2} \conj{0}}^2}
				- \half \frac{\xt_{\conj{2} \conj{1}}}{\xt_{\conj{2} \conj{1}}^2} \right)
				- \half \frac{\xt_{\conj{2} \conj{0}} \cdot \xt_{21}}{\xt_{\conj{2} \conj{0}}^2 \xt_{21}^2} \right)
				\nonumber
				\\
				&\qquad \qquad 
				+ \frac{z_2^2}{2}
				\left(
				\frac{\xt_{\conj{2} \conj{0}} \cdot \xt_{21}}{\xt_{\conj{2} \conj{0}}^2 \xt_{21}^2}
				+
				\frac{\xt_{{2} {0}} \cdot \xt_{\conj{2} \conj{1}}}{\xt_{{2} {0}}^2 \xt_{\conj{2} \conj{1}}^2}
				\right)
				\Bigg]
				\nonumber
				\\
				&\qquad 
				+	
				z_0^2
				\Bigg[
				\left(2 z_1 (z_1 + z_2) + z_2^2\right)
				\left(
				\frac{\xt_{21}}{\xt_{21}^2} \cdot
				\left( \frac{\xt_{\conj{2} \conj{1}}}{\xt_{\conj{2} \conj{1}}^2}
				- \half \frac{\xt_{\conj{2} \conj{0}}}{\xt_{\conj{2} \conj{0}}^2} \right)
				- \half \frac{\xt_{{2} {0}} \cdot \xt_{\conj{2} \conj{1}}}{\xt_{{2} {0}}^2 \xt_{\conj{2} \conj{1}}^2} \right)
				\nonumber
				\\
				&\qquad \qquad 
				+ \frac{z_2^2}{2}
				\left(
				\frac{\xt_{\conj{2} \conj{0}} \cdot \xt_{21}}{\xt_{\conj{2} \conj{0}}^2 \xt_{21}^2}
				+
				\frac{\xt_{{2} {0}} \cdot \xt_{\conj{2} \conj{1}}}{\xt_{{2} {0}}^2 \xt_{\conj{2} \conj{1}}^2}
				\right)
				\Bigg]
				\Bigg\}
				\left[ 1 - S_{\contrip}^{(3)\dagger} \right] \left[ 1 - S_{012}^{(3)} \right]
				\label{eq:ddis-FL-qqbarg-nlo}
			\end{align}
			for the longitudinal, and
			\begin{align}
				\notag
				& F_{T, \, q \bar q  g}^{D(4) \, \textrm{NLO}} (\xbj, Q^2, \mx, t)
				=
				2 \nc Q^2
				\as \cf
				\sum_f e_f^2
				\int_0^1 \frac{\ud z_0}{z_0}
				\int_0^1 \frac{\ud z_1}{z_1}
				\int_0^1 \frac{\ud z_2}{z_2}
				\nonumber
				\\
				& \qquad \times
				\delta(z_0 \! + \! z_1 \! + \! z_2 \! - \! 1)
				\int_{\xt_0} \int_{\xt_1} \int_{\xt_2} \int_{\cxt_0} \int_{\cxt_1} \int_{\cxt_2}
				\ical_{\textrm{F.S.}} (\mx, t)
				\nonumber
				\\
				& \qquad \times
				z_0 z_1 \frac{Q^2}{X_{012} X_{\conj{0}\conj{1}\conj{2} }}
				\besk_1\left(QX_{012}\right) \besk_1\left(Q X_{\conj{0}\conj{1}\conj{2} }\right)
				\nonumber
				\\
				& \qquad \times
				\Big\lbrace
				\Upsilon^{(|a|^2)}_{\textrm{reg.}} + \Upsilon^{(|b|^2)}_{\textrm{reg.}} 
				+ \Upsilon^{(a')}_{\textrm{inst.}} + \Upsilon^{(b')}_{\textrm{inst.}} + \Upsilon^{(ab)}_{\textrm{interf.}}
				\Big\rbrace
				\left[ 1 - S_{\contrip}^{(3)\dagger} \right] \left[ 1 - S_{012}^{(3)} \right]
				\label{eq:ddis-FT-qqbarg-nlo}
			\end{align}
			for the transverse polarization, where the $\Upsilon$ terms are defined in Eqs.~\eqref{eq:ddis-qqbarg-nlo-upsilon-a^2}--\eqref{eq:ddis-qqbarg-nlo-upsilon-ab}. The dipole scattering amplitudes are understood to be evaluated at the rapidity:
			\vspace{-1ex}
			\begin{align}
				S_{012}^{(3)} & \equiv \left\langle S_{012}^{(3)} \right\rangle_{Y_2^{+}},
				\\
				\Ytwoplus & = \log \left( z_2 \frac{x_0 Q^2}{\xbj Q_0^2}\right)
			\end{align}
			analogously to NLO DIS~\cite{oma3,Beuf:2017bpd}.
			Both NLO accuracy $q \bar q g$ contributions~\eqref{eq:ddis-FL-qqbarg-nlo}, \eqref{eq:ddis-FT-qqbarg-nlo} to the structure functions in the form presented here are new and previously unpublished. Contrasting to the previous results known in the literature, $F_{L, \, q \bar q g}^D$ has not been known in any approximation, and the $F_{T, \, q \bar q g}^D$ calculated above supersedes the large-$Q^2$ limit approximation Eq.~\eqref{eq:ddis-ft-wusthoff}. Some analytical work is left to be done in future
			work however: the final state integrals~\eqref{eq:ddis-final-state-int} need to be performed, and ideally the transverse squared wavefunction~\eqref{eq:ddis-qqbarg-psiT^2} would be symmetrized in the quark-antiquark exchange, as was done for NLO DIS in Ref.~\cite{Beuf:2017bpd}. Though we do note that the symmetrization of~\eqref{eq:ddis-FL-qqbarg-nlo} is done and the two terms in the curly braces could be combined using the symmetry of the phase space, which would combine the terms into one and produce an overall factor of two. Further work is also to see how the result Eq.~\eqref{eq:ddis-FT-qqbarg-nlo} reduces to the large-$Q^2$ limit result Eq.~\eqref{eq:ddis-ft-wusthoff}, and how the adjoint dipole structure emerges.

%
%
%
%
%
%

\chapter{Conclusions and outlook}
	\label{ch:conclusions}

The work presented in this thesis builds on a decade of next-to-leading order Color Glass Condensate theory progress. It culminates in the state-of-the-art accuracy theory calculation and data comparison of the deep inelastic scattering cross sections from the CGC effective field theory. This paves the way towards accurate understanding of gluon saturation in QCD, and high accuracy theory calculations will be needed in the forthcoming precision small-$x$ era to be kicked off by the Electron-Ion Collider in early 2030s.

In Article~\cite{oma2}, new light-front perturbation theory tools are developed to facilitate the calculations of loop contributions which are present at next-to-leading order and beyond in the perturbation theory. The developed formalism is built on the four-dimensional helicity scheme of dimensional regularization, and the resulting calculation rules should be automatable for computational analytic calculation of observables at NLO accuracy and beyond. These tools are used to calculate the next-to-leading order DIS cross sections in the CGC formalism, which were verified to agree with the known results from the literature~\cite{Beuf:2016wdz,Beuf:2017bpd}.

The Articles \cite{oma1} and \cite{oma3} work towards the first numerical evaluation and data comparison of the next-to-leading order DIS cross sections to HERA data. First in Article~\cite{oma1} we show that at NLO the kinematics of the scattering are intimately related to the factorization of the large soft gluon logarithm into the Balitsky-Kovchegov renormalization group evolution. This brought the perturbative calculation of the cross sections under control and reasonable NLO corrections were seen, which would make comparisons between theory and data possible in the future.

Article~\cite{oma3} proceeds to combine the NLO DIS impact factors evaluated in the first article with enhanced BK equations known in the literature, which include in their prescriptions some of the most important beyond leading order corrections to the BK equation. 
Together these yield the state-of-the-art theory accuracy evaluation of the DIS cross sections calculated in the CGC formalism. We fit the initial shape of the dipole amplitudes using these NLO cross sections to the combined HERA data and found excellent description of the data. We found that the three different resummation approaches to the BK evolution described the available data comparably, and that new data from future experiments might be able to discern between the prescriptions.
The determined dipole amplitudes can be used to calculate predictions of other observables, the data are available at~\cite{oma3:zenodo}. As the first application, the dipole amplitudes have been used as an ingredient to calculate exclusive production of longitudinally polarized heavy vector mesons in next-to-leading order accuracy~\cite{Mantysaari:2021ryb}. The results of Article~\cite{oma3} are discussed in more detail in Sec.~\ref{sec:summary-oma3}, and the theoretical uncertainties of the results are discussed in Sec.~\ref{sec:fits-uncert}.

Some new and unpublished results are presented in this thesis as well. We derive a new form for the NLO DIS loop contribution in Sec.~\ref{sec:dip-z2}, which makes it possible to evaluate the dipole amplitudes of the NLO contributions at the same consistent rapidity scale, which was not possible previously as discussed in~\cite{oma1, oma3}. While this distinction is a beyond NLO effect, it could be numerically important for phenomenology.

In Sec.~\ref{sec:assess-nlobk} we estimate the impact of the NLO BK evolution on the fits of Article~\cite{oma3} by computing NLO BK evolved dipole amplitudes using the initial conditions determined in~\cite{oma3}. These NLO BK evolved dipole amplitudes are used to compute reduced cross sections which are compared to HERA data. The enhanced BK equations are found to approximate the NLO BK equation reasonably well in this simple data comparison, which is promising for the prospect of a full NLO+NLL accuracy fit.

Lastly, we calculate analytically for the first time the tree-level NLO contribution to the diffractive DIS structure functions for both longitudinally and transversely polarized virtual photon in Sec.~\ref{sec:ddis-nlo}. However, final state emissions of gluons were not included and some finalization work is left to be done, as discussed in Sec.~\ref{sec:ddis-nlo-qqbarg}. This $q \bar q g$ contribution has previously been known only for the transversely polarized photon and in leading $\log(Q^2)$ accuracy valid at large $Q^2$ --- this is discussed in Ch.~\ref{ch:ddis}.
Once the full NLO accuracy diffractive DIS cross sections become available, they will provide a key opportunity to study saturation and test the universality of the dipole amplitude.

The CGC theory field is progressing with strides towards next-to-leading order accuracy in multiple fronts.
NLO accuracy calculations of scattering processes are becoming available, with the longitudinal NLO DIS cross section for massive quarks being available~\cite{Beuf:2021qqa}, and the calculation of the transverse case is ongoing. Other processes are advancing to NLO as well: the single inclusive hadron production~\cite{Ducloue:2016shw, Iancu:2016vyg, Ducloue:2017mpb, Ducloue:2017dit, Liu:2019iml, Liu:2020mpy, Chirilli:2011km, Chirilli:2012jd, Altinoluk:2014eka, Watanabe:2015tja, Stasto:2014sea, Kang:2014lha}, exclusive light~\cite{Boussarie:2016bkq} and heavy vector meson production~\cite{Mantysaari:2021ryb}, dijets in $pA$ collisions~\cite{Iancu:2020mos} and dijets in DIS~\cite{Roy:2019hwr, Caucal:2021ent}.
Improvements to the description of the scattering process off the color-field are being studied as well. The NLO BK equation has been derived in the target momentum fraction prescription~\cite{Ducloue:2019ezk}, the behavior of which will be important to compare to the projectile momentum fraction picture equation~\cite{Balitsky:2008zza}. Finite-$\nc$ corrections to the NLO BK equation have been calculated~\cite{Lappi:2020srm}, and corrections to the eikonal approximation are being calculated for the gluon and quark propagators in next-to-eikonal accuracy~\cite{Altinoluk:2014oxa, Altinoluk:2015gia, Altinoluk:2020oyd, Altinoluk:2021lvu}.
The NLO DIS cross sections for massive quarks will be of great interest from the perspective of this thesis, since they will make possible to account for the contribution of the charm quark in the HERA data, which will improve the accuracy of the analysis. Overall, the general progress of the field towards NLO calculations of observables is bringing about the era of precision saturation phenomenology.
	
	\printbibliography[notcategory=paperit,title=References,heading=bibintoc]
	
\end{document}